\documentclass[12pt]{article}
\usepackage{amsmath,amssymb} 
\usepackage{graphicx}
\usepackage{float}
\usepackage{subcaption}   

\DeclareMathOperator{\sech}{sech}
\usepackage{amsfonts}
\usepackage{enumerate}
\usepackage{ytableau}
\usepackage{hyperref}
\numberwithin{equation}{section}
\usepackage{cite}
\usepackage{bm}
\usepackage{xcolor}
\usepackage{tikz}
\usepackage{blkarray}  
\begin{document}

\def\eqa{\begin{eqnarray}}
\def\eqae{\end{eqnarray}}
\def\be{\begin{equation}}
\def\ee{\end{equation}}
\def\bea{\begin{eqnarray}}
\def\eea{\end{eqnarray}}
\def\ba{\begin{array}}
\def\ea{\end{array}}
\def\bd{\begin{displaymath}}
\def\ed{\end{displaymath}}
\def\eg{{\it e.g.~}}
\def\ie{{\it i.e.~}}
\def\Tr{{\rm Tr}}
\def\tr{{\rm tr}}
\def\>{\rangle}
\def\<{\langle}
\def\a{\alpha}
\def\b{\beta}
\def\c{\chi}
\def\e{\epsilon}
\def\f{\phi}
\def\vf{\varphi}
\def\tvf{\tilde{\varphi}}
\def\g{\gamma}
\def\h{\eta}
\def\j{\psi}
\def\k{\kappa}
\def\l{\lambda}
\def\m{\mu}
\def\n{\nu}
\def\w{\omega}
\def\p{\pi}
\def\q{\theta}
\def\r{\rho}
\def\s{\sigma}
\def\t{\tau}
\def\u{\upsilon}
\def\x{\xi}
\def\z{\zeta}
\def\D{\Delta}
\def\F{\Phi}
\def\G{\Gamma}
\def\J{\Psi}
\def\L{\Lambda}
\def\W{\Omega}
\def\P{\Pi}
\def\Q{\Theta}
\def\S{\Sigma}
\def\U{\Upsilon}
\def\X{\Xi}
\def\nab{\nabla}
\def\pa{\partial}
\newcommand{\lra}{\leftrightarrow}

\newcommand{\bc}{{\mathbb{C}}}
\newcommand{\br}{{\mathbb{R}}}
\newcommand{\bz}{{\mathbb{Z}}}
\newcommand{\bp}{{\mathbb{P}}}
\newcommand{\dd}{\mathrm{d}}
\newcommand{\ii}{\mathrm{i}}
\newcommand{\SL}{\mathrm{SL}}
\newcommand{\SU}{\mathrm{SU}}
\newcommand{\SO}{\mathrm{SO}}
\newcommand{\cO}{\mathcal O}
\def\nn{\nonumber \\}
\newcommand{\AdS}{\mathrm{AdS}}
\newcommand{\red}{\textcolor[RGB]{255,0,0}}
\newcommand{\blue}{\textcolor[RGB]{0,0,255}}
\newcommand{\green}{\textcolor[RGB]{0,255,0}}
\newcommand{\cyan}{\textcolor[RGB]{0,255,255}}
\newcommand{\magenta}{\textcolor[RGB]{255,0,255}}
\newcommand{\yellow}{\textcolor[RGB]{255,255,0}}
\newcommand{\sky}{\textcolor[RGB]{135, 206, 235}}
\newcommand{\orange}{\textcolor[RGB]{255, 127, 0}}

\title{\textbf{Holographic Spread Complexity of a Probe Particle with General Initial Data}}
\vspace{14mm}
\author{Zekun Qian, Zhehan Li and Jia Tian 
\footnote{wukongjiaozi@ucas.ac.cn}
}
\date{}
\maketitle

\begin{center}
	{\it State Key Laboratory of Quantum Optics and Quantum Optics Devices, Institute of Theoretical Physics, Shanxi University, Taiyuan 030006, P.~R.~China\\
	}
\vspace{10mm}
\end{center}

\makeatletter
\def\blfootnote{\xdef\@thefnmark{}\@footnotetext}
\makeatother

\begin{abstract}
We investigate the holographic spread complexity of a massive particle probing global AdS$_3$ with general initial data. By constructing an adapted Hamiltonian, we map each bulk geodesic to a boundary Krylov dynamical pair, whose dynamics are governed solely by the conserved chiral charges. We find that, for generic initial data, the original identification $\dot{C}\sim p_\rho$ fails; instead, both the complexity and its growth rate depend nonlinearly on the bulk Noether charges. We systematically analyze this nonlinear encoding via several perturbative expansions. We further investigate two refined variants: the symmetry-resolved spread complexity, which restores the bulk radial \mbox{$\pi$}-periodicity, and the chiral spread complexity, which admits an exact closed form approaching the standard complexity at small angular momentum. Finally, by directly quantizing an effective AdS\mbox{$_2$} theory, we obtain a new holographic interpretation of spread complexity in terms of the lapse function.
\end{abstract}
\baselineskip 18pt
\newpage

\tableofcontents

\section{Introduction}
Quantum complexity measures the minimum resources required to prepare, simulate, or distinguish quantum states. It therefore probes intrinsic aspects of quantum dynamics that may not be fully captured by correlation measures such as entanglement entropy. In the context of the AdS/CFT correspondence~\cite{Maldacena:1997re,Gubser:1998bc,Witten:1998qj}, holographic complexity has emerged as an important probe, offering a bridge between the microscopic growth of quantum structure and bulk gravitational dynamics. Notably, the black hole interior can continue to grow after the entanglement entropy of a finite boundary interval has saturated~\cite{Hartman:2013qma}. The growth of quantum complexity has been proposed as a boundary description of this continued interior growth~\cite{Susskind:2014rva}. Numerous holographic complexity proposals have been put forward~\cite{Susskind:2014rva,Brown:2015bva,Couch:2016exn,Belin:2021bga,Belin:2022xmt}. In particular, the ``complexity=anything'' (CAny) proposal~\cite{Belin:2021bga,Belin:2022xmt} suggests that an infinite class of classical gravitational observables can serve as holographic duals to complexity, as they exhibit late-time linear growth in the thermofield double state and the switchback effect under shock-wave perturbations. In contrast, within the regime of probe dynamics, the growth of boundary complexity admits a more intuitive interpretation as the momentum of objects falling into the bulk~\cite{Susskind:2014jwa,Susskind:2018tei,Susskind:2019ddc,Brown:2018kvn,Susskind:2020gnl,Magan:2018nmu,Barbon:2020uux,Barbon:2020olv,Barbon:2019tuq,Lin:2019kpf,Lin:2019qwu,Ageev:2018msv,Ageev:2018nye}. On the field theory side, while various definitions of quantum complexity exist, practical calculations demand a measure that is both unambiguously defined and computationally tractable. Moreover, rather than relying on qualitative dynamical comparisons, one aims for a precise, quantitative matching between boundary quantum complexity and bulk gravitational observables.

Krylov complexity for operators~\cite{Parker:2018yvk} and spread complexity for states~\cite{Balasubramanian:2022tpr} provide such a framework. Starting from a Hamiltonian $H$ and a seed state or operator, the Lanczos algorithm constructs the minimal Krylov subspace explored by the time evolution, mapping the dynamics onto an effective hopping problem on a
semi-infinite chain. The entire problem is determined by the survival amplitude and its associated spectral measure. This also makes the tools of orthogonal polynomials and spectral analysis directly available~\cite{Muck:2022xfc,Balasubramanian:2025xkj,Caputa:2021sib,Lunt:2025dcc,Murugan:2026rfa,Muck:2026top,Qu:2025lanczos}. The Krylov framework has developed into a broad research program in quantum dynamics, with recent reviews systematizing the methods of Krylov space and their applications across quantum information, quantum field theory, and gravity~\cite{Nandy:2024evd,Rabinovici:2025otw,Baiguera:2025dkc,Jeong:2026gdc}.

Within holography, concrete progress has primarily been achieved along two main avenues. The first explicit bulk realization of the Krylov complexity was established~\cite{Rabinovici:2023yex} within the DSSYK/JT setting~\cite{Berkooz:2018jqr,Berkooz:2018qkz,Garcia-Garcia:2017pzl,Jackiw:1984je,Teitelboim:1983ux}, where spread complexity was shown~\cite{Lin:2022rbf,Rabinovici:2023yex,Jian:2020qpp,Ambrosini:2024sre,Aguilar-Gutierrez:2025pqp} to match the renormalized length of the wormhole connecting the two boundaries. Further developments can be found in~\cite{Ambrosini:2025hvo,Heller:2024ldz,Das:2024tnw,Balasubramanian:2024lqk,Xu:2024gfm,Alfinito:2026cky,Aguilar-Gutierrez:2026nmd,Aguilar-Gutierrez:2026ogo,Fu:2025kkh,Heller:2025ddj,Miyaji:2025yvm,Jeong:2026iac}. The key advantage here is that lower-dimensional gravity theory admits a quantum description, such as the DSSYK model, allowing spread complexity to be computed explicitly. However, an analogous microscopic description for higher-dimensional gravitational theories is not available, making a direct generalization challenging. To proceed, one possible strategy is to take the semiclassical gravitational partition function as an input for computing the spread complexity, and subsequently match the results against a bulk ansatz numerically~\cite{Balasubramanian:2026azk,Bhattacharya:2026ybq}. 

A second avenue focuses on probe dynamics. It was first explicitly demonstrated in~\cite{Caputa:2024sux} that the growth rate of spread complexity can be identified with the proper radial momentum of a free particle. The scope and validity of this identification were further scrutinized in~\cite{Fan:2024iop,He:2024pox,Li:2025fqz,Li:2026pdh}. In particular, in our previous work~\cite{Li:2025fqz,Li:2026pdh}, we proposed that spread complexity and its growth rate are more naturally related to the bulk Noether charges. Furthermore, exact multi-time Krylov correlators~\cite{Alfinito:2026vah} or semiclassical Lanczos data~\cite{Qu:2026dmv} have also been found to correspond to geometric observables. This probe approach has since been extended to $\mathcal N=4$ SYM, the Anabal\'on--Ross soliton, confining backgrounds, quiver theories, and charged or extended objects~\cite{Fatemiabhari:2025cyy,Anabalon:2021tua,Fatemiabhari:2025usn,Fatemiabhari:2026rob,Nunez:2026vhw,Fatemiabhari:2025poq,Fatemiabhari:2026goj,Nastase:2026lhz,Fatemiabhari:2026eem,Chatzis:2026ekd,Chatzis:2026oou}. The effect of confinement on Krylov state complexity has also been studied in the Ising chain with a longitudinal field~\cite{Jiang:2025wpj}. Related investigations have also explored renormalization-group flows, deformed backgrounds, Coulomb-branch trajectories, Lin--Maldacena geometry and BMN models, giant gravitons, orbifold SCFTs, and Lifshitz or hyperscaling-violating spacetimes~\cite{Nunez:2026rgflow,Roychowdhury:2026eta,Zoakos:2026coulomb,Roychowdhury:2026lin,Roychowdhury:2026bmnstate,Roychowdhury:2026planewave,Roychowdhury:2026bmnspectral,Graef:2026probe,Baume:2026chaos,Fadafan:2026lifshitz}.

Despite this progress, several conceptual and technical limitations persist in the holographic description of spread complexity. First, the proper-momentum dictionary is not a statement about a geodesic alone; rather, it depends crucially on the choice of dual quantum state~\cite{Li:2026pdh}. Merely computing the radial momentum of a bulk probe is insufficient to claim a holographic complexity; one must also specify the precise boundary state under consideration. Second, evaluating spread complexity for generic states is challenging; most existing analyses are restricted to early-time expansions, which only capture limited dynamical information. The only solvable construction in~\cite{Caputa:2024sux} is confined to dynamics on an AdS$_2$ slice governed by a single rank-1 $\SU(1,1)$ conformal symmetry, whereas closed-form field-theory results for more general states remain scarce. Third, the presence of conserved charges introduces additional structure. Cyclic motion along isometric directions can nontrivially contribute to spread complexity, and projecting onto fixed-charge sectors and constructing the Krylov basis do not commute~\cite{Nastase:2026lhz,Chatzis:2026ekd,Chatzis:2026oou,Caputa:2025ozd,Caputa:2025mii,Das:2026gko}. Addressing these issues requires a controlled setting in which general initial data, explicit spectral control, and charge-resolved constructions can be analyzed together.

In this paper, we study a simple setting: a massive particle probing global AdS$_3$ with general initial position and velocities. On the bulk side, the initial data uniquely specify a timelike geodesic. On the boundary side, they correspond to a product of two chiral Perelomov coherent states~\cite{Perelomov:1986uhd}, thereby extending the analysis beyond the single $\SU(1,1)$ sector.  

Equipped with an exact mapping between the geodesic initial data and the CFT state parameters, we are able to perform a systematic and precise holographic matching. For example, we explicitly derive how a non-vanishing initial radial momentum modifies the original proper-momentum dictionary. The harmonic content of the first small-$\jmath$ correction rules out a linear relation between standard spread complexity and the bulk Noether charges at leading semiclassical order. Nevertheless, analytic control remains possible. We introduce three perturbative regimes that together cover a wide range of physical configurations. Our perturbative results indicate that spread complexity is generically encoded as a nonlinear function of the bulk Noether charges.

The presence of conserved angular momentum further motivates refined Krylov constructions. By decomposing the seed state into fixed-charge sectors, we compute the corresponding symmetry-resolved spread complexity~\cite{Caputa:2025ozd,Caputa:2025mii}. The symmetry-resolved spread complexities are $\pi$-periodic, in contrast to the $2\pi$-periodic standard spread complexity, and thus reflect the radial periodicity of the bulk motion. Furthermore, we introduce a ``chiral spread complexity'' defined by summing the independent complexities of the two chiral $\SU(1,1)$ sectors. This quantity admits a simple, exact closed-form expression in terms of the particle's initial data and bulk Noether observables. 

Finally, we take a bulk-oriented route by quantizing the radial theory in a fixed-angular-momentum sector and evaluating the spread complexity of the associated effective AdS$_2$ coherent state. Within this reduced framework, we uncover a novel holographic interpretation of spread complexity in terms of the lapse function. This provides a concrete demonstration that choosing different initial states yields distinct holographic dictionaries for spread complexity.

The remainder of the paper is organized as follows. Section~\ref{sec:geodesics} solves the classical geodesic problem with general initial data. Section~\ref{sec:CFT} constructs the dual coherent-state description and its adapted Hamiltonian. Section~\ref{sec:the spread complexity} analyzes the standard spread complexity, detailing its early-time, small-$\jmath$, small-$|z|$, cold--hot, and semiclassical limits. Section~\ref{sec:res} investigates the symmetry-resolved spread complexity, while Section~\ref{sec:chiral} introduces the chiral construction. Section~\ref{sec:effectAdS2} develops the fixed-charge effective AdS$_2$ description and establishes its lapse-function interpretation. We conclude with a summary in Section~\ref{sec:summary}. Technical details regarding the Noether-charge dictionary, orthogonal polynomials, perturbation theory, and spectral transformations are collected in the appendices. 
\section{Free-particle geodesics in global AdS$_3$}
\label{sec:geodesics}
We consider a massive particle probe moving in global AdS$_3$, whose metric is
\begin{equation}
  \dd s^2
  =
  -\cosh^2\rho\,\dd t^2
  +\dd\rho^2
  +\sinh^2\rho\,\dd\phi^2 .
  \label{eq:global-metric}
\end{equation}
It is convenient to describe particle worldlines using the embedding coordinates
\begin{align}
   X^{-1}&=\cosh\rho\cos t,
  &
  X^0&=\cosh\rho\sin t,
  \nonumber\\
  X^1&=\sinh\rho\cos\phi,
  &
  X^2&=\sinh\rho\sin\phi ,
\end{align}
which satisfy
\begin{equation}
 \eta_{AB}X^A X^B=-1,\quad  \eta_{AB}=\text{diag}(-1,-1,1,1)\,.
\end{equation}
Let $\tau$ denote proper time such that the timelike normalization condition is
\begin{equation}
  \cosh^2\rho\left(\frac{\dd t}{\dd\tau}\right)^2
  -\left(\frac{\dd\rho}{\dd\tau}\right)^2
  -\sinh^2\rho\left(\frac{\dd\phi}{\dd\tau}\right)^2
  =1\,.
  \label{eq:proper-time-normalization}
\end{equation}
In the embedding space, the geodesic equation takes the simple form
\begin{equation}
  \frac{\dd^2X^A}{\dd \tau^2}+X^A=0 .
\end{equation}
We fix the proper-time origin by setting $t(0)=0$. The initial data are then
\begin{equation}
  \rho(0)=\rho_0,
  \qquad
  \phi(0)=\phi_0,
  \qquad
  q_\rho=\left.\frac{\dd\rho}{\dd\tau}\right|_{\tau=0},
  \qquad
  q_\phi=\left.\sinh\rho\,\frac{\dd\phi}{\dd\tau}\right|_{\tau=0} .
  \label{eq:bulk-initial-data}
\end{equation}
These data specify the initial position and spatial velocity and determine the geodesic.
\begin{figure}[H]
  \centering
  \begin{minipage}[c]{0.43\textwidth}
    \centering
    \includegraphics[width=1\linewidth]{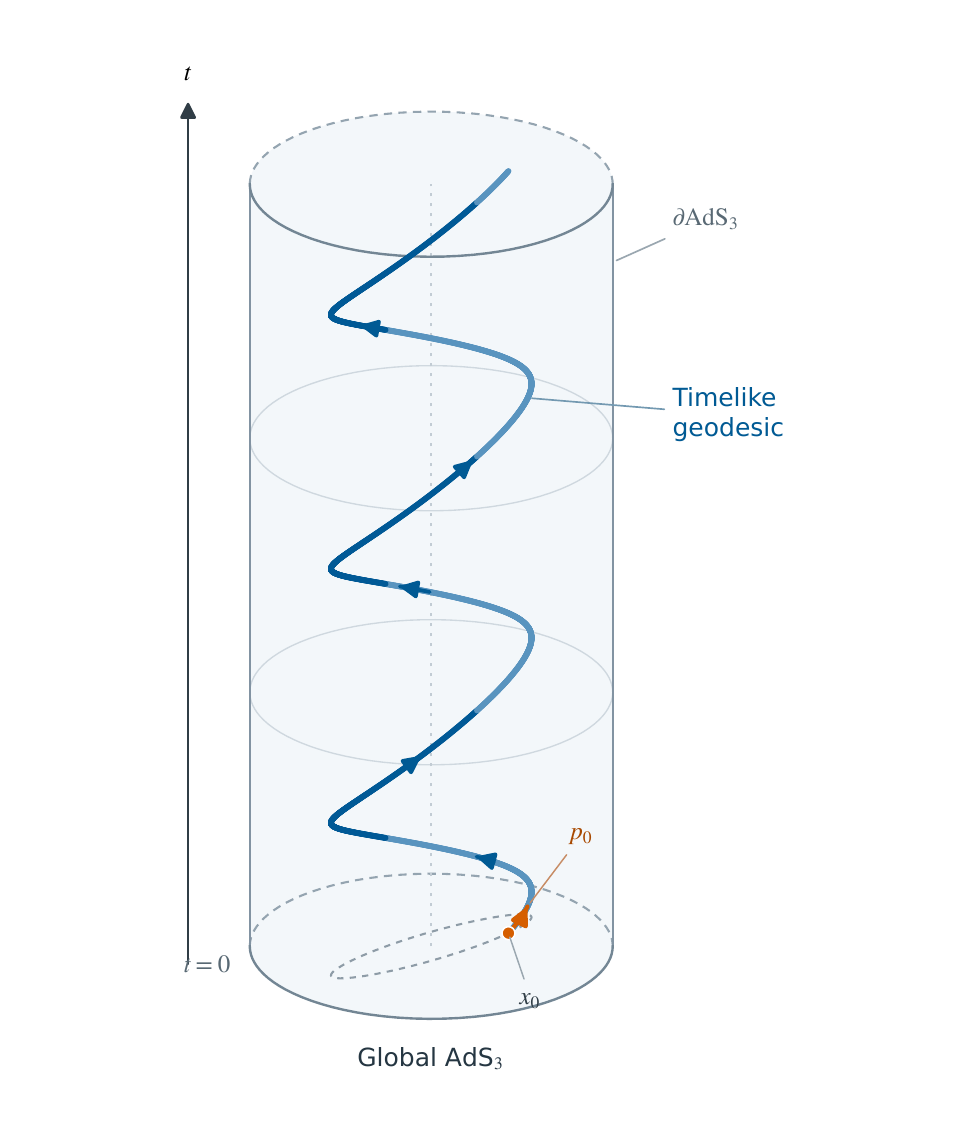}
  \end{minipage}\hfill
  \begin{minipage}[c]{0.53\textwidth}
\captionsetup{font=small,justification=raggedright,singlelinecheck=false}
    \caption{A generic timelike geodesic with radial oscillations and rotations in global $\AdS_3$, shown on a conformally compactified cylinder with time increasing upward. $x_0$ and $p_0$ mark the initial position and momentum.}
    \label{fig:global-geodesic}
  \end{minipage}
\end{figure}
The solution with the initial data \eqref{eq:bulk-initial-data} is
\begin{align}
  X^{-1}(\tau)&=c_{\rho_0}\cos \tau+q_\rho s_{\rho_0} \sin \tau,
  \\
  X^{0}(\tau)&=\gamma_0\sin \tau,
  \\
  X^1(\tau)+\ii X^2(\tau)
  &=
  e^{\ii\phi_0}
  \left[
    s_{\rho_0}\cos \tau+(c_{\rho_0} q_\rho+\ii q_\phi)\sin \tau
  \right],
\end{align}
where we introduce the abbreviations $s_{\rho_0}\equiv\sinh\rho_0,\,c_{\rho_0}\equiv\cosh\rho_0$ and define the Lorentz factor $\gamma_0=\sqrt{1+q_\rho^2+q_\phi^2}$. In global coordinates, the geodesics are therefore
\begin{align}
\sinh^2\rho(\tau)
  &=
  \left(s_{\rho_0}\cos \tau+c_{\rho_0} q_\rho\sin \tau\right)^2
  +q_\phi^2\sin^2\tau ,
  \\
  \tan\bigl(\phi(\tau)-\phi_0\bigr)
  &=
  \frac{q_\phi\sin \tau}
  {s_{\rho_0}\cos \tau+c_{\rho_0} q_\rho\sin \tau},
  \\
  \tan t(\tau)
  &=
  \frac{\gamma_0\sin \tau}
  {c_{\rho_0}\cos \tau+s_{\rho_0} q_\rho\sin \tau}.
\end{align}
Two special cases will be useful below. When $q_\phi=0$, the geodesic lies in an $\AdS_2$ subspace at fixed $\phi$. When $q_\rho=0$ and $q_\phi^2=s_{\rho_0}^2$, it is circular, with $\rho(\tau)=\rho_0$. The purely radial case was studied in~\cite{Caputa:2024sux,Li:2025fqz}, where the spread complexity growth rate has the same time dependence as the proper radial momentum. As we show below, this relation does not hold generically when radial or angular initial momentum is non-vanishing\footnote{This observation was also made in~\cite{Fatemiabhari:2026eem}.}.

We derive the particle's canonical momenta from the action
\begin{equation}
  S_{\mathrm{p}}=-m\int\dd t\,\sqrt{\cosh^2\rho-\dot\rho^{\,2}
  -\sinh^2\rho\,\dot\phi^{\,2}},
\end{equation}
which take the form
\begin{equation}
p_\rho=
m\frac{\dd\rho}{\dd\tau},
\qquad
p_\phi
=m\sinh^2\rho\,\frac{\dd\phi}{\dd\tau}.
\label{eq:coordinate-time-canonical-momenta}
\end{equation}
The aforementioned mismatch for $q_\phi \neq 0$ was recently addressed in~\cite{Chatzis:2026oou,Chatzis:2026ekd}, whose authors suggest that, when the action possesses a cyclic variable (namely $\phi$), the appropriate momentum to compare against the spread complexity growth rate is obtained via the Routhian reduction. The corresponding Routhian, capturing the effective radial AdS$_2$ dynamics, reads
\begin{equation}
\mathcal{R}(\rho,\dot{\rho})=-\sqrt{m^2+\frac{p_\phi^2}{\sinh^2\rho}}\sqrt{\cosh^2\rho-\dot{\rho}^2}.
\end{equation}
Differentiating $\mathcal{R}$ with respect to $\dot{\rho}$ recovers the momentum conjugate to the original radial coordinate $\rho$,
\begin{equation}
\frac{\partial \mathcal{R}}{\partial \dot{\rho}}=\frac{\sqrt{m^2+p_\phi^2/\sinh^2\rho}}{\sqrt{\cosh^2\rho-\dot{\rho}^2}}\,\dot{\rho}=m\frac{\dd\rho}{\dd\tau}=p_\rho,
\end{equation}
upon applying the proper-time normalization condition~\eqref{eq:proper-time-normalization}. The Routhian prescription of Refs.~\cite{Chatzis:2026oou,Chatzis:2026ekd} instead uses a radial coordinate defined from the reduced kinetic term. For the Routhian above, writing $M(\rho)=\sqrt{m^2+p_\phi^2/\sinh^2\rho}$ and $\dd y=M(\rho)\dd\rho$ gives the conjugate momentum $P_y=p_\rho/M(\rho)$. At nonzero $p_\phi$, the conversion factor depends on $\rho$. A direct comparison for rotating global AdS probes finds differences at higher orders in time~\cite{Fatemiabhari:2026eem}. Even though $p_\rho$ does not generally match the spread-complexity rate, in Section~\ref{sec:effectAdS2} we will directly quantize this effective $\AdS_2$ dynamics and establish a new holographic interpretation for the spread complexity defined within the effective $\AdS_2$ theory.

\section{The CFT description}
\label{sec:CFT}
The bulk initial data $(\rho_0,\phi_0,q_\rho,q_\phi)$ uniquely specify a corresponding initial state $|\psi_0\rangle$ in the dual CFT. 
To construct this state, we perform a Wick rotation $t\to-\ii t_E$, mapping the boundary cylinder to the complex plane via
\be
z=e^{t_E+\ii\phi},
\qquad
\bar z=e^{t_E-\ii\phi}.
\ee
Quantum states are then prepared via the Euclidean path integral on the unit disk.  It was proposed in~\cite{Caputa:2024sux} and further elaborated in~\cite{Li:2025fqz,Li:2026pdh} by directly quantizing the free particle theory in AdS$_2$ that a bulk particle in the semiclassical limit is dual to a CFT Perelomov coherent state\footnote{From the canonical quantization perspective, the bulk state of the quantized particle is likewise described by a coherent state.}~\cite{Perelomov:1986uhd}. Specifically, this state is created by inserting a scalar primary operator inside the unit disk:
\be 
|\psi_0\rangle=\frac{1}{\mathcal{N}}\cO(z)|\Omega\rangle,
\ee 
where $|\Omega\rangle$ denotes the CFT vacuum state and $\mathcal{N}$ is a normalization constant. Such locally excited states were previously introduced in holographic CFTs to model local quench protocols~\cite{Nozaki:2014hna,Caputa:2014vaa,Caputa:2026hxt}. 
When both $q_\rho$ and $q_\phi$ vanish, Euclidean path integral arguments show that the insertion point $z_0$ is related to the bulk initial position $\rho_0$ by~\cite{Li:2025fqz}
\be 
|z|=\tanh\frac{\rho_0}{2}. \label{eq:z0r0}
\ee 
Moreover, in this restricted case, the space of geodesics can be identified with the coset space $\SU(1,1)/\text{U}(1)$, which coincides with the manifold of states created by inserting a scalar primary operator in the unit disk:
\bea
\cO(z,\bar{z})|\Omega\rangle  \,  \leftrightarrow \, \text{geodesic} \, \leftrightarrow \, \text{coset } \SU(1,1)/\text{U}(1).
\eea
Here, the $\SU(1,1)$ group corresponds to the diagonal subgroup of the full $\SU(1,1)_+\times \SU(1,1)_{-}$ global conformal symmetry.
Now for general initial data, this correspondence naturally lifts to
\be 
\cO(z_+,z_-)|\Omega\rangle  \,  \leftrightarrow \, \text{geodesic} \, \leftrightarrow \, \text{coset } \frac{\SU(1,1)_+\times \SU(1,1)_-}{\text{U}(1)\times \text{U}(1)}. \label{eq:cosetp}
\ee 
The identification can be realized explicitly as follows. We start from a reference geodesic with $\rho_0=\phi_0=q_\rho=q_\phi=0$ (equivalent to $X^1+\ii X^2=0$), which is dual to the asymptotic lowest-weight state $\cO(0,0)|\Omega\rangle\equiv |h,0\rangle\otimes |\bar{h},0\rangle$ in radial quantization with $\bar h=h$ for a scalar primary. Acting with an isometry transformation $g_+\otimes g_-\in \SU(1,1)_+\times \SU(1,1)_-$ on the spacetime coordinates,
\bea  
 \mathbf X(t,\rho,\phi)
  =
  \begin{pmatrix}
    X^{-1}+\ii X^0 & X^1+\ii X^2\\
    X^1-\ii X^2 & X^{-1}-\ii X^0
  \end{pmatrix}
  \to g_+ \mathbf{X}g_-^{T},
\eea  
we obtain a new geodesic 
\begin{equation}
  \mathbf X_{\mathrm{geo}}(\tau)
  =
  g_+\mathbf X_0(\tau+\delta)g_-^{T},
  \qquad
  \mathbf X_0(\tau)=\begin{pmatrix}e^{\ii \tau}&0\\0&e^{-\ii \tau}\end{pmatrix}.
  \label{eq:section3-ads3-transformed-geodesic}
\end{equation}
Here, $\delta$ accounts for the residual time-translation freedom of the stabilizer $\mathrm{U}(1)\times\mathrm{U}(1)$, and is fixed by enforcing the convention $t(0)=0$.  
Parameterizing an $\SU(1,1)/\mathrm{U}(1)$ coset representative (with $\sigma=\pm$) as
\begin{equation}
  g_\sigma=
  \begin{pmatrix}
    \cosh(\rho_\sigma/2)
    &
    e^{-\ii t_\sigma}\sinh(\rho_\sigma/2)
    \\
    e^{\ii t_\sigma}\sinh(\rho_\sigma/2)
    &
    \cosh(\rho_\sigma/2)
  \end{pmatrix},
  \label{eq:coset}
\end{equation}
we find
\be
\mathbf X_{\mathrm{geo}}(\tau)=
\begin{pmatrix}
\begin{gathered}
\displaystyle
\cosh\frac{\rho_+}{2}\cosh\frac{\rho_-}{2}e^{\ii(\tau+\delta)}\\[-2pt]
\displaystyle {}+{}
\sinh\frac{\rho_+}{2}\sinh\frac{\rho_-}{2}
 e^{-\ii(t_++t_-+\tau+\delta)}
\end{gathered}
&
\begin{gathered}
\displaystyle
\cosh\frac{\rho_+}{2}\sinh\frac{\rho_-}{2}
 e^{\ii(t_-+\tau+\delta)}\\[-2pt]
\displaystyle
{}+{}
\sinh\frac{\rho_+}{2}\cosh\frac{\rho_-}{2}
 e^{-\ii(t_++\tau+\delta)}
\end{gathered}
\\[8pt]
\begin{gathered}
\displaystyle
\sinh\frac{\rho_+}{2}\cosh\frac{\rho_-}{2}
 e^{\ii(t_++\tau+\delta)}\\[-2pt]
\displaystyle
{}+{}
\cosh\frac{\rho_+}{2}\sinh\frac{\rho_-}{2}
 e^{-\ii(t_-+\tau+\delta)}
\end{gathered}
&
\begin{gathered}
\displaystyle
\cosh\frac{\rho_+}{2}\cosh\frac{\rho_-}{2}e^{-\ii(\tau+\delta)}\\[-2pt]
\displaystyle {}+{}
\sinh\frac{\rho_+}{2}\sinh\frac{\rho_-}{2}
 e^{\ii(t_++t_-+\tau+\delta)}
\end{gathered}
\end{pmatrix}.
\label{eq:ads3-transformed-geodesic-matrix}
\ee
Matching the entries of \eqref{eq:ads3-transformed-geodesic-matrix} against the initial-data solution yields the identification
\(\cosh\rho_\sigma=c_{\rho_0}\gamma_0-\sigma s_{\rho_0}q_\phi\) and the matching condition
\begin{equation}
     \sinh\rho_\sigma e^{\ii t_\sigma}
    =
    e^{-\ii\sigma\phi_0}
    \left(
      s_{\rho_0}\gamma_0-\sigma c_{\rho_0}q_\phi-\ii q_\rho
    \right).
  \label{eq:identification}
\end{equation}
Meanwhile, the bulk isometry induces a projective conformal transformation on the boundary coordinates,
which maps the reference state $\cO(0,0)|\Omega\rangle$ to
\be 
\cO(0,0)|\Omega\rangle \quad \to \quad |\psi_0\rangle=\cO(z_+,z_-)|\Omega\rangle,\quad z_\sigma=e^{-\ii t_\sigma}\tanh\frac{\rho_\sigma}{2}.
\ee 
In summary, a bulk geodesic characterized by the initial data $(\rho_0,\phi_0,q_\rho,q_\phi)$ is dual to the time-evolved CFT state
\be 
|\psi(t)\rangle=e^{-\ii t H}\cO(z_+,z_-)|\Omega\rangle\equiv e^{-\ii t H} |z_+\rangle\otimes|z_-\rangle, 
\ee 
subject to the matching conditions~\eqref{eq:identification}. In the semiclassical limit, $2h\simeq m$ by the matching in \eqref{eq:xi-from-energy}, and the unnormalized charges $mL^{\rm bulk}_{AB}$ scale linearly with $h$. 

As demonstrated in~\cite{Li:2025fqz,Li:2026pdh}, this holographic dictionary is further supported by matching the bulk Noether charges associated with the AdS Killing vectors against the expectation values of the corresponding CFT symmetry generators; the details are summarized in Appendix~\ref{app:embedding-charge-dictionary}.

\section{The spread complexity}
\label{sec:the spread complexity}
The definition of spread complexity requires only an initial (or seed) state $|\psi_0\rangle$ and the Hamiltonian $H$. Using the Lanczos algorithm, one constructs an orthogonal basis $\{|K_n\rangle\}$, often called the Krylov chain, spanning the Krylov subspace $\mathcal{K}=\operatorname{span}\{H^k|\psi_0\rangle \mid k=0,1,2,\dots\}$. In this basis, the Hamiltonian takes a tridiagonal form:
\be 
H|K_n\rangle=a_n|K_n\rangle+b_{n}|K_{n-1}\rangle+b_{n+1}|K_{n+1}\rangle,
\ee 
where the Lanczos coefficients $a_n$ and $b_n$ encode the complete dynamical information within the Krylov subspace.

For the time-evolved state $|\psi(t)\rangle=e^{-\ii H t}|\psi_0\rangle$, the transition amplitudes and the spread complexity are defined as
\bea 
\varphi_n(t)=\langle K_n|\psi(t)\rangle,\qquad C(t)=\sum_{n=0}^\infty n|\varphi_n(t)|^2.
\eea 
In the energy representation, the Krylov basis states $|K_n\rangle$ correspond to orthogonal polynomials of degree $n$ with respect to the spectral measure $\mu(E)=|\langle\psi_0|E\rangle|^2$. Consequently, the theory of orthogonal polynomials provides powerful analytical tools~\cite{Muck:2022xfc,Balasubramanian:2025xkj}, particularly when the system possesses symmetries: the Krylov basis is governed by classical or semiclassical orthogonal polynomials. For example, for a probe particle in global AdS$_2$, the dynamics enjoys an $\SU(1,1)$ symmetry; accordingly, the associated orthogonal polynomials are the Meixner polynomials, and the spectral measure follows a negative binomial ($\mathrm{NB}$) distribution. We summarize the relevant properties of the $\mathrm{NB}$ distribution and Meixner polynomials in Appendix~\ref{app:nb-meixner}, which will be used extensively in the subsequent analysis. 
 
In our AdS$_3$ setup, the symmetry group is $\SU(1,1)_+\times \SU(1,1)_-$. We denote the generators of each chiral factor $\SU(1,1)_\sigma$ by $L^{\sigma}_n,\, n=0,\pm 1$ which satisfy the commutation relations
\be 
[L_n^\sigma,L_m^\sigma]=(n-m)L_{n+m}^\sigma.
\ee 
Writing a group element as
\be 
g_\sigma=e^{\ii \tau \sum_n \epsilon_n L_n^\sigma}
\ee 
its action on a primary state is determined by
\be 
g_\sigma \cO(z_\sigma(0))|\Omega\rangle=g_\sigma \cO(z_\sigma(0))g_\sigma^{\dagger}|\Omega\rangle \propto \cO(z_\sigma(\tau))|\Omega\rangle,\qquad \frac{\dd z_\sigma(\tau)}{\dd\tau}=\ii \sum_n \epsilon_n z_\sigma^{n+1}.
\ee 
Thus, the coset representative \eqref{eq:coset} corresponds to the unitary operator
\be 
g_\sigma=e^{-\frac{\rho_\sigma}{2}(e^{\ii t_\sigma}L_{+1}^\sigma -e^{-\ii t_\sigma}L_{-1}^\sigma )}.
\ee 
The initial CFT state is a direct product of two generalized Perelomov coherent states:
\bea 
&&|\psi_0\rangle=|z_+\rangle\otimes|z_-\rangle,\\
&&|z_\sigma\rangle=g_\sigma |h,0\rangle=(1-|z_\sigma|^2)^h \sum_{n=0}^\infty z_\sigma^n \sqrt{\frac{(2h)_n}{n!}}|h,n\rangle .\label{eq:expansion}
\eea  
The global Hamiltonian and angular-momentum operators are given by
\begin{equation}
  H=L_0^{+}+L_0^{-},
  \qquad
  P=L_0^{+}-L_0^{-},
  \qquad
  [H,P]=0.
\end{equation}
With this orientation, the dictionary in Appendix~\ref{app:embedding-charge-dictionary} gives $\langle P\rangle/h=-2\jmath$, where $\jmath\equiv p_\phi/m$. Because the system has the conserved charge $P$, the full Hilbert space decomposes into a direct sum of fixed-$P$ sectors according to the eigenvalues of $P$. This structure naturally allows one to define the symmetry-resolved spread complexity~\cite{Caputa:2025ozd}, which we will investigate in a later section.

As shown in ~\cite{Li:2025fqz}, we can more conveniently construct the Krylov basis and evaluate the spread complexity by passing to an equivalent system via a unitary transformation. Specifically, the unitary operator $g_+^{-1}\otimes g_-^{-1}$ maps the seed state $|\psi_0\rangle$ to the reference lowest-weight state $|h,0\rangle\otimes|h,0\rangle$, while transforming $L_0^\sigma$ into $g_\sigma^{-1}L_0^\sigma g_\sigma$. In this adapted representation, the dynamical pair transforms as
\be 
(H,|\psi_0\rangle)\to (G,|h,0\rangle\otimes |h,0\rangle ),
\ee 
where $G=G_++G_-$ with
\begin{align}
G_\sigma
&=g_\sigma^{-1}L_0^\sigma g_\sigma
=\cosh\rho_\sigma\,L_0^\sigma
+\frac{\sinh\rho_\sigma}{2}
\left(
 e^{\ii t_\sigma}L_{+1}^\sigma
 +e^{-\ii t_\sigma}L_{-1}^\sigma
\right)\nonumber\\
&=(c_{\rho_0}\gamma_0-\sigma s_{\rho_0}q_\phi)L_0^\sigma\nonumber\\
&\quad+\frac{1}{2}e^{-\ii\sigma\phi_0}
\left(s_{\rho_0}\gamma_0-\sigma c_{\rho_0}q_\phi-\ii q_\rho\right)L_{+1}^\sigma+\frac{1}{2}e^{\ii\sigma\phi_0}
\left(s_{\rho_0}\gamma_0-\sigma c_{\rho_0}q_\phi+\ii q_\rho\right)L_{-1}^\sigma.
\label{eq:global-Gsigma-e-jmath}
\end{align}
In going to the second expression, we have used the matching conditions~\eqref{eq:identification} to express the coefficients in terms of the bulk initial data $(\rho_0,\phi_0,q_\rho,q_\phi)$. This unitary transformation preserves all moments of the Hamiltonian, $\langle h,0|\otimes \langle h,0|G^m|h,0\rangle\otimes|h,0\rangle=\langle \psi_0|H^m|\psi_0\rangle$.
Thus, the reference state $|h,0\rangle\otimes |h,0\rangle$ under $G$ shares the same spectral measure as $|\psi_0\rangle$ under $H$, leaving all Lanczos coefficients and the spread complexity invariant. One of the primary advantages of this representation is that all the bulk initial data are explicitly encoded within the transformed Hamiltonian $G$. In other words, there is a one-to-one correspondence between bulk geodesics and the operators $G$. Nevertheless, in contrast to the simpler $\SU(1,1)/\mathrm{U}(1)$ case, this transformation alone does not immediately yield a closed-form Krylov basis, due to the subtleties explained below. 

\begin{figure}[H]
 \centering
 \includegraphics[width=0.87\linewidth]{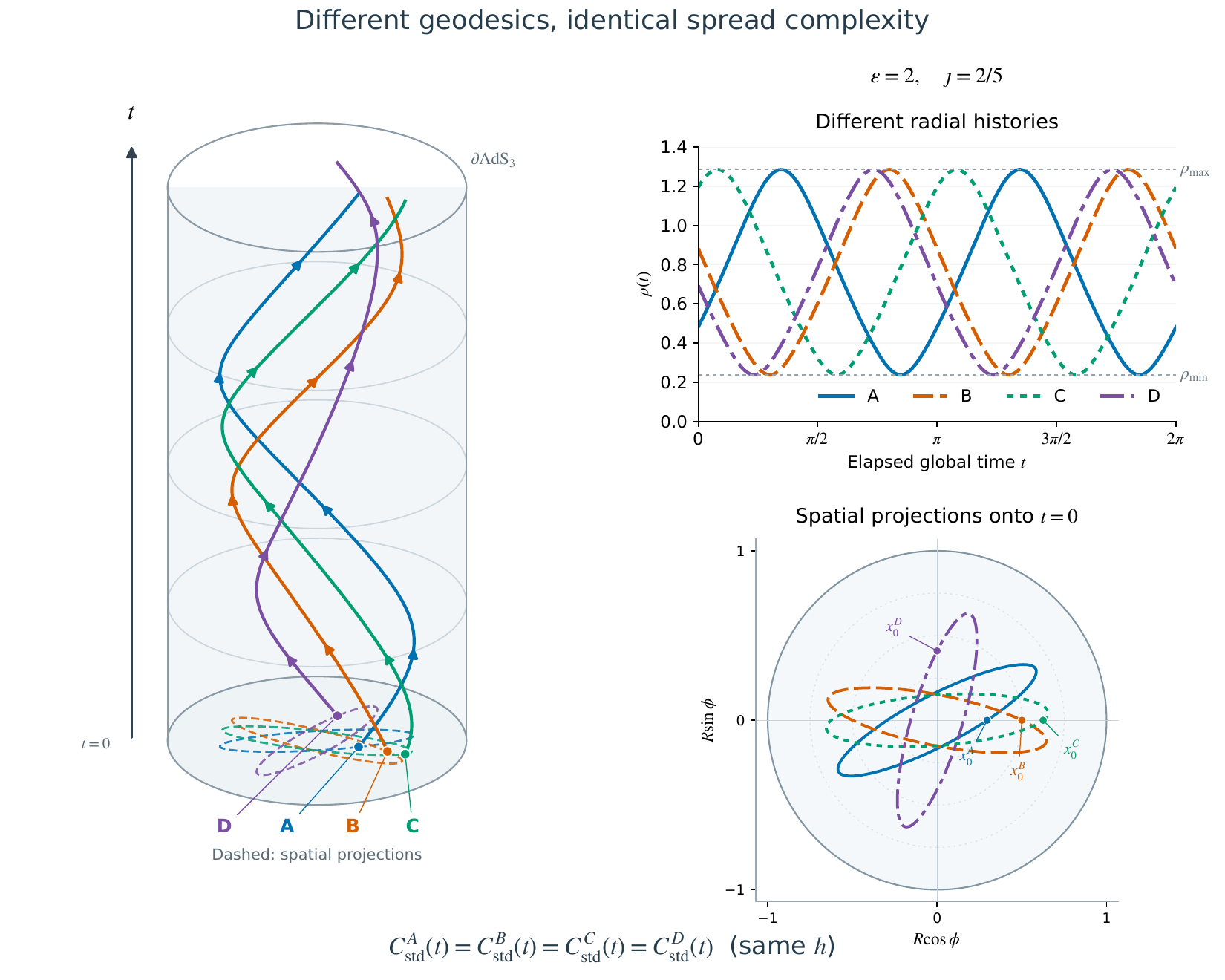}
 \caption{Four timelike geodesics in global $\AdS_3$ with identical
 standard spread complexity at the same $h$ and elapsed global time;
 all have $\e=2$ and $\jmath=2/5$.
 Upper right: the radial projection of the geodesics.
 Lower right: the angular projection of the geodesics,
 $(R\cos\phi,R\sin\phi)$ with $R=(2/\pi)\arctan(\sinh\rho)$.}
 \label{fig:global-equal-complexity-geodesics}
\end{figure}

To simplify the problem further, we use another key property highlighted in~\cite{Li:2025fqz}: the spread complexity is invariant under the stabilizer group transformation. Conjugating $G$ by the $\mathrm{U}(1)\times \mathrm{U}(1)$ stabilizer element 
\begin{equation}
 U=\exp\!\left[-\ii\sum_{\sigma=\pm1}
 t_\sigma(L_0^\sigma-h)\right],
\end{equation}
and using the identity $U L_{\pm1}^\sigma U^\dagger
=e^{\pm\ii t_\sigma}L_{\pm1}^\sigma$, we rotate the Hamiltonian $G$ into the real, canonical form
\begin{equation}
 \widetilde G\equiv U^\dagger G U
 =\sum_{\sigma=\pm1}\left[
 (\e-\sigma\jmath)L_0^\sigma
 +\frac{\sqrt{(\e-\sigma\jmath)^2-1}}{2}
 (L_{+1}^\sigma+L_{-1}^\sigma)\right],
 \label{eq:rotated-Hamiltonian}
\end{equation}
where we have introduced dimensionless particle energy and angular momentum:
\begin{equation}
 L_{-10}^{\rm bulk}=\frac{E}{m}\equiv \e= c_{\rho_0}\gamma_0,
 \qquad L_{12}^{\rm bulk}=\frac{p_\phi}{m}\equiv\jmath=s_{\rho_0}q_\phi.
\end{equation}
It immediately implies that distinct geodesics sharing the same chiral parameters $c_{\rho_\sigma}=\e-\sigma\jmath$ give rise to the same spread complexity, as illustrated in  Fig.~\ref{fig:global-equal-complexity-geodesics}. In other words, the spread complexity depends solely on the charges $(\epsilon,\jmath)$. This behavior echoes the ``Complexity=Anything'' proposal~\cite{Belin:2021bga,Belin:2022xmt}, where a given quantum complexity can be holographically dual to a family of bulk observables.
Notice that the conserved quantities $\e$ and $\jmath$ completely govern the equations of motion:
\begin{equation}
 \left(\frac{\dd\rho}{\dd\tau}\right)^2
 =\frac{\e^2}{\cosh^2\rho}-1
 -\frac{\jmath^2}{\sinh^2\rho},
 \qquad \frac{\dd\phi}{\dd\tau}=\frac{\jmath}{\sinh^2\rho},\qquad
 \frac{\dd t}{\dd\tau}=\frac{\e}{\cosh^2\rho}.
 \label{eq:equal-complexity-radial-motion}
\end{equation}
Thus, the family of geodesics sharing an identical spread complexity can be parameterized as
\begin{equation}
 \rho_i(t)=\tilde{\rho}(t+\delta_i),
 \qquad \phi_i(t)=\tilde{\phi}(t+\delta_i)+\alpha_i,
 \label{eq:equal-complexity-orbit-phases}
\end{equation}
where the pair $(\tilde{\rho},\tilde{\phi})$ denotes the reference geodesic dual to the canonical Hamiltonian \eqref{eq:rotated-Hamiltonian} $\tilde{G}$, and $\delta_i, \alpha_i$ are arbitrary shift constants. This already conflicts with the original conjecture relating the spread complexity rate to the proper radial momentum~\cite{Caputa:2024sux}: shifting $\delta_i$ alters the instantaneous radial momentum, whereas the spread complexity remains unchanged. In contrast, the refined proposal introduced in~\cite{Li:2025fqz,Li:2026pdh} is still valid. As we will demonstrate below, both the spread complexity and its growth rate can still be expressed in terms of the bulk charge observables $L_{AB}^{\rm bulk}(t)$ on the phase space.

For a generic initial state, the spread complexity usually does not admit a simple closed-form expression. Let us first consider two special cases.

\subsection{Two special cases}
When $\phi_0=q_\phi=0$, the matching conditions
\eqref{eq:identification} give identical chiral parameters for both sectors:
\be 
\rho_+=\rho_-\equiv\rho_{\rm c},\quad \cosh\rho_c=c_{\rho_0}\gamma_0
\quad
t_+=t_-\equiv t_{\rm c},\quad \sinh\rho_c e^{\ii t_c}=s_{\rho_0}\gamma_0-\ii q_\rho.
\ee 
In this case, the bulk motion is purely radial and confined to an AdS$_2$ slice, which is governed by the diagonal $\SU(1,1)$ subgroup generated by $L_n^{\mathrm{diag}}=L_n^++L_n^-$. The spread complexity and its growth rate have simple closed forms~\cite{Balasubramanian:2022tpr}:
\bea  
C_{\mathrm{std}}(t)&=&4h\sinh^2\rho_{\rm c}\sin^2\frac{t}{2}
=4h (\sinh^2\rho_0+q_\rho^2\cosh^2\rho_0)\sin^2\frac{t}{2},\label{eq:cr}\\
\frac{\dd C_{\mathrm{std}}(t)}{\dd t}&=&2h\sinh^2\rho_{\rm c}\sin t
=2h (\sinh^2\rho_0+q_\rho^2\cosh^2\rho_0)\sin t ,
\eea  
where the subscript ``$\mathrm{std}$'' denotes the standard spread complexity, distinguishing it from other variants introduced later. Note that the result \eqref{eq:cr} also follows directly from \eqref{eq:app-nb-complexity} via the orthogonal-polynomial method for $\mathrm{NB}(4h,\tanh^2(\rho_{\mathrm{c}}/2))$.

Thus, a non-zero initial radial momentum $q_\rho$ only changes the overall factor of the spread complexity and its growth rate. On the other hand, the proper radial momentum now becomes
\be 
p_\rho(t)=m(q_\rho\cos t-\gamma_0\sinh\rho_0 \sin t ),
\ee  
which implies that for $q_\rho\neq 0$,
\be 
\frac{\dd C_{\mathrm{std}}(t)}{\dd t} \not\propto p_\rho(t), \qquad \text{but rather} \qquad \frac{\dd C_{\mathrm{std}}(t)}{\dd t}\propto p_\rho(t)- p_\rho(0)\cos t.
\ee 
Nevertheless, following~\cite{Li:2025fqz,Li:2026pdh}, we can still relate it to the bulk Noether charge observables.
For this diagonal Krylov chain, the spread-complexity operator in the adapted frame $(G,|h,0\rangle\otimes |h,0\rangle)$ is the total level operator,
\be 
\hat{C}_{\mathrm{std}}=L_0^{\mathrm{diag}}-2h,
\ee 
where the constant shift ensures that the seed state has vanishing initial complexity.  Under time evolution generated by $G$, the Heisenberg equation of motion gives
\be 
\frac{\dd\hat{C}_{\mathrm{std}}}{\dd t}
=\ii [G,\hat{C}_{\mathrm{std}}]
=-\frac{1}{2\ii}
\sum_{\sigma=\pm1}\sinh\rho_\sigma
\left(
e^{\ii t_\sigma}L_{+1}^\sigma
-e^{-\ii t_\sigma}L_{-1}^\sigma
\right).
\label{eq:diagonal-complexity-operator-rate}
\ee 
In the semiclassical limit, the normalized expectation values $\ell_n^\sigma = \langle L_n^\sigma \rangle / h$ map directly to the classical bulk Noether charges derived in Appendix~\ref{app:embedding-charge-dictionary}:
\begin{equation}
    \ell_0^\sigma
    =
    L_{-1,0}^{\mathrm{bulk}}-\sigma L_{12}^{\mathrm{bulk}},
    \qquad 
    \ell_{\pm 1}^\sigma
    =
  L_{01}^{\mathrm{bulk}}+\sigma L_{-1,2}^{\mathrm{bulk}}\pm\ii \left(\sigma L_{02}^{\mathrm{bulk}}-L_{-1,1}^{\mathrm{bulk}} \right).
\end{equation}
Taking the expectation value of \eqref{eq:diagonal-complexity-operator-rate} and applying this dictionary, we arrive at
\be
\dot C_{\mathrm{std}}(t)=-2h\sinh\rho_{\mathrm{c}}\left[\sin t_{\mathrm{c}}\, L_{01}^{\mathrm{bulk}}(t)
-\cos t_{\mathrm{c}}\, L_{-1,1}^{\mathrm{bulk}}(t)\right],\label{eq:spreadP}
\ee  
where the time-dependent bulk Noether observables $L_{AB}^{\mathrm{bulk}}(t)$ are defined in \eqref{eq:timebulk}.

As a second special case, consider the non-trivial circular orbit with $\rho_0>0$,
$q_\rho=0$, and $q_\phi^2=s_{\rho_0}^2$. The matching conditions \eqref{eq:identification} simplify to
\be 
\begin{aligned}
\varsigma&\equiv \frac{q_\phi}{s_{\rho_0}}=\pm1,\quad
 \rho_{\varsigma}=0,
\quad
\rho_{-\varsigma}=2\rho_0,\\
{t_{-\varsigma}}&{=\varsigma\phi_0,
\quad
z_{\varsigma}=0,
\quad
z_{-\varsigma}=e^{-\ii\varsigma\phi_0}\tanh\rho_0} ,
\end{aligned}
\label{eq:circular-chiral-matching}
\ee 
indicating that the circular orbit corresponds to a single chiral sector ($\sigma=-\varsigma$). Thus, the spread complexity is still given by \eqref{eq:app-nb-complexity}:
\be
C_{\mathrm{std}}(t)
=C_{-\varsigma}(t)
=2h\sinh^2(2\rho_0)\sin^2\frac{t}{2}
=8h s_{\rho_0}^2c_{\rho_0}^2\sin^2\frac{t}{2},\label{eq:ccir}
\ee
and the non-vanishing growth rate is
\be
\frac{\dd C_{\mathrm{std}}(t)}{\dd t}
=4h s_{\rho_0}^2c_{\rho_0}^2\sin t.
\ee
Because the orbit is circular, the proper radial momentum vanishes identically, $p_\rho(t)=m\,\dd\rho/\dd\tau=0$,
whereas the complexity growth rate does not at generic times. In this case, only the $\sigma=-\varsigma$ sector contributes to the rate $\langle\ii[H,\widehat N_{-\varsigma}]\rangle_t$, where $\widehat N_\sigma=g_\sigma(L_0^\sigma-h)g_\sigma^{-1}$ is the Krylov level operator in the original frame, as in \eqref{eq:chiral-complexity-physical-level-operator}. Substituting the circular
matching conditions \eqref{eq:circular-chiral-matching}, and using the charge-matching dictionary as in \eqref{eq:spreadP}, we obtain the bulk expression:
\be 
\begin{aligned}
\dot C_{\mathrm{std}}(t)
&={}
-h\sinh(2\rho_0)
\Big[
\sin(\varsigma\phi_0)
\left(
L_{01}^{\mathrm{bulk}}(t)
-\varsigma L_{-1,2}^{\mathrm{bulk}}(t)
\right)
\\
&\hspace{42mm}
-\cos(\varsigma\phi_0)
\left(
\varsigma L_{02}^{\mathrm{bulk}}(t)
+L_{-1,1}^{\mathrm{bulk}}(t)
\right)
\Big].
\end{aligned}
\label{eq:circular-spread-rate-noether-charge}
\ee 

Interestingly, the radial and circular geodesics exhibit the same spread complexity whenever their parameters satisfy
\be
2\sinh^2\rho_0+2q_\rho^2\cosh^2\rho_0\big|_{\rm radial}
=\sinh^2(2\rho_0)\big|_{\rm circular}.
\ee
In terms of the conserved quantities $\jmath$ and $\e$, this condition is equivalent to
\begin{equation}
 (\e_{\rm r},\jmath_{\rm r})
 =\left(\sqrt{1+2|\jmath_{\rm c}|(1+|\jmath_{\rm c}|)},0\right),
 \qquad
 (\e_{\rm c},\jmath_{\rm c})
 =(1+|\jmath_{\rm c}|,\jmath_{\rm c}).
 \label{eq:radial-circular-charge-pair}
\end{equation}
In fact, up to the chiral exchange $\jmath\to-\jmath$, these radial--circular
pairs \eqref{eq:radial-circular-charge-pair} are the only ones that exhibit exact degeneracy in $C_{\rm std}$ for distinct charges
$(\e,|\jmath|)$. This can be verified by matching the first few coefficients in the early-time expansion of $C_{\rm std}$.

\begin{figure}[H]
  \centering
  \includegraphics[width=\textwidth]{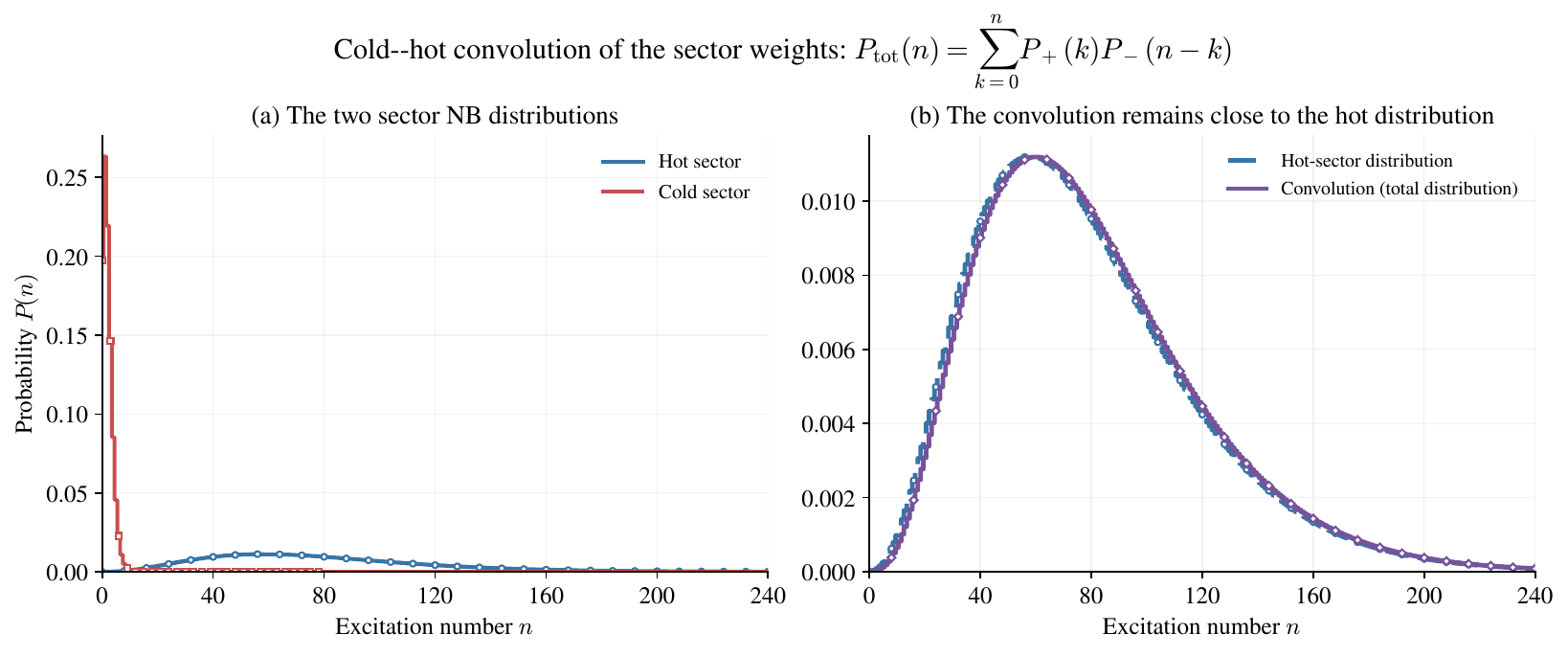}
  \caption{Negative binomial sector weights and their convolution at $h=2$ and $\epsilon-\jmath=2$, $\epsilon+\jmath=40$ (so $\epsilon=21$ and $\jmath=19$). Substituting these parameters into $|z_\pm|^2=(\epsilon\mp\jmath-1)/(\epsilon\mp\jmath+1)$ gives $|z_+|^2=1/3$ for the cold sector and $|z_-|^2=39/41$ for the hot sector.}
  \label{fig:cold-hot-nb-convolution}
\end{figure}

\begin{figure}[H]
  \centering
  \includegraphics[width=\textwidth]{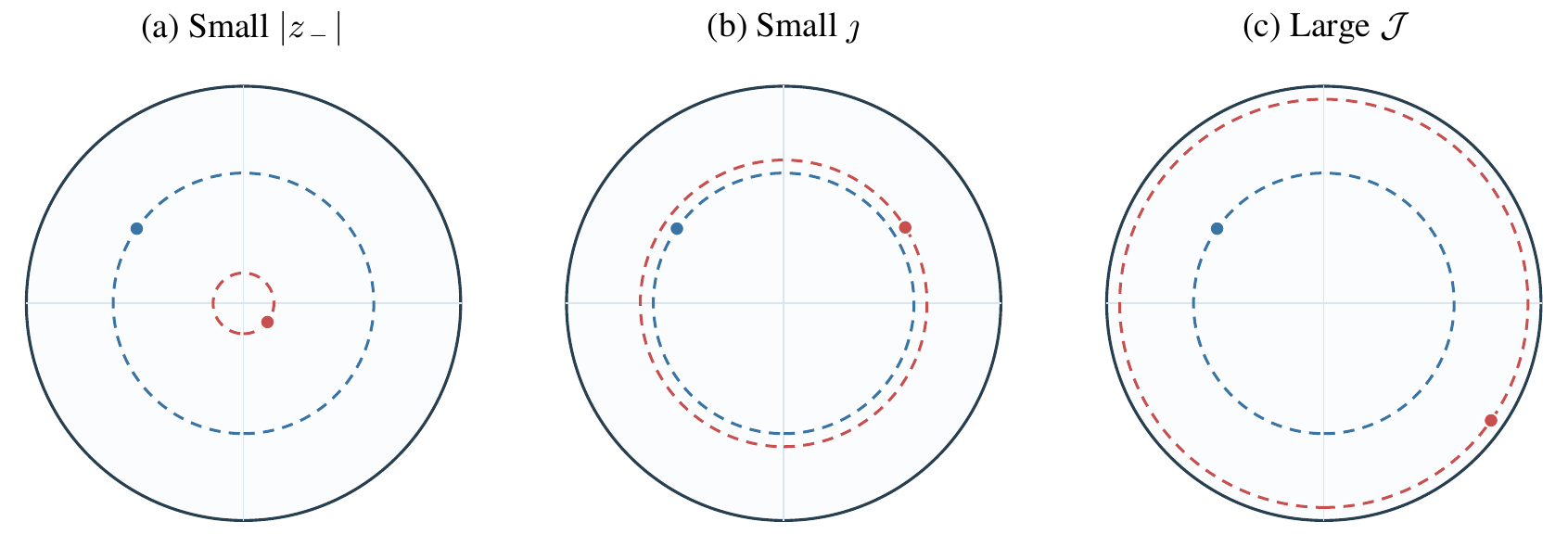}
  \caption{Coherent-state parameters $z_\pm$ in the three perturbative regimes. Blue and red points represent $z_+$ and $z_-$, respectively; dashed circles show their moduli, and the solid boundary is $|z|=1$. The common blue radius provides a visual reference across the panels. (a) The small-$|z_-|$ expansion approaches the circular configuration $z_-=0$, with $z_+$ fixed inside the disk. (b) At fixed $\epsilon>1$, the small-$\jmath$ expansion approaches $|z_+|=|z_-|$. (c) For $\jmath>0$, the large-$\mathcal J$ expansion at fixed $\mathcal{D}=\epsilon-\mathcal J>1$, with $\mathcal J=|\jmath|$, keeps $|z_+|^2=q_*=(\mathcal{D}-1)/(\mathcal{D}+1)$ fixed while $|z_-|^2\to1$.}
  \label{fig:coherent-state-perturbative-limits}
\end{figure}

\begin{figure}[H]
  \centering
  \includegraphics[width=\textwidth]{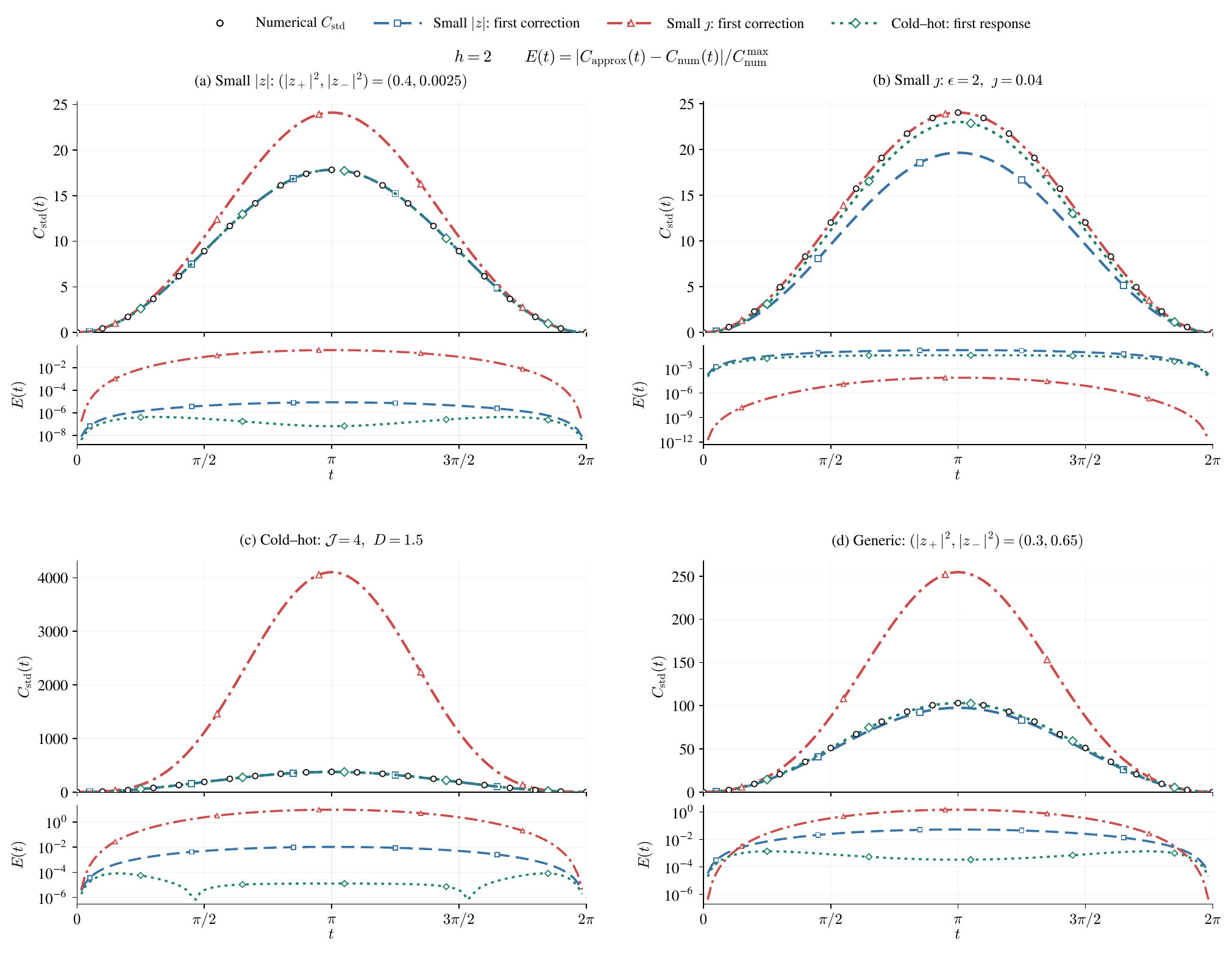}
  \caption{First-order approximations to $C_{\rm std}(t)$ at $h=2$: small-$|z|$ (blue squares), small-$\jmath$ (red triangles), and cold--hot (green diamonds), compared with numerical results (black circles). Parameters: (a) $(|z_+|^2,|z_-|^2)=(2/5,1/400)$; (b) $(\epsilon,\jmath)=(2,1/25)$; (c) $(\mathcal J,\mathcal{D})=(4,3/2)$, with $\mathcal J=|\jmath|$ and $\mathcal{D}=\epsilon-\mathcal J$; (d) $(|z_+|^2,|z_-|^2)=(3/10,13/20)$. Lower panels show $E(t)=|C_{\rm approx}-C_{\rm num}|/C_{\rm num}^{\max}$ on a logarithmic scale, omitting revival endpoints.}
  \label{fig:first-order-std-comparison}
\end{figure}

Beyond these two special cases, a simple closed-form expression for the spread complexity is generally unavailable, except at the special conformal dimension $h=1/2$. Nevertheless, we demonstrate below that the two solvable trajectories serve as natural starting points for systematic perturbative expansions, providing valuable analytical control. Specifically, we can perform a small-$\jmath$ expansion around the purely radial geodesic, or a small-$|z_-|$ (or small-$|z_+|$) expansion around the circular geodesic. Both approaches rely on standard quantum perturbation theory adapted to the Krylov subspace, as detailed in Appendix~\ref{app:per}. In addition, we develop a large-$\jmath$ expansion scheme, which we refer to as the cold--hot approximation. This regime goes beyond conventional quantum perturbation theory. Configurations $(a)$ and $(c)$ of Fig.~\ref{fig:coherent-state-perturbative-limits} share the same structure: in both, $|z_+|$ and $|z_-|$ are widely separated. A more direct and physical perspective comes from examining the spectral measure. Given a dynamical pair $(G,|\psi_0\rangle)$, the Krylov basis, Lanczos coefficients, and spread complexity are completely determined by the spectral measure.
Because our seed state is a tensor product of two chiral coherent states, the total spectral measure is the convolution of two individual chiral probability distributions. Suppose one sector is ``hot'', characterized by a broad excitation distribution, while the other is ``cold'', with excitations confined to a narrow range. The convolution of a broad distribution with a narrow one closely resembles the broad distribution itself, shifted and slightly broadened by the narrow kernel. For intuition, if two distributions are Gaussian $X_{\rm hot}\sim \mathcal{N}(\mu_h,\sigma_h^2),\,X_{\rm cold}\sim \mathcal{N}(\mu_c,\sigma_c^2)$, their convolution is $\mathcal{N}(\mu_c+\mu_h,\sigma_h^2+\sigma_c^2)$. In Fig.~\ref{fig:cold-hot-nb-convolution}, we illustrate this behavior for the convolution of a hot and a cold NB distribution.
Thus, rather than a standard operator perturbation, this approach is akin to a small-noise or narrow-kernel expansion in probability theory.
The three perturbative regimes are summarized schematically in Fig.~\ref{fig:coherent-state-perturbative-limits}.
Deferring the technical derivations to the next subsection, we summarize here the resulting analytical expressions for the standard spread complexity through first order in each regime, with $\mathcal J=|\jmath|$ and $u=\sin^2(t/2)$:
\begin{align}
 \text{small-}|z_-|:\qquad
 C_{\mathrm{std}}(t)
 &\simeq 2h\bigl[(\epsilon+\mathcal J)^2-1\bigr]u
 \nonumber\\[-2pt]
 &\quad+8h\,
 \frac{u\bigl[1+\bigl((\epsilon+\mathcal J)^2-1\bigr)u\bigr]}
      {1+2(\epsilon+\mathcal J-1)u}\,|z_-|^2,
 \\[5pt]
 \text{small-}\jmath:\qquad
 C_{\mathrm{std}}(t)
 &\simeq 4h\Bigl[(\epsilon^2-1)u
 +\bigl(u+(\epsilon^2-1)u^2\bigr)\jmath^2\Bigr],
 \\[5pt]
 \text{cold--hot}:\qquad
 C_{\mathrm{std}}(t)
 &\simeq 2h\bigl[(\epsilon+\mathcal J)^2-1\bigr]u
 \nonumber\\[-2pt]
 &\quad+4h(\epsilon-\mathcal J-1)\,
 \frac{u\bigl[1+\bigl((\epsilon+\mathcal J)^2-1\bigr)u\bigr]}
      {1+2(\epsilon+\mathcal J-1)u}.
\end{align}
For the small-$|z_-|$ expansion, we take $\jmath<0$; the opposite sign follows by exchanging the two chiralities.
The three first-order approximations are compared with the numerical standard spread complexity in Fig.~\ref{fig:first-order-std-comparison}.  Within its respective regime of validity, each first-order approximation exhibits excellent agreement with the numerics. The cold--hot approximation also remains accurate for generic parameters, cf. panel (d).

In addition to these three perturbative schemes, we will also compare with the standard early-time and semiclassical (large-$h$) approximations. We elaborate on the details of these approximations and their potential physical meanings in the next subsection.

\subsection{Perturbative and asymptotic approximations}
We first explain why it is challenging to obtain a closed-form expression for the spread complexity. For general initial data, one must consider the full $\SU(1,1)_+\times \SU(1,1)_-$ group. As a result, the Krylov subspace cannot be identified with the entire module spanned by $|h,0\rangle\otimes|h,0\rangle$ and its descendants; meanwhile, the Krylov basis vectors become complicated linear combinations of the states. Moreover, the expansion coefficients depend explicitly on the parameters in $G$, in contrast to the diagonal case where the Krylov basis is the same as long as the Hamiltonian is a linear combination of the $\SU(1,1)$ generators. To illustrate these difficulties, we present the first three Krylov states:
\bea  
|K_0\rangle &=&|h,0\rangle\otimes|h,0\rangle\equiv |0,0\rangle ,\nn
|K_1\rangle &=&\frac{1}{N_1}(s_{\rho_+}e^{-\ii t_+}|1,0\rangle+s_{\rho_-}e^{-\ii t_-}|0,1\rangle),\nn
|K_2\rangle &=&\frac{1}{N_2}\left\{\sqrt{\frac{2h+1}{2}}(s_{\rho_+}^2e^{-2\ii t_+}|2,0\rangle+s_{\rho_-}^2e^{-2\ii t_-}|0,2\rangle)+\sqrt{2h}s_{\rho_+}s_{\rho_-}e^{-\ii(t_++t_-)}|1,1\rangle  \right.\nn
&+&\left. \frac{c_{\rho_+}-c_{\rho_-}}{s_{\rho_+}^2+s_{\rho_-}^2}\left(s_{\rho_+}s_{\rho_-}^2e^{-\ii t_+}|1,0\rangle-s_{\rho_-}s_{\rho_+}^2e^{-\ii t_-}|0,1\rangle\right) \right\} \label{eq:krylov-basis}
\eea  
where $|n,m\rangle\equiv|h,n\rangle\otimes|h,m\rangle$, and $N_1$ and $N_2$ are normalization factors. Note that because there are multiple states $\Phi_k=\{|n,m\rangle|n+m=k\}$ with the same level $k=n+m$, the Gram-Schmidt orthogonalization procedure cannot remove mixing between different levels; namely, the Krylov basis vector $|K_n\rangle$ can contain not only the sector $\Phi_n$ but also lower sectors $\Phi_k$ with $k<n$. This implies that the complexity operator $\hat{C}_{\rm std}$ cannot be expressed as a simple combination of the $\SU(1,1)_+\times \SU(1,1)_-$ generators. In other words, the standard spread complexity cannot be directly computed from a linear combination of the expectation values of the generators. This is the primary technical obstacle preventing a simple closed-form expression for the standard spread complexity.

Another perspective on this difficulty comes from the spectral measure. As mentioned earlier, in the energy representation, the Krylov basis vectors are generated by orthogonal polynomials,
\be 
|K_n\rangle=p_n(G)|\psi_0\rangle,\quad \langle K_m|K_n\rangle=\sum_E \mu(E)p_m(E)p_n(E)=\delta_{nm},
\ee 
with respect to the measure $\mu(E)=|\langle \psi_0|E\rangle|^2$. In our case, the total spectral measure is the convolution of the two individual ones. To see this, consider the moment generating function of the Hamiltonian:
\be 
Z(p)=\langle \psi_0|e^{p (L_0^++L_0^-)}|\psi_0\rangle=Z_+(p)Z_-(p),\quad Z_\sigma(p)=e^{hp}\left(\frac{1-|z_\sigma|^2}{1-e^p |z_\sigma|^2}\right)^{2h}.\label{eq:moment-gf}
\ee 
Each $Z_\sigma$ corresponds to a $\text{NB}(2h,|z_\sigma|^2)$:
\bea  
&&P_\sigma(H_\sigma=h+n)=\frac{\Gamma(n+2h)}{n!\Gamma(2h)}q_\sigma^n (1-q_\sigma)^{2h},\nn
&&q_1\equiv |z_+|^2=\frac{\epsilon-\jmath-1}{\epsilon-\jmath+1},\quad q_2\equiv |z_-|^2=\frac{\epsilon+\jmath-1}{\epsilon+\jmath+1}.
\eea 
Thus, $Z(p)$ corresponds to the convolution of two NB distributions:
\bea  
&&P(H=2h+n)=[(1-q_1)(1-q_2)]^{2h}\sum_{k=0}^n\frac{\Gamma(k+2h)\Gamma(n-k+2h)}{k!(n-k)!\Gamma(2h)^2}q_1^kq_2^{n-k},\nn
&&=[(1-q_1)(1-q_2)]^{2h}\,\frac{\Gamma(n+2h)}{n!\,\Gamma(2h)}\,q_2^{\,n}\;{}_2F_1\!\left(-n,\,2h;\,1-n-2h;\,\frac{q_1}{q_2}\right),
\eea  
which can be cast into a more compact form in terms of the Gegenbauer polynomials:
\be
P(H=2h+n)=\left[(1-q_1)(1-q_2)\right]^{2h}\,(q_1 q_2)^{n/2}\,C_n^{(2h)}\!\left(\frac{q_1+q_2}{2\sqrt{q_1q_2}}\right)
\label{eq:sw}
\ee
When $q_1=q_2=q$, it reduces to a single $\mathrm{NB}(4h,q)$ as expected. For generic $q_1\neq q_2$, however, the nontrivial weight does not correspond to any known classical or semiclassical orthogonal polynomials that would allow an analytical construction of the Krylov basis. We expect this obstruction to be a generic feature of symmetry groups of higher rank. 
Below we present our results using the approximation methods mentioned previously, which may also be useful for studying spread complexity in broader setups involving high-rank symmetry groups. 

\subsubsection{Early-time expansion}  
The simplest approximation is the early-time limit, where expanding $C_{\rm std}$ in powers of $t$ requires only a finite truncation of the Krylov basis. 

The general procedure is to first use the generating function $\log Z(p)$ to compute the cumulants $\kappa_n=\left.\partial_p^n\log Z(p)\right|_{p=0}=\kappa_n^++\kappa_n^-$. The first six cumulants are given by
\begin{align}
\kappa_1&=h\sum_{i=1}^2 \frac{1+q_i}{1-q_i},\qquad
\kappa_2=2h\sum_{i=1}^2\frac{q_i}{(1-q_i)^2},\nonumber\\
\kappa_3&=2h\sum_{i=1}^2\frac{q_i(1+q_i)}{(1-q_i)^3},\qquad
\kappa_4=2h\sum_{i=1}^2\frac{q_i(1+4q_i+q_i^2)}{(1-q_i)^4},\nonumber\\
\kappa_5&=2h\sum_{i=1}^2\frac{q_i(1+11q_i+11q_i^2+q_i^3)}{(1-q_i)^5},\nonumber\\
\kappa_6&=2h\sum_{i=1}^2\frac{q_i(1+26q_i+66q_i^2+26q_i^3+q_i^4)}{(1-q_i)^6}.
\end{align}  
From these cumulants, one can recursively determine the Lanczos coefficients and subsequently evaluate the spread complexity~\cite{Balasubramanian:2022tpr}:
\be 
C_{\rm std}=\kappa_2 t^2+\left(\frac{\kappa_4}{6}-\frac{\kappa_3^2}{4\kappa_2}\right)t^4+h B_6(h)t^6+\mathcal{O}(t^8).
\ee 
The expansion through $\mathcal{O}(t^4)$ was previously obtained in~\cite{Fatemiabhari:2026eem}. From these expressions, we see that although the initial state is a direct product and cumulants are additive, mixing between the two sectors already appears at the subleading order in the small-$t$ expansion. Since all cumulants scale linearly with $h$, the terms up to $\mathcal{O}(t^4)$ depend on $h$ solely through an overall prefactor. However, the $t^6$ term generically breaks this simple scaling. To isolate its $h$-dependence, we define the rescaled cumulants $\nu_n=\kappa_n/h$. We then find
\begin{align}
B_6(h)&=B_6^{(\infty)}-\frac{(\nu_2^2\nu_5-4\nu_2\nu_3\nu_4+3\nu_3^3)^2}{36\nu_2^3(2h\nu_2^3+\nu_2\nu_4-\nu_3^2)},
\label{eq:global-std-t6}\\
B_6^{(\infty)}&=\frac{\nu_6}{60}-\frac{\nu_3\nu_5}{8\nu_2}-\frac{\nu_4^2}{18\nu_2}
+\frac{5\nu_3^2\nu_4}{12\nu_2^2}-\frac{\nu_3^4}{4\nu_2^3}.
\nonumber
\end{align}
The $h$-dependence of $B_6$ is a $1/h$ effect, $B_6(h)=B_6^{(\infty)}+\mathcal O(1/h)$, which goes beyond the semiclassical description. 

While the early-time expansion is straightforward, truncating the series at a finite order inevitably obscures global analytic properties of the time evolution. For instance, since the dynamics in global AdS is periodic, the exact spread complexity must also exhibit periodicity, which is lost in any finite-order truncation.

\subsubsection{Small-$\jmath$ expansion}
We begin with the small-$\jmath$ expansion and focus on the leading-order correction with the general perturbation theory in Appendix~\ref{app:per}.
For fixed $\e>1$ and $|\jmath|<\e-1$, we expand the Hamiltonian $\widetilde G$ defined in \eqref{eq:rotated-Hamiltonian} with respect to $\jmath$ to the second order:
\begin{align}
 \widetilde G
 ={}&\sum_{\sigma=\pm1}\Bigg[
 \e L_0^\sigma
 +\frac{\sqrt{\e^2-1}}{2}
   (L_{+1}^\sigma+L_{-1}^\sigma)\nonumber\\
 &\quad+\sigma\jmath\left(
 -L_0^\sigma-\frac{\e}{2\sqrt{\e^2-1}}
   (L_{+1}^\sigma+L_{-1}^\sigma)\right)
 -\frac{\jmath^2}{4(\e^2-1)^{3/2}}
   (L_{+1}^\sigma+L_{-1}^\sigma)
 \Bigg]+\mathcal O(\jmath^3).
 \label{eq:small-j-rotated-G-expansion}
\end{align}
 Interchanging the two chiral sectors exchanges $z_+$ and $z_-$, maps $\jmath$ to $-\jmath$, and leaves both the spectral measure and the spread complexity invariant. Therefore, \(C(t)\) is an even function of \(\jmath\), and we set \(g=\jmath^2\). Below, we present three different approaches, each with its own advantages, for carrying out this calculation.

\noindent{\textbf{Quantum perturbation theory.}} 
To obtain the perturbative correction to the spread complexity, we must first determine the perturbative expansion of both the Krylov basis and the transition amplitudes.
It is more convenient to work in the energy representation using the orthogonal polynomials. Let \(\mu_g(m)\) be the deformed spectral weight and \(p_n^{(g)}(E=2h+m)\) be the corresponding orthonormal polynomials, which obey
\begin{equation}
 \sum_{m=0}^{\infty}\mu_g(m)\,
 p_n^{(g)}(2h+m)p_k^{(g)}(2h+m)\equiv\langle p_n^{(g)},p_k^{(g)}\rangle_g=\delta_{nk},
\label{eq:orthogonal}
\end{equation}
with \(p_0^{(g)}=1\). Starting from the perturbed spectral weight \(\mu_g(m)=\mu_0(m)\left[1+g\mathcal V_1(m)+\mathcal O(g^2)\right] \), we want to derive the perturbed orthogonal polynomials \(p_n^{(g)}=p_n^{(0)}+g p_n^{(1)}+\mathcal O(g^2) \). Substituting these expansions into \eqref{eq:orthogonal} and using the orthogonality relation \(\langle p_n^{(0)},p_k^{(0)}\rangle_0=\delta_{nk}\), we obtain the condition to determine $p_n^{(1)}$:
\be 
\langle p_k^{(0)},p_n^{(1)}\rangle_0
+\langle p_k^{(1)},p_n^{(0)}\rangle_0+\mathcal V_{kn}=0, \label{eq:pn1}
\ee 
where $\mathcal{V}_{kn}$ is the matrix element of $\mathcal{V}_1$ in the unperturbed basis,  defined as 
\be 
\mathcal V_{kn}\equiv
\sum_{m=0}^{\infty}\mu_0(m)\,
p_k^{(0)}\mathcal V_1(m)p_n^{(0)}=\langle p_k^{(0)}|\mathcal{V}_1|p_n^{(0)}\rangle_0. \label{eq:matrixel}
\ee 
Because \(p_k^{(1)}\) has degree at most \(k\), the second inner product in \eqref{eq:pn1} vanishes when \(k<n\), fixing the coefficient of \(p_k^{(0)}\) in \(p_n^{(1)}\) to be \(-\mathcal V_{kn}\). At \(k=n\), normalization gives the coefficient \(-\mathcal V_{nn}/2\), while no \(p_k^{(0)}\) terms appear for \(k>n\) since \(p_n^{(1)}\) has degree at most \(n\). Thus, \eqref{eq:pn1} is solved by
\be 
p_n^{(1)}
=-\frac12\mathcal V_{nn}p_n^{(0)}
-\sum_{k=0}^{n-1}\mathcal V_{kn}p_k^{(0)}. \label{eq:pn1s}
\ee
Having determined the perturbed polynomials, we next compute the perturbed Krylov transition amplitudes $\Phi_n(t)=\langle K_n|e^{-\ii t\tilde{G}}|0,0\rangle=\Phi_n^{(0)}+g\Phi_n^{(1)}+\mathcal{O}(g^2)$. In the energy representation, they can be written as
\bea 
 \Phi_n(g,t)&=&\sum_{m=0}^{\infty}\mu_g(m)\,
 p_n^{(g)}(2h+m)e^{-\ii t(2h+m)},\\
 &=&\sum_m \mu_0(m)(1+g \mathcal{V}_1(m))(p_n^{(0)}+g p_n^{(1)})e^{-\ii t(2h+m)}\nn 
 &=&\Phi_n^{(0)}+g\sum_m(\mu_0(m) p_n^{(1)}+\mu_0(m)\mathcal{V}_1(m)p_n^{(0)})e^{-\ii t(2h+m)}+\mathcal{O}(g^2)
 \label{eq:small-j-amplitude-spectral-sum}
\eea 
Substituting \eqref{eq:pn1s} into \eqref{eq:small-j-amplitude-spectral-sum} and expanding $\mathcal{V}_1(m)p_n^{(0)}$ in terms of $p_k^{(0)}$ yields
\begin{equation}
 \Phi_n^{(1)}(t)
 =\frac{\mathcal V_{nn}}{2}\Phi_n^{(0)}(t)
 +\sum_{k=n+1}^{\infty}\mathcal V_{kn}\Phi_k^{(0)}(t),
 \label{eq:small-j-amplitude}
\end{equation}
from which the perturbed spread complexity $C(t)=\sum_{n=0}^\infty n|\Phi_n(t)|^2
 =C_0(t)+g C_1(t)+\mathcal O(g^2)$ follows, with the first-order correction given by
\be 
 C_1(t)=2\operatorname{Re}\sum_{n=0}^{\infty}
 n\big(\Phi_n^{(0)}(t)\big)^*\Phi_n^{(1)}(t)\,. \label{eq:c1t}
\ee 
The above formulation of the first-order correction is completely general. Specializing now to our model, we first substitute the matching condition
\be 
q_\sigma\equiv |z_\sigma|^2
=\frac{\e-\sigma\jmath-1}{\e-\sigma\jmath+1},
\ee
into the spectral weight in \eqref{eq:sw} and expand in $g$, obtaining
\begin{equation}
 \mathcal V_1(m)=\frac{2m(m-1)}{(4h+1)(\e^2-1)^2}
 -\frac{2m}{(\e+1)(\e^2-1)}
 +\frac{2h}{(\e+1)^2}.
 \label{eq:spec-small-j-V1}
\end{equation}
Since $\mathcal{V}_1(m)$ is quadratic in $m$, only matrix elements $\mathcal{V}_{kn}$ with $|k-n|\leq 2$ are non-vanishing. Consequently, the sum in \eqref{eq:small-j-amplitude} truncates at $k=n+2$, and \eqref{eq:pn1s} simplifies to
\bea  
&&p_n^{(1)}=-\frac{1}{2}\mathcal{V}_{nn}p_n^{(0)}-\mathcal{V}_{n-1,n}p_{n-1}^{(0)}-\mathcal{V}_{n-2,n}p_{n-2}^{(0)}.
\label{eq:3pn}
\eea  
An efficient way to evaluate $\mathcal{V}_{kn}$ is to apply the three-term recurrence relation:
\be 
E p_n^{(0)}(E)=b_{n+1}^{(0)}p_{n+1}^{(0)}(E)
+a_n^{(0)}p_n^{(0)}(E)+b_n^{(0)}p_{n-1}^{(0)}(E).
\ee

Recall that the undeformed orthogonal polynomials are Meixner polynomials corresponding to the negative-binomial measure $\mathrm{NB}(4h,q=\frac{\e-1}{\e+1})$, with the Lanczos coefficients $a_n^{(0)}$ and $b_{n}^{(0)}$  given in \eqref{eq:app-nb-lanczos}. Alternatively, in the operator formalism at $g=0$, the generator $\tilde{G}=\e \mathcal L_0+\sqrt{\e^2-1}(\mathcal L_{+1}+\mathcal L_{-1})/2$ acts on the basis $|n\rangle\equiv |2h,n\rangle$. As for the CFT generators $L_n^\sigma$, $\mathcal L_{-1}$ raises the level and $\mathcal L_{+1}$ lowers it:
\be 
\mathcal L_0|n\rangle=(2h+n)|n\rangle,\quad \mathcal L_{-1}|n\rangle=\sqrt{(n+1)(n+4h)}\,|n+1\rangle,\quad 
\mathcal L_{+1}=\mathcal L_{-1}^{\dagger}, \label{eq:lact} 
\ee 
which directly gives $a_n^{(0)}=\e(2h+n)$ and
$b_n^{(0)}=\sqrt{\e^2-1}\sqrt{n(n+4h-1)}/2$.
Writing $m=E-2h$ in $\mathcal{V}_1(m)$ and applying the recurrence relation to $p_n^{(0)}$, we obtain
\begin{align}
 \mathcal V_{nn}
&=\frac{n\left[(3\e^2-1)n+4h(\e^2-1)-2\e^2\right]}
 {(4h+1)(\e^2-1)^2},\nn 
 \mathcal V_{n-1,n}&=\frac{4\e(n-1)b_n^{(0)}}{(4h+1)(\e^2-1)^2},\quad
 \mathcal V_{n-2,n}=\frac{2b_{n-1}^{(0)}b_n^{(0)}}{(4h+1)(\e^2-1)^2} .
 \label{eq:small-j-V-bands}
\end{align}
The recurrence also allows us to eliminate the
lower-degree polynomials in \eqref{eq:3pn}:
\begin{equation}
 \begin{aligned}
 b_n^{(0)}p_{n-1}^{(0)}
 &=(E-a_n^{(0)})p_n^{(0)}-b_{n+1}^{(0)}p_{n+1}^{(0)},\\
 b_{n-1}^{(0)}b_n^{(0)}p_{n-2}^{(0)}
 &=\left[(E-a_{n-1}^{(0)})(E-a_n^{(0)})-(b_n^{(0)})^2\right]p_n^{(0)}\\
 &\quad-(E-a_{n-1}^{(0)})b_{n+1}^{(0)}p_{n+1}^{(0)} .
 \end{aligned}
 \label{eq:small-j-lower-elimination}
\end{equation}
leading to an alternative expression for the first-order deformed polynomials:
\begin{equation}
 \begin{aligned}
 p_n^{(1)}(E)={}&\left[
 \frac{n\left[(3\e^2-1)n+4h(\e^2-1)-2\e^2\right]}
 {2(4h+1)(\e^2-1)^2}-\mathcal V_1(E-2h)\right]p_n^{(0)}(E)\\
 &+\frac{4\e n\,b_{n+1}^{(0)}}{(4h+1)(\e^2-1)^2}\,p_{n+1}^{(0)}(E)
 +\frac{2b_{n+1}^{(0)}b_{n+2}^{(0)}}{(4h+1)(\e^2-1)^2}\,p_{n+2}^{(0)}(E) .
 \end{aligned}
 \label{eq:small-j-polynomial-response-explicit}
\end{equation}
Finally, substituting $\Phi_n^{(0)}(t)$ from \eqref{eq:app-nb-amplitudes} into \eqref{eq:small-j-amplitude} to obtain $\Phi_n^{(1)}(t)$ and plugging the result into \eqref{eq:c1t} yields
\begin{equation}
 C_1(t)=
 4h\big[u+(\e^2-1)u^2\big],\quad u=\sin^2\frac{t}{2},
 \label{eq:small-j-C1-all-level}
\end{equation}
which is a polynomial in $u$ of degree two. While the standard perturbative approach is straightforward, it becomes increasingly complicated at higher orders. To address this, we introduce two alternative approaches.

\noindent{\textbf{Rational spectral transformation.}} As seen from the perturbative procedure above, the main technical task is to determine how the Krylov basis (or equivalently, the corresponding orthogonal polynomials) adapts to a deformation of the spectral measure. This is a classic problem in the theory of orthogonal polynomials (see, e.g.,~\cite{Ismail_2005}). When the measure is multiplied by a rational function, the new orthogonal polynomials can be constructed systematically from the unperturbed ones through a rational spectral transformation, which we briefly review in Appendix~\ref{app:christoffel}. This framework applies directly to our setting. Truncating the deformed spectral weight at first order in $g$, we have
\be 
\mu_g(m) \simeq \mu_0(m)\big[1+g\mathcal{V}_1(m)\big] \equiv \mu_0(m){\mathcal{M}}(m),
\ee 
where the linear deformation factor is a quadratic polynomial in $m$:
\bea  
{\mathcal{M}}(m)&=&1+g\left[ 
\frac{2m(m-1)}{(4h+1)(\e^2-1)^2}
-\frac{2m}{(\e+1)(\e^2-1)}+\frac{2h}{(\e+1)^2}\right]\nn
&=&\frac{2g}{(4h+1)(\e^2-1)^2}(m-m_+)(m-m_-),
\eea 
with roots $m_\pm$ determined by
\be 
m_++m_-=(4h+1)\epsilon-4h,\qquad m_+m_-=(4h+1)\left[\frac{(\e^2-1)^2}{2g}+h(\e-1)^2\right].
\ee 
Because $\mathcal{M}(m)$ has two zeros, the deformed orthogonal polynomials are generated by a double-Christoffel transformation. Let $\Pi_n(m)$ denote the monic reference Meixner polynomial with norm $\mathcal N_n=\langle\Pi_n,\Pi_n\rangle_0$, so that 
$p_n^{(0)}(2h+m)=\Pi_n(m)/\sqrt{\mathcal N_n}$ and
$\mathcal N_{n+1}/\mathcal N_n=(b_{n+1}^{(0)})^2$. The deformed monic polynomial is then given by
\begin{equation}
 \widehat\Pi_n(m)=
 \frac{\Pi_{n+2}(m)+A_n(g)\Pi_{n+1}(m)+B_n(g)\Pi_n(m)}
 {(m-m_+)(m-m_-)} ,
 \label{eq:small-j-rst-christoffel}
\end{equation}
where
\begin{equation}
 A_n(g)=-\frac{1}{D_n(g)}
 \begin{vmatrix}
 \Pi_n(m_+)&\Pi_{n+2}(m_+)\\
 \Pi_n(m_-)&\Pi_{n+2}(m_-)
 \end{vmatrix},\qquad
 B_n(g)=\frac{D_{n+1}(g)}{D_n(g)} .
 \label{eq:small-j-rst-root-system}
\end{equation}
with
\begin{equation}
 D_n(g)=\begin{vmatrix}
 \Pi_n(m_+)&\Pi_{n+1}(m_+)\\
 \Pi_n(m_-)&\Pi_{n+1}(m_-)
 \end{vmatrix} .
 \label{eq:small-j-rst-root-determinant}
\end{equation}
Moreover, the deformed norm follows from \eqref{eq:christoffel-multiple-norm} as
\begin{equation}
 \widehat{\mathcal N}_n=\frac{2g}{(4h+1)(\e^2-1)^2} B_n\mathcal N_n .
 \label{eq:small-j-rst-norm}
\end{equation}
Using \eqref{eq:small-j-rst-norm} to normalize
\eqref{eq:small-j-rst-christoffel}, we obtain an exact identity
\begin{equation}
 \begin{aligned}
 {\mathcal M_g}(m)\widehat p_n^{(g)}(2h+m) 
 ={}&\sqrt{c_n(g)}\,p_n^{(0)}(2h+m)\\[-2pt]
 &+\frac{b_{n+1}^{(0)}}{\sqrt{c_n(g)}}\left[
 d_n(g)p_{n+1}^{(0)}(2h+m)
 +\tfrac{2g\,b_{n+2}^{(0)}}{(4h+1)(\e^2-1)^2}\,p_{n+2}^{(0)}(2h+m)
 \right].
 \end{aligned}
 \label{eq:small-j-rst-normalized}
\end{equation}
where the coefficients expand as
\begin{equation}
 \begin{aligned}
 &c_n(g):=\frac{2g\,B_n(g)}{(4h+1)(\e^2-1)^2}=1+gv_n+\mathcal O(g^2),\\
 &d_n(g):=\frac{2g\,A_n(g)}{(4h+1)(\e^2-1)^2}=gw_n+\mathcal O(g^2),\\
&v_n=\frac{n[(3\e^2-1)n+4h(\e^2-1)-2\e^2]}{(4h+1)(\e^2-1)^2},\quad 
 w_n=\frac{4\e n}{(4h+1)(\e^2-1)^2} .
 \end{aligned}
 \label{eq:small-j-rst-regular-coefficients}
\end{equation}
Extracting the $\mathcal{O}(g)$ coefficient from \eqref{eq:small-j-rst-normalized} and restoring $E=2h+m$, we arrive at
\begin{equation}
 \begin{aligned}
 p_n^{(1)}(E)={}&\left[\frac{v_n}{2}-\mathcal V_1(E-2h)\right]p_n^{(0)}(E)\\[-2pt]
 &+w_n b_{n+1}^{(0)}p_{n+1}^{(0)}(E)
 +\frac{2b_{n+1}^{(0)}b_{n+2}^{(0)}}{(4h+1)(\e^2-1)^2}p_{n+2}^{(0)}(E) ,
 \end{aligned}
 \label{eq:small-j-rst-polynomial-response}
\end{equation}
which agrees identically with \eqref{eq:small-j-polynomial-response-explicit}. More importantly, this approach generalizes systematically to arbitrary order: at order $g^{M}$, the deformation polynomial ${\mathcal{M}}(m)$ possesses $2M$ roots, allowing the deformed orthogonal polynomials to be constructed directly via the multi-point Christoffel formula \eqref{eq:christoffel-multiple}. We note that the Christoffel transformation has also been recently employed to study spread complexity in~\cite{Chowdhury:2026fjb}.  While this method systematically constructs the deformed Krylov basis, to make contact with bulk observables, it is more transparent to switch to the Heisenberg picture and derive an explicit operator expression for the spread complexity. 

\noindent{\textbf{Jacobi flow.}} In this third approach, we keep the Krylov basis fixed to the unperturbed $p_n^{(0)}$ and encode the deformation of spectral measure entirely into a flow of the Jacobi operator:
\begin{align}
 J(g)&=\sum_{n=0}^{\infty} a_n(g)|n\rangle\langle n|
  +\sum_{n=0}^{\infty} b_{n+1}(g)\left(|n+1\rangle\langle n|+|n\rangle\langle n+1|\right)=\sum_{k=0}^{\infty} g^k J_k,
 \label{eq:pert-jacobi-definition}
\end{align}
with $J_0=\tilde{G}$. We define the spread complexity operator as $\hat{C}=\mathcal{N}=\sum_{n=0}^{\infty} n|n\rangle\langle n|$. In the Heisenberg picture, the time-evolved operator $\hat{C}(t)=e^{\ii t J}\hat{C}e^{-\ii t J}=\hat{C}_0(t)+g \hat{C}_1(t)+\mathcal{O}(g^2)$ obeys the Heisenberg equation of motion:
\be 
\frac{\dd\hat{C}(t)}{\dd t}=\ii [J(g),\hat{C}(t)] \quad \Longrightarrow \quad \frac{\dd\hat{C}_1(t)}{\dd t}=\ii [J_0,\hat{C}_1(t)]+\ii[J_1,\hat{C}_0(t)].
\ee 
This immediately provides an operator expression for the growth rate of spread complexity. Integrating this equation, or equivalently, differentiating $e^{\ii t J(g)}\hat{C}e^{-\ii t J(g)}$ directly with respect to $g$, yields an explicit integral representation for the first-order complexity operator:
\be 
\hat{C}_{1}(t)=\frac{\partial }{\partial g}\left(e^{\ii t J(g)}\hat{C}e^{-\ii t J(g)}\right)\Big|_{g=0}=\ii \int_0^t \dd s\,\big[J_{1,I}(s),\hat{C}_{I}(t)\big],\label{eq:c1h}
\ee 
where $\mathcal{O}_{I}(s)\equiv e^{\ii s J_0}\mathcal{O}e^{-\ii s J_0}$ denotes an operator in the interaction picture.

In general, the expansion operators $J_k$ can be constructed recursively from the Jacobi flow \eqref{eq:spec-flow}. At leading order,  $J_1$ is specified by the first-order correction of the Lanczos coefficients, $a_n(g)=a_n^{(0)}+g a_n^{(1)}+\mathcal{O}(g^2),b_n(g)=b_n^{(0)}+g b_n^{(1)}+\mathcal{O}(g^2)$. Having already determined the perturbed polynomials $p_n^{(1)}$, we can read off $a_n^{(1)}$ and $b_n^{(1)}$ from the $\mathcal{O}(g)$ terms of the three-term recurrence relation:
\begin{align}
 E p_n^{(1)}={}&b_{n+1}^{(0)}p_{n+1}^{(1)}+a_n^{(0)}p_n^{(1)}+b_n^{(0)}p_{n-1}^{(1)}+b_{n+1}^{(1)}p_{n+1}^{(0)}+a_n^{(1)}p_n^{(0)}+b_n^{(1)}p_{n-1}^{(0)}.
 \label{eq:small-j-recursion-first-order}
\end{align}
Expressed in terms of the matrix elements $\mathcal{V}_{kn}$, they take the simple forms
\begin{align}
	b_{n+1}^{(1)}
 &=\frac{b_{n+1}^{(0)}}2
   \left(\mathcal V_{n+1,n+1}-\mathcal V_{nn}\right),\\
  a_n^{(1)}
 &=b_{n+1}^{(0)}\mathcal V_{n,n+1}-b_n^{(0)}\mathcal V_{n-1,n}. \label{eq:lan1}
\end{align}
Using the matrix elements $(J_1)_{nn}=a_n^{(1)}$ and $(J_1)_{n+1,n}=b_{n+1}^{(1)}$ together with the $\SU(1,1)$ action \eqref{eq:lact}, we can reconstruct the operator $J_1$ in closed algebraic form:
\begin{align}
 J_1={}&\frac{3\e}{(4h+1)(\e^2-1)}\mathcal L_0^2
 +\frac{\sqrt{\e^2-1}(3\e^2-1)}{2(4h+1)(\e^2-1)^2}
 \{\mathcal L_0,\mathcal L_x\}\nonumber\\
 &-\frac{\e}{\e^2-1}\mathcal L_0
 -\frac{\sqrt{\e^2-1}\e^2}{(\e^2-1)^2}\mathcal L_x
 +\frac{2h\e(1-2h)}{(4h+1)(\e^2-1)}\mathbf 1,
 \label{eq:small-j-J1-diagonal-generators}
\end{align}
where $\{A,B\}=AB+BA$ is the anticommutator and
$\mathcal L_x\equiv (\mathcal L_{+1}+\mathcal L_{-1})/2$. Then we can evaluate the integrand of \eqref{eq:c1h}
\begin{align}
\big[J_{1,I}(s),\hat C_I(t)\big]
={}&\frac{\ii}{(4h+1)(\e^2-1)^{3/2}}
\Bigg[
\left\{
3\e\sqrt{\e^2-1}\,\mathcal L_0(s)
+\frac{3\e^2-1}{2}\mathcal L_x(s),
\,X(s,t)
\right\} \nonumber\\[4pt]
&\qquad
-\frac{3\e^2-1}{2}
\left\{\mathcal L_0(s),Y(s,t)\right\} \nonumber\\[4pt]
&\qquad
-(4h+1)\e\sqrt{\e^2-1}\,X(s,t)
+(4h+1)\e^2Y(s,t)
\Bigg],
\end{align}
where
\begin{align}
X(s,t)
&=\e\sqrt{\e^2-1}\,
  \bigl[1-\cos(t-s)\bigr]\mathcal L_y(s)
  -\sqrt{\e^2-1}\,\sin(t-s)\mathcal L_x(s),\\
Y(s,t)
&=\Bigl[1+(\e^2-1)\bigl(1-\cos(t-s)\bigr)\Bigr]\mathcal L_y(s)
  +\sqrt{\e^2-1}\,\sin(t-s)\mathcal L_0(s),\\
\mathcal L_y &\equiv\frac{\mathcal L_{-1}-\mathcal L_{+1}}{2\ii},\qquad \mathcal L_a(s)\equiv e^{\ii J_0 s}\mathcal L_a e^{-\ii J_0 s},\qquad a=0,x,y.
\end{align}
Taking the expectation values in the reference state $|0\rangle$ and completing the time integration reproduces our earlier result \eqref{eq:small-j-C1-all-level}. 

Finally, invoking the holographic dictionary between the boundary expectation value $\langle \mathcal L_n\rangle$ and bulk Noether charges $L_{AB}^{\rm bulk}$, and taking the semiclassical limit where the correlation functions factorize, we conclude that the first-order correction to holographic spread complexity takes the schematic form
\bea  
C_1(t)&\sim & \int^t \left(\mathcal{G}_1(t-s)L^{\rm bulk}(s)+\mathcal{G}_2(t-s)L^{\rm bulk}(s)L^{\rm bulk}(s)\right)ds,
\eea  
where $L^{\rm bulk}$ denotes the unperturbed bulk Noether charges. Here the expression is schematic: the kernels $\mathcal G_{1,2}(t-s)$ follow directly from the commutator above upon large-$h$ factorization, though their explicit forms are not needed for our purposes. What is physically significant is the non-local convolution structure involving both linear and quadratic combinations of the bulk charges along the history of the state. Differentiating with respect to $t$, we obtain the first-order correction to the complexity growth rate
\begin{align}
\frac{\dd C_1(t)}{\dd t}\sim {}& \mathcal{G}_1(0)L^{\rm bulk}(t)+\mathcal{G}_2(0) L^{\rm bulk}(t)L^{\rm bulk}(t) \nonumber\\
&+\int_0^t \dd s\,\Big[\dot{\mathcal{G}}_1(t-s)L^{\rm bulk}(s)+\dot{\mathcal{G}}_2(t-s)L^{\rm bulk}(s)L^{\rm bulk}(s)\Big],
\end{align}
which splits into an instantaneous contribution built from the bulk Noether charges at time $t$, plus a convolution term.

Using the methods above, we can systematically compute higher-order corrections. For instance, the second- and third-order corrections read
\begin{align}
 \frac{C_2(t)}{h}&=-\left[\left(12+\frac{16}{\e^2-1}\right)u^2
      +64\frac{2h+1}{4h+1}\e^2u^3+8\frac{2h+1}{4h+1}(\e^2-1)(5\e^2-1)u^4\right],\label{eq:small-j-C2}\\
\frac{C_3(t)}{h}&=
\frac{16\e^2}{(\e^2-1)^2}u^2
+\frac{256(2h+1)^2\e^2(3\e^2-1)}
{(4h+1)^2(\e^2-1)^2}u^3\nonumber\\
&\quad+\frac{16(2h+1)\left[(320h+263)\e^4
-(128h+112)\e^2+1\right]}
{(4h+1)^2(\e^2-1)}u^4\nonumber\\
&\quad+\frac{128(h+1)(2h+1)}{(4h+1)^2}
\left[
(2\e^2-1)(25\e^2-1)u^5
+(\e^2-1)(21\e^4-14\e^2+1)u^6
\right].\label{eq:small-j-C3}
\end{align}
We have verified that $C_m$ is a polynomial in $u$ of degree $2m$ for $1\le m\le80$ at the checked parameter values. The spread complexity can be organized as a double expansion:
\be 
C_{\rm std}(t)=\sum_{m=0}^{\infty} g^m C_m(u),\qquad C_m(u)=\sum_{d=1}^{\max(1,2m)}c_{m,d}\,u^d .
\ee 
$c_{m,1}=0$ for $m\geq2$. This follows from the short-time expansion $C(t)=b_1^2t^2+\mathcal O(t^4)$: since $b_1^2$ is strictly linear in $g$ and $u=t^2/4+\mathcal O(t^4)$, every perturbative order beyond the first must begin at $\mathcal{O}(u^2)$, in agreement with \eqref{eq:small-j-C2} and \eqref{eq:small-j-C3}.

\begin{figure}[htbp]
 \centering
 \includegraphics[width=0.98\linewidth]{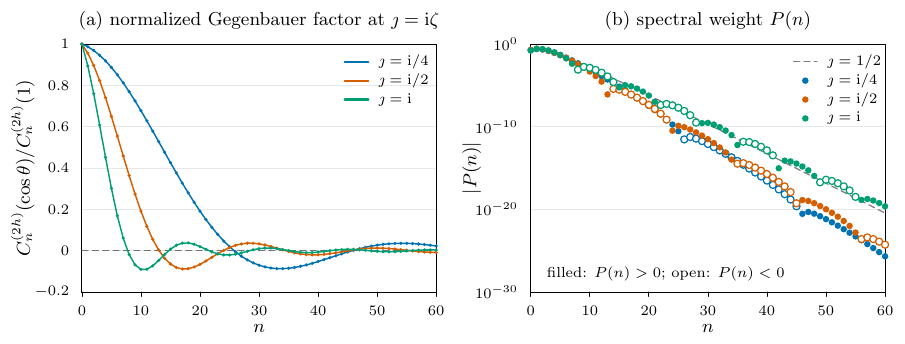}
 \caption{The seed spectral weight \eqref{eq:sw} at imaginary angular momentum $\jmath=\ii\zeta$, for $h=1$ and $\e=2$. (a) The Gegenbauer factor of \eqref{eq:sw} normalized by its aligned value ($\jmath=0$, $x=1$), $C_n^{(2h)}(\cos\theta)/C_n^{(2h)}(1)$, oscillates around zero. (b) $|P(n)|$ on a logarithmic scale; filled (open) markers denote $P(n)>0$ ($P(n)<0$). The first negative weight occurs at $n=26,\,14,\,8$ for $\zeta=1/4,\,1/2,\,1$. For real $\jmath=1/2$ (dashed) all weights are positive.}
 \label{fig:small-j-imaginary-weight}
\end{figure}

A natural question is whether this perturbative series converges. There is a simple physical reason to expect that it does not. Since $g=\jmath^2$, continuing to negative $g<0$ corresponds to taking an imaginary angular momentum $\jmath=\ii\zeta$ ($\zeta\in\mathbb{R}$). In this regime, the two chiral parameters $q_1$ and $q_2$ become complex conjugates of each other, and the spectral weights \eqref{eq:sw} remain real but no longer positive-definite; instead, they oscillate with the level $n$, as illustrated in Fig.~\ref{fig:small-j-imaginary-weight}.
Such a signed spectral measure fails to define a valid unitary Krylov chain: $b_n^2$ inevitably turns negative at some finite depth. Indeed, in the exact Lanczos recursion, the first non-positive $b_n^2$ is encountered at a depth $n_*\propto1/\zeta$. Because a power series in $g$ with a finite radius of convergence around $g=0$ would have to remain analytic for both signs of $g$, the breakdown of positivity for $g<0$ mirrors Dyson's classic instability argument for QED~\cite{Dyson:1952tj}, suggesting that the radius of convergence is zero.

\begin{figure}[htbp]
 \centering
 \includegraphics[width=0.98\linewidth]{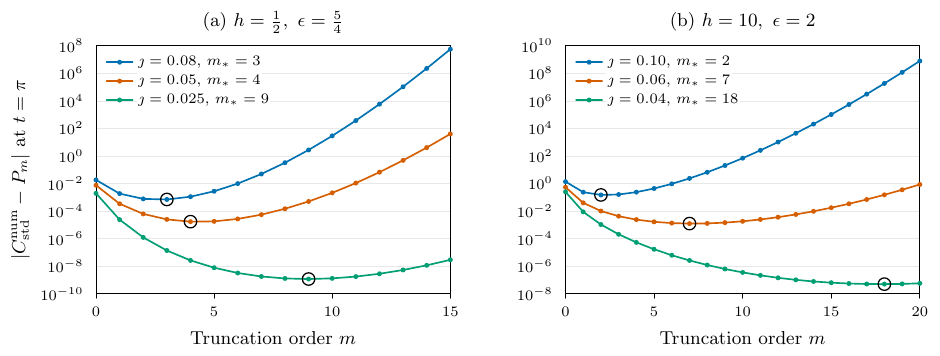}
 \caption{Truncation error $|C_{\rm std}^{\rm num}-P_m|$ at $t=\pi$ as a function of the truncation order $m$, where $P_m=\sum_{k=0}^m g^k C_k(t)$ is the partial sum with $g=\jmath^2$, shown for three values of $\jmath$ in each panel. Circles mark the optimal truncation order $m_*$ for each curve. The numerical values $C_{\rm std}^{\rm num}$ are computed using spectral and Krylov cutoffs $(N,K)=(420,140)$ in (a) and $(2400,850)$ in (b), whose numerical errors lie well below the plotted range.} 
 \label{fig:small-j-order-comparison}
\end{figure}

Numerical evidence confirming this expectation is presented in Fig.~\ref{fig:small-j-order-comparison}, which shows that the partial sums exhibit an optimal truncation order $m_*$ that grows as $g$ decreases, as expected for an asymptotic series. We find numerical indications that the Borel resummation of this series reproduces the exact $C_{\rm std}(t)$. A comprehensive analysis of this asymptotic behavior, together with analogous results in the Poincar\'e and Rindler patches, will be presented in an upcoming paper.

\subsubsection{Small-$|z|$ expansion}
\label{sec:near-circular-expansion}
In this subsection, we apply the same quantum perturbation framework to study the small-$|z|$ expansion around the circular orbit.  For definiteness, we set $\varsigma=1$ in
\eqref{eq:circular-chiral-matching}, so that $z_+=0$ and
$z_-=e^{-\ii\phi_0}\tanh\rho_0$; the case $\varsigma=-1$ follows by exchanging
the two chiralities. Because only $|z_+|^2$ enters the spectral measure, we take $|z_+|^2\equiv \eta$ as the perturbation parameter and expand the spread complexity as
\begin{equation}
 C_{\rm std}(t)=C_0(t)+\eta\,C_1(t)+O(\eta^2),\qquad
 C_0(t)=8h\,s_{\rho_0}^2c_{\rho_0}^2\,u ,\qquad u=\sin^2\frac t2,
 \label{eq:nc-expansion}
\end{equation}
where $C_0$ is the circular-orbit result \eqref{eq:ccir}. Setting $q_1=|z_+|^2=\eta$ and $q_2=|z_-|^2=\tanh^2\rho_0$ in \eqref{eq:sw} and expanding to the first order in $\eta$, we find the deformed measure $\mu_\eta(m)$ \begin{align}
 \mu_\eta(m)&=\nu(m)\bigl[1+\eta\mathcal V_1(m)\bigr]+O(\eta^2),\nonumber\\
 \mathcal V_1(m)&
 =2h\left[\frac{m}{(m+2h-1)\tanh^2\rho_0}-1\right],
 \label{eq:nc-QPT-insertion}
\end{align}
where $m=H-2h$ and $\nu(m)= c_m\tanh^{2m}\!\rho_0/c_{\rho_0}^{4h}$, $c_m\equiv (2h)_m/m!$, is the undeformed spectral measure $\mathrm{NB}(2h,|z_-|^2)$ of Appendix~\ref{app:nb-meixner}. In contrast to \eqref{eq:spec-small-j-V1}, $\mathcal V_1(m)$ is no longer a polynomial in $m$. Instead, it can be rewritten compactly in terms of the shift of $\nu(m)$ as
\be 
\mathcal{V}_1(m)=2h\left[\frac{\nu(m-1)}{\nu(m)}-1\right],\qquad \nu(-1)=0,
\ee 
so $\mathcal V_1(0)=-2h$, including at $h=1/2$, where the rational expression in \eqref{eq:nc-QPT-insertion} is indeterminate at $m=0$. The shift form gives the matrix elements of $\mathcal V_1$:
\begin{equation}
 \mathcal V_{kn}=2h\Bigl[\sum_m\nu(m)\,p_k^{(0)}(m+1)p_n^{(0)}(m+1)-\delta_{kn}\Bigr].
\end{equation}
Using the Meixner shift identity \eqref{eq:app-meixner-shift}, which here takes the form
\begin{equation}
 c_nM_n(m+1)=c_nM_n(m)-\frac1{s_{\rho_0}^2}\sum_{j=0}^{n-1}c_jM_j(m),
\end{equation}
together with the orthogonality relation $\sum_m\nu M_jM_k=\delta_{jk}/(c_j\tanh^{2j}\rho_0)$, we obtain the matrix elements in closed form:
\begin{equation}
 \mathcal V_{nn}=\frac{2h}{s_{\rho_0}^2}\,(1-\Upsilon_n),\qquad
 \mathcal V_{kn}=-\frac{2h}{s_{\rho_0}^2}\,(-\tanh\rho_0)^{k-n}
 \sqrt{\frac{c_n}{c_k}}\;\Upsilon_n\quad(k>n),
 \label{eq:nc-V}
\end{equation}
where $\Upsilon_n$ is the Meixner function
\begin{align}
 \Upsilon_n&=1-\frac{1}{s_{\rho_0}^2c_n}\sum_{j=0}^{n-1}c_j\tanh^{2(n-j)}\!\rho_0
 ={}_2F_1\!\left(-n,1;2h;\frac1{c_{\rho_0}^2}\right),\nonumber\\
 \sum_{n=0}^\infty c_n\Upsilon_n w^n&=\frac{(1-w)^{1-2h}}{1-w\tanh^2\rho_0}.
 \label{eq:nc-QPT-Upsilon-generating}
\end{align}
Substituting \eqref{eq:nc-V} into \eqref{eq:lan1} yields the first-order corrections to the Lanczos coefficients:
\begin{equation}
 a_n^{(1)}=2h\bigl[(n+1)\Upsilon_n-n\Upsilon_{n-1}\bigr],\qquad
 b_n^{(1)}=\frac{h}{s_{\rho_0}^2}\,b_n^{(0)}\bigl(\Upsilon_{n-1}-\Upsilon_n\bigr),
 \label{eq:nc-QPT-jacobi-response}
\end{equation}
with $\Upsilon_{-1}=0$. Since $\Upsilon_n$ is not a polynomial in $n$, the
first-order Jacobi operator $J_1$ is no longer a polynomial in the $\SU(1,1)$
generators, making the operator representation of $\hat{C}_1(t)$ cumbersome; we therefore bypass the Jacobi-flow route here.

For the same reason, the sum over $k$ in \eqref{eq:small-j-amplitude} does not
truncate. The unperturbed circular-orbit transition amplitudes are given by~\eqref{eq:app-nb-amplitudes} with the parameters $2h$ and $\tanh^2\rho_0$:
\begin{align}
 \Phi_k^{(0)}(t)&=\sqrt{c_k}\,\zeta(t)^k\,\Phi_0^{(0)}(t),\nonumber\\
 \zeta(t)&=-\tanh\rho_0\,\frac{1-e^{-\ii t}}{1-e^{-\ii t}\tanh^2\rho_0},\qquad
 \Phi_0^{(0)}(t)=\left(\frac{1}{c_{\rho_0}^2\bigl(1-e^{-\ii t}\tanh^2\rho_0\bigr)}\right)^{2h},
 \label{eq:nc-circular-amplitudes}
\end{align}
up to the constant phase from the energy shift $2h$. Inserting \eqref{eq:nc-V} and \eqref{eq:nc-circular-amplitudes} into \eqref{eq:small-j-amplitude}, the inner sum over $k$ reduces to a geometric series in $-\zeta(t)\tanh\rho_0$, while the outer sum over $n$ in \eqref{eq:c1t} evaluates via the generating function \eqref{eq:nc-QPT-Upsilon-generating}. We obtain
\begin{equation}
 C_1(t)=\frac{8h\,u\,(1+\sinh^22\rho_0\,u)}{1+4s_{\rho_0}^2u}
 =\frac{C_0\,(2h+C_0)}{s_{\rho_0}^2\,(2hc_{\rho_0}^2+C_0)} .
 \label{eq:nc-C1}
\end{equation}
Unlike the leading small-$\jmath$ correction \eqref{eq:small-j-C1-all-level}, which is a polynomial in $u$, $C_1(t)$ here is a rational function of $C_0$ (and hence of $u$), reflecting the non-polynomial nature of the perturbation $\mathcal V_1(m)$ in \eqref{eq:nc-QPT-insertion}.

For $h>1/2$, we can construct the perturbed orthogonal polynomials by a rational spectral transformation. Passing to the shifted negative-binomial probability measure $\widetilde\nu=\mathrm{NB}(2h-1,\tanh^2\rho_0)$ absorbs the pole in $\mathcal V_1(m)$ through the identity
\begin{equation}
 \nu(m)=\frac{m+2h-1}{(2h-1)c_{\rho_0}^2}\,\widetilde\nu(m),\qquad
 \nu(m)\bigl[1+\eta\mathcal V_1(m)\bigr]
 =\widetilde\nu(m)\,{\mathcal M_\eta}(m), 
 \label{eq:nc-rst-weight}
\end{equation}
When its slope $\ell_\eta$ is nonzero, the linear multiplier ${\mathcal M_\eta}(m)$ has a single zero at $m=m_\eta$: 
\begin{equation}
 \begin{aligned}
 {\mathcal M_\eta}(m) 
 &=\frac{(1-2h\eta)(m+2h-1)+2h\eta m/\tanh^2\rho_0}
 {(2h-1)c_{\rho_0}^2}
 =\ell_\eta(m-m_\eta),\\
 \ell_\eta&=\frac{1+2h\eta/s_{\rho_0}^2}{(2h-1)c_{\rho_0}^2},\qquad
 m_\eta=-\frac{(2h-1)(1-2h\eta)}{1+2h\eta/s_{\rho_0}^2} .
 \end{aligned}
\end{equation}
From Appendix~\ref{app:nb-meixner}, the monic Meixner polynomial for $\widetilde\nu$ and its squared norm are given by
\begin{equation}
 \widetilde M_n(m)=(-1)^n(2h-1)_n\,s_{\rho_0}^{2n}
 M_n(m;2h-1,\tanh^2\rho_0),\qquad
 \mathcal N_n^{(2h-1)}=n!(2h-1)_n\,(s_{\rho_0}c_{\rho_0})^{2n} .
\end{equation}
Setting $\varrho_n(\eta)=\widetilde M_{n+1}(m_\eta)/\widetilde M_n(m_\eta)$, the Christoffel formula \eqref{eq:christoffel-multiple} and its norm \eqref{eq:christoffel-multiple-norm} at $M=1$ give
\begin{equation}
 \begin{aligned}
 p_n^{(\eta)}(m)&=\frac{\widetilde M_{n+1}(m)-\varrho_n(\eta)\widetilde M_n(m)}
 {(m-m_\eta)\sqrt{\widehat{\mathcal N}_n(\eta)}},\\
 \widehat{\mathcal N}_n(\eta)&=-\ell_\eta\varrho_n(\eta)\mathcal N_n^{(2h-1)} .
 \end{aligned}
 \label{eq:nc-rst-polynomial}
\end{equation}
Expanding $p_n^{(\eta)}(m)=p_n^{(0)}(m)+\eta p_n^{(1)}(m)+\mathcal O(\eta^2)$, we can write the first-order correction entirely in the unperturbed orthonormal basis as
\begin{equation}
 \begin{aligned}
 p_n^{(1)}(m)={}&-\frac{h}{s_{\rho_0}^2}(1-\Upsilon_n)\,p_n^{(0)}(m)\\
 &+\frac{2h}{s_{\rho_0}^2}\sum_{k=0}^{n-1}
 (-\tanh\rho_0)^{n-k}\sqrt{\frac{c_k}{c_n}}\,
 \Upsilon_k\,p_k^{(0)}(m),
 \end{aligned}
 \label{eq:nc-rst-polynomial-response}
\end{equation}
where the sum is empty for $n=0$. This agrees with \eqref{eq:pn1s} and the matrix elements \eqref{eq:nc-V}.

At second order, for fixed $h>0$ and $0<\tanh^2\rho_0<1$, we obtain
\begin{align}
 C_2(t)={}&C_1(t)+\frac{C_1(t)}{(1+4s_{\rho_0}^2u)^2}
 \Biggl[1-\frac{\tanh^2\rho_0\,C_1(t)}{2h(1+4s_{\rho_0}^2u)}\nonumber\\
 &\hspace{2em}\times{}_2F_1\!\left(1,2;2h+1;
 \frac{\tanh^2\rho_0\,C_1(t)}{2h(1+4s_{\rho_0}^2u)}\right)\Biggr].
 \label{eq:nc-C2}
\end{align}

\begin{figure}[htbp]
 \centering

 \includegraphics[width=0.98\linewidth]{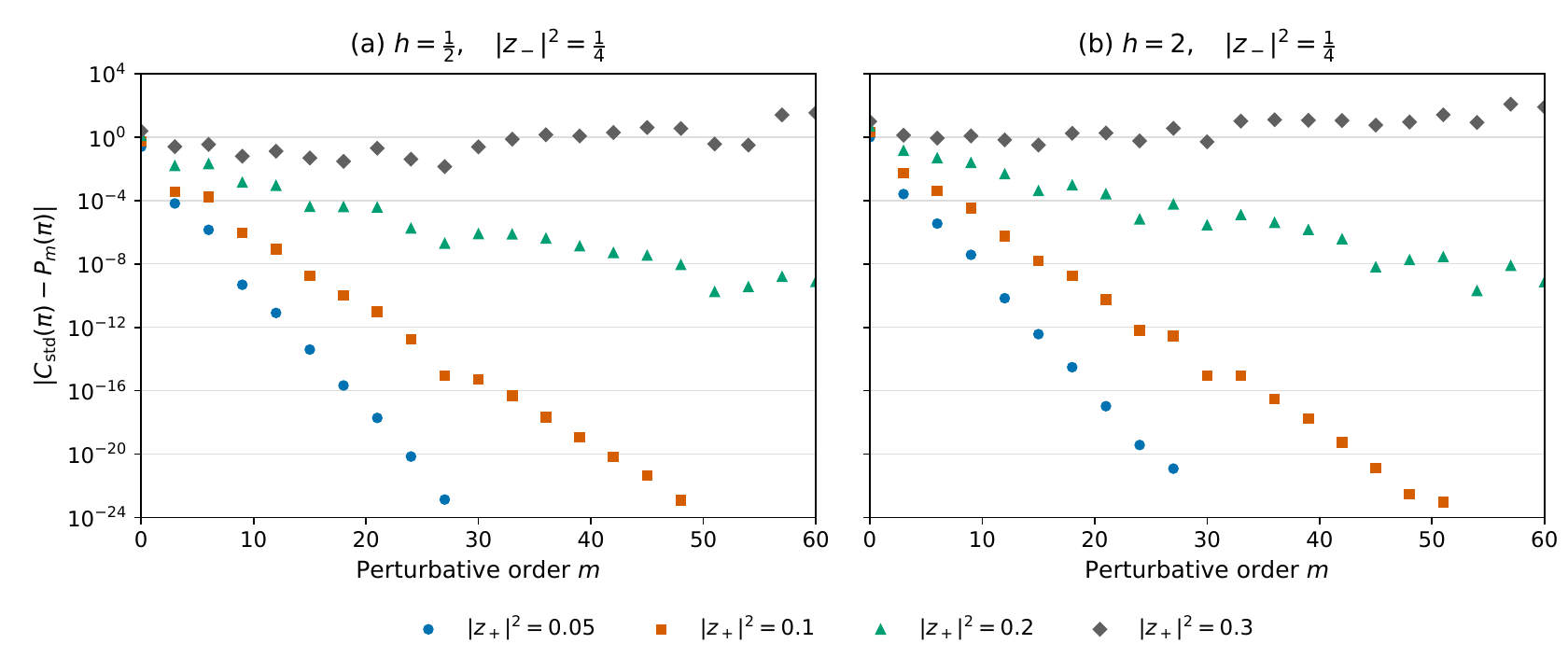}
 \caption{Truncation error $|C_{\rm std}-P_m|$ at $t=\pi$ of the small-$|z|$ expansion, $P_m=\sum_{n=0}^{m}C_n\,|z_+|^{2n}$, as a function of perturbative order $m$, at $|z_-|^2=\tanh^2\rho_0=1/4$ for $h=1/2$ (a) and $h=2$ (b). Circles, squares, triangles and diamonds denote $|z_+|^2=0.05,\,0.1,\,0.2,\,0.3$, respectively.}
 \label{fig:small-z-order}
\end{figure}

The numerical partial sums in Fig.~\ref{fig:small-z-order} indicate a finite radius of convergence close to $|z_-|^2=\tanh^2\rho_0$, consistent with the real-axis positivity boundary $\eta=-\tanh^2\rho_0$ of the exact convolution. For $|z_+|^2<|z_-|^2$, the error decreases with truncation order, with no optimal truncation, in contrast with Fig.~\ref{fig:small-j-order-comparison}; for $|z_+|^2=0.3>|z_-|^2$, the partial sums do not converge. This is consistent with a radius of convergence close to $|z_-|^2=\tanh^2\rho_0$.
\subsubsection{Cold--hot approximation}
We now consider the large-spin regime $\mathcal J=|\jmath|\to\infty$ at fixed gap $\mathcal{D}=\epsilon-\mathcal J>1$, as introduced earlier. For definiteness, take $\jmath>0$; the opposite sign exchanges the two chiralities. In terms of $\mathcal{J}$ and $\mathcal{D}$, the parameters $q_1=|z_+|^2$ and $q_2=|z_-|^2$ take the form
\begin{equation}
 q_1=\frac{\mathcal{D}-1}{\mathcal{D}+1},\qquad
 q_2=\frac{2\mathcal J+\mathcal{D}-1}{2\mathcal J+\mathcal{D}+1}.
 \label{eq:std-ch-nb-parameters}
\end{equation}
As $\mathcal J\to\infty$, $q_1$ remains fixed while $q_2\to 1$. The total-level spectral measure is the convolution of a ``cold'' distribution $\omega=\mathrm{NB}(2h,q_1)$ and a ``hot'' distribution $\nu=\mathrm{NB}(2h,q_2)$:
\begin{equation}
 \mu(n)=\sum_{a\geq0}\omega(a)\,\nu(n-a).
 \label{eq:convolution}
\end{equation}
To develop a systematic expansion, we expand the shifted hot weight $\nu(n-a)$ via the finite-difference Newton series
\begin{equation}
 \nu(n-a)=\sum_{k=0}^{a}\binom{a}{k}(\Delta^k\nu)(n),
 \label{eq:ch-newton}
\end{equation}
where $\Delta$ denotes the backward difference operator
\begin{equation}
 (\Delta\nu)(n)=\nu(n-1)-\nu(n),\qquad \nu(n)=0\ \text{for}\ n<0.
 \label{eq:shift}
\end{equation}
Averaging \eqref{eq:ch-newton} over the cold occupation $a\equiv n_{\rm c}$ yields the factorial-moment expansion
\begin{equation}
 \mu(n)=\sum_{k=0}^{\infty}\frac{F_k}{k!}\,(\Delta^k\nu)(n),
 \qquad
 \frac{F_k}{k!}=\mathbb E\left[\binom{n_{\rm c}}{k}\right]=\frac{(2h)_k}{k!}\left(\frac{q_1}{1-q_1}\right)^k=\frac{(2h)_k}{k!}\left(\frac{\mathcal{D}-1}{2}\right)^k,
 \label{eq:ch-factorial-expansion}
\end{equation}
where $F_k=\mathbb E[n_{\rm c}(n_{\rm c}-1)\cdots(n_{\rm c}-k+1)]$ are the factorial moments of the cold distribution. Because each difference $\Delta^k\nu$ is suppressed by $(1-q_2)^k\sim \mathcal O(\mathcal J^{-k})$, this expansion is controlled for $\mathcal J\gg 1$ at fixed $\mathcal D$.  We emphasize that this expansion technique is not restricted to discrete distributions, but applies equally well to convolutions involving continuous measures. Since $F_k/k!$ grows exponentially with base $(\mathcal D-1)/2$, this heuristic suggests convergence for $\mathcal D<3$. 

Keeping the leading correction in \eqref{eq:ch-factorial-expansion} gives
\begin{equation}
 \mu(n)=\nu(n)+h(\mathcal{D}-1)\bigl[\nu(n-1)-\nu(n)\bigr]+\cdots,
 \qquad \nu(-1)=0 .
 \label{eq:std-ch-measure-first-order}
\end{equation}
This takes the same form as the first-order deformation in \eqref{eq:nc-QPT-insertion}, with $2h\eta$ replaced by $h(\mathcal{D}-1)$. Thus, we can use the same trick of shifting the hot reference measure to $\widetilde\nu=\mathrm{NB}(2h-1,q_2)$ to obtain a linear multiplier:
\begin{equation}
 \begin{aligned}
 {\mathcal M}(n)&=\frac{1-q_2}{2h-1}
 \left[\bigl(1-h(\mathcal{D}-1)\bigr)(n+2h-1)
 +\frac{h(\mathcal{D}-1)}{q_2}n\right],\\
 \mu(n)&={\mathcal M}(n)\widetilde\nu(n)+\cdots .
 \end{aligned}
 \label{eq:std-ch-single-zero-weight}
\end{equation}
Just as in \eqref{eq:nc-rst-weight}, the single-point Christoffel construction applies. Matching to the first-order result \eqref{eq:nc-C1} gives
\begin{equation}
 \begin{aligned}
 C_{\rm std}(t)&\simeq C_{\rm hot}(t)+ \frac{\mathcal{D}-1}{2}\widetilde{C}_1(t),\\
 C_{\rm hot}(t)&=2h\bigl[(\e+\mathcal J)^2-1\bigr]u,\qquad 
 \widetilde{C}_1(t)=
 \frac{8h u\bigl[1+\bigl((\e+\mathcal J)^2-1\bigr)u\bigr]}
 {1+2(\e+\mathcal J-1)u},
 \end{aligned}
 \label{eq:cold-hot}
\end{equation}
which remains valid throughout the regime $\mathcal{J}\gg 1$.

\begin{figure}[htbp]
 \centering
 \includegraphics[width=0.98\linewidth]{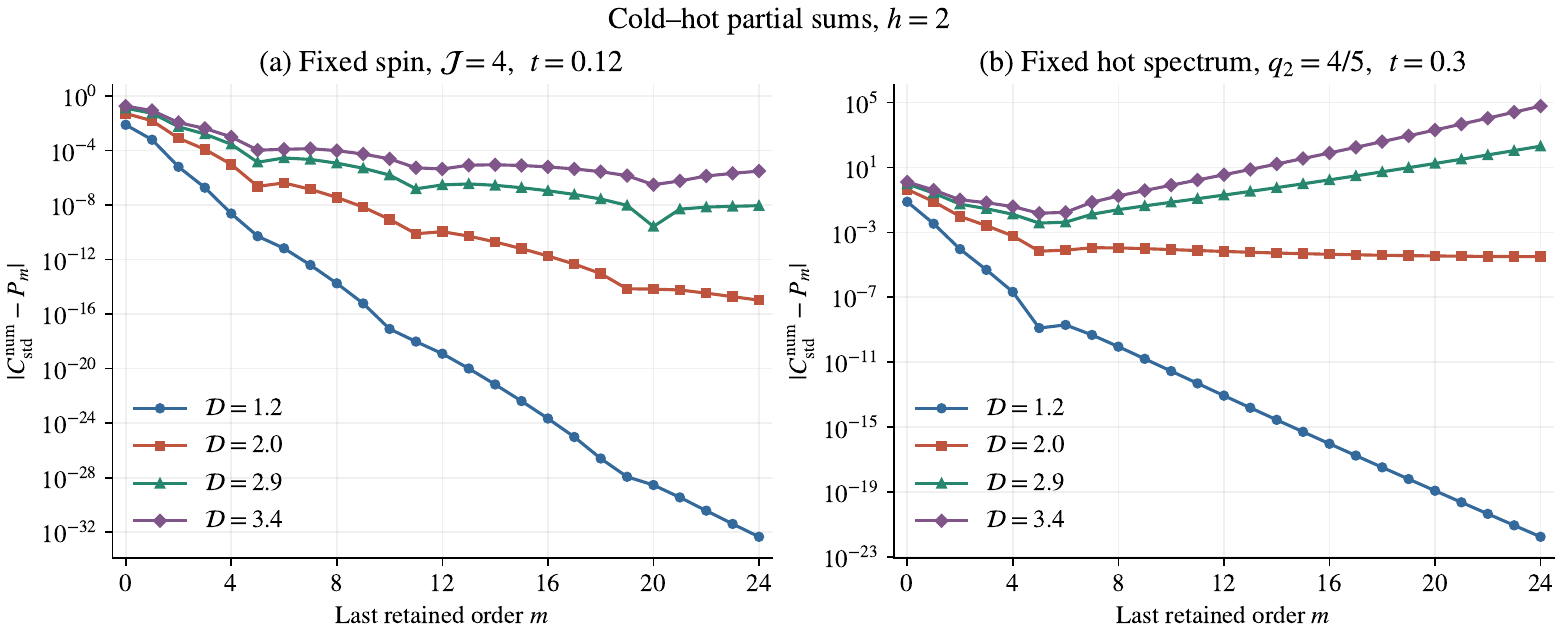}
 \caption{Absolute error $|C_{\rm std}^{\rm num}-P_m|$ of the cold--hot partial sums $P_m=C_{\rm hot}+\sum_{k=1}^{m}\widetilde C_k c^k$, with $c=(\mathcal D-1)/2$, through order $m=24$ at $h=2$. Circles, squares, triangles and diamonds denote $\mathcal D=1.2,\,2.0,\,2.9,\,3.4$, respectively. (a) Fixed $\mathcal J=4$ and $t=0.12$, with $q_2$ given by \eqref{eq:std-ch-nb-parameters}. (b) Fixed $q_2=4/5$ and $t=0.3$.}
 \label{fig:cold-hot-order}
\end{figure}

The similarity between the deformed measure here and that of the small-$|z|$ expansion is not a coincidence: both approximations expand the same underlying convolution $\mathrm{NB}(2h,q_2)*\mathrm{NB}(2h,q_1)$. Whereas the small-$|z|$ expansion sends $q_1=\eta\to0$ at fixed $q_2$, the cold--hot expansion takes $q_2\to1$ at fixed $q_1$. Specifically, we have the parameter identification
\begin{equation}
 \frac{\eta}{1-\eta}=\frac{\mathcal{D}-1}{2},\qquad
 q_2=\frac{2\mathcal J+\mathcal{D}-1}{2\mathcal J+\mathcal{D}+1}.
 \label{eq:std-ch-small-z-parameter-map}
\end{equation}
 At fixed $h,q_2,u$, \eqref{eq:ch-factorial-expansion} implies that the two perturbative series are related by
\begin{equation}
 C_{\rm std}=C_0+\sum_{m=1}^{\infty}C_m\eta^m
 =C_0+\sum_{k=1}^{\infty}\widetilde C_k\left(\frac{\eta}{1-\eta}\right)^k,
 \qquad
 \widetilde C_k=\sum_{m=1}^k(-1)^{k-m}\binom{k-1}{m-1}C_m .
 \label{eq:std-ch-small-z-coefficient-map}
\end{equation}
At fixed $q_2$ this is an expansion in $(\mathcal D-1)/2$. The role of $\mathcal J\gg1$ is to make $q_2\to1$, where the coefficients $\widetilde C_k$ simplify.
In particular, $\widetilde C_1=C_1$, $\widetilde C_2=C_2-C_1$, and $\widetilde C_3=C_3-2C_2+C_1$. The large-$\mathcal J$ asymptotics can therefore be extracted directly by sending $q_2\to 1$ in these finite linear combinations, without having to resum the small-$|z|$ series at finite $q_1$. Fig.~\ref{fig:cold-hot-order} compares the finite-order cold--hot partial sums with the numerically computed standard spread complexity for several values of the gap. The results indicate that, analogous to the small-$|z|$ expansion, the expansion \eqref{eq:std-ch-small-z-coefficient-map} appears to converge for $\mathcal D\lesssim3$.

\subsubsection{Semiclassical limit} 
For the perturbative coefficients computed above, the finite-$h$ dependence of $C_m(t)/h$, for instance in \eqref{eq:small-j-C2} and \eqref{eq:small-j-C3}, drops out at leading order for $h\gg1$. Furthermore, since $C_{\rm std}\propto h$, the relevant Krylov levels scale as $n\sim h$ at fixed $\e$, $\jmath$, and physical
time. Assuming that the Lanczos coefficients admit the smooth limits
\begin{equation}
 y\equiv\frac nh,\qquad
 A(y)=\lim_{h\to\infty}\frac{a_{\lfloor hy\rfloor}}h,\qquad
 B(y)=\lim_{h\to\infty}\frac{b_{\lfloor hy\rfloor+1}}h,
 \label{eq:global-semiclassical-Jacobi-scaling}
\end{equation}
the quantum dynamics on the Krylov chain reduces to semiclassical dynamics.  Starting from the discrete Schr\"odinger equation for the transition amplitude $\Phi_n(t)$
\begin{equation}
 \ii\partial_t\Phi_n
 =a_n\Phi_n+b_n\Phi_{n-1}+b_{n+1}\Phi_{n+1},
 \label{eq:global-semiclassical-Jacobi-evolution}
\end{equation}
and introducing the WKB ansatz $\Phi_n(t)=f(y,t)e^{\ii hS(y,t)}$ on the lattice of spacing $1/h$, we obtain
\begin{equation}
 \Phi_{n\pm1}(t)\simeq e^{\pm\ii p}\Phi_n(t),
 \qquad p=\partial_yS.
\end{equation}
Substituting this ansatz into \eqref{eq:global-semiclassical-Jacobi-evolution} at leading order in $1/h$ yields the Hamilton--Jacobi equation
\begin{equation}
 \partial_tS+H_K(y,\partial_yS)=0,\qquad
 H_K(y,p)=A(y)+2B(y)\cos p.
 \label{eq:global-semiclassical-Krylov-symbol}
\end{equation}
The corresponding classical characteristics obey Hamilton's equations:
\begin{equation}
 \dot y_{\rm cl}=-2B(y_{\rm cl})\sin p,\qquad
 \dot p=-A'(y_{\rm cl})-2B'(y_{\rm cl})\cos p.
 \label{eq:global-semiclassical-Hamilton-equations}
\end{equation}
Because $H_K$ has no explicit time dependence, its value is conserved along each characteristic. Since the seed spectral distribution has mean and variance
\begin{equation}
 E_0=\frac{\langle H\rangle}{h}=2\e,\qquad
 \operatorname{Var}\!\left(\frac Hh\right)
 =\frac{\e^2+\jmath^2-1}{h},
 \label{eq:global-semiclassical-seed-energy}
\end{equation}
it concentrates sharply on the energy shell
\begin{equation}
 A(y_{\rm cl})+2B(y_{\rm cl})\cos p=E_0
\end{equation}
as $h\to\infty$. Eliminating $p$ between this energy constraint and \eqref{eq:global-semiclassical-Hamilton-equations} yields
\begin{equation}
 \dot y_{\rm cl}^{\,2}
 =4B(y_{\rm cl})^2-\big[E_0-A(y_{\rm cl})\big]^2.
\end{equation}
On the outgoing branch ($\dot y_{\rm cl}>0$), we have
\begin{equation}
 \dot y_{\rm cl}
 =\sqrt{4B(y_{\rm cl})^2-\big[E_0-A(y_{\rm cl})\big]^2},
 \qquad
 t=\int_0^{y_{\rm cl}(t)}
 \frac{\dd y}{\sqrt{4B(y)^2-\big[E_0-A(y)\big]^2}}.
 \label{eq:semi-e}
\end{equation}
The square root is real within the classically allowed region; it vanishes at a turning point, beyond which the returning branch takes the negative root. Note that subtracting the constant $2h$ from the Hamiltonian simply shifts $A(y)\to A(y)-2$ and $E_0\to E_0-2$, leaving \eqref{eq:semi-e} invariant. Thus, once the continuum limits $A(y)$ and $B(y)$ are known, the semiclassical spread complexity $y_{\rm cl}(t)=\lim_{h\to\infty}C(t)/h$ follows directly from \eqref{eq:semi-e}. In particular, when $A(y)$ and $B(y)$ admit an expansion in the perturbative parameter $g$, \eqref{eq:semi-e} can be solved order by order. 
 At the unperturbed point $\jmath=0$, the exact Meixner coefficients give
\begin{equation}
 A_0(y)=\e(y+2),\qquad
 B_0(y)=\frac{\sqrt{\e^2-1}}2\sqrt{y(y+4)},
\end{equation}
so that $4B_0^2-(E_0-A_0)^2=y\big[4(\e^2-1)-y\big]$, and integrating \eqref{eq:semi-e} yields the outgoing trajectory
\begin{equation}
 y_{\rm cl}(t)=4(\e^2-1)\sin^2\frac t2,\qquad 0<t<\pi,
\end{equation}
in exact agreement with $C_0(t)/h$. To first order, \eqref{eq:lan1} gives
\begin{align}
 a_n={}&\e(n+2h)
 +\jmath^2\frac{\e n(3n+8h-1)}{(4h+1)(\e^2-1)}
 +\mathcal O(\jmath^4),\nonumber\\
 b_n={}&\frac{\sqrt{\e^2-1}}2\sqrt{n(n+4h-1)}
 \left[1+\jmath^2
 \frac{(3\e^2-1)(n-1)+\frac12(4h+1)(\e^2-1)}
 {(4h+1)(\e^2-1)^2}+\mathcal O(\jmath^4)\right],
 \label{eq:global-semiclassical-jmath-Lanczos}
\end{align}
from which the continuum limits \eqref{eq:global-semiclassical-Jacobi-scaling} yield
\begin{align}
 A(y)={}&\e(y+2)
 +\jmath^2\frac{\e y(3y+8)}{4(\e^2-1)}
 +\mathcal O(\jmath^4),\nonumber\\
 B(y)={}&\frac{\sqrt{\e^2-1}}2\sqrt{y(y+4)}
 \left[1+\jmath^2
 \frac{(3\e^2-1)y+2(\e^2-1)}{4(\e^2-1)^2}
 +\mathcal O(\jmath^4)\right].
 \label{eq:global-semiclassical-jmath-symbol}
\end{align}
Expanding the outgoing branch equation \eqref{eq:semi-e}, we obtain
\begin{align}
 \dot y_{\rm cl}^{\,2}={}&y_{\rm cl}[4(\e^2-1)-y_{\rm cl}]
 \nonumber\\
 &+\jmath^2\left[4y_{\rm cl}+3y_{\rm cl}^2
 -\frac{y_{\rm cl}^3}{2(\e^2-1)}\right]
 +\mathcal O(\jmath^4).
 \label{eq:global-semiclassical-jmath-speed}
\end{align}
To solve perturbatively, we take
$y_{\rm cl}(t)=4(\e^2-1)u+\jmath^2y_1(u)+\mathcal O(\jmath^4)$,
with the previously defined $u=\sin^2(t/2)$. Using
$\dot u^{\,2}=u(1-u)$, the order-$\jmath^2$ equation is
\begin{equation}
 2u(1-u)\frac{dy_1}{du}-(1-2u)y_1
 =4u+4(\e^2-1)u^2(3-2u).
 \label{eq:global-semiclassical-jmath-ray-ODE}
\end{equation}
The solution analytic at $u=0$ is $y_1=4[u+(\e^2-1)u^2]$.
The homogeneous term proportional to $\sqrt{u(1-u)}$ is excluded by the initial quadratic  behavior
$y_{\rm cl}(t)=(\e^2+\jmath^2-1)t^2+\mathcal O(t^4)$.
Thus the formal outgoing solution is
\begin{equation}
 y_{\rm cl}(t)=4(\e^2-1)u
 +4\jmath^2[u+(\e^2-1)u^2]+\mathcal O(\jmath^4),
 \label{eq:global-semiclassical-jmath-ray}
\end{equation}
which reproduces $C_0(t)/h+\jmath^2C_1(t)/h$. Continuing this procedure to higher order in $\jmath$, one can reproduce $\lim_{h\to\infty}C_n(t)\equiv C_n^{\rm cl}(t)$. Similarly, one can carry out a semiclassical analysis for the small-$|z|$ approximation; since the procedure and result are analogous, we omit the details here.

More broadly, the semiclassical limit potentially admits a complete solution via the Riemann-Hilbert formulation for the underlying orthogonal polynomials, combined with the Deift-Zhou nonlinear steepest-descent method. As demonstrated in~\cite{Lunt:2025dcc} and~\cite{Murugan:2026rfa}, the asymptotic Lanczos coefficients, and hence their continuum limits $A(y)$ and $B(y)$, can be extracted directly from the spectral weight. Under a mild one-cut assumption, this problem reduces to a classical logarithmic Coulomb gas (charge-equilibrium) problem. A detailed investigation along these lines lies beyond the scope of the present work and will be reported elsewhere.  
 
Before introducing other variants of spread complexity, let us
summarize the main results of this section:
\begin{enumerate}
  \item Different geodesics can give rise to the same spread
    complexity.  This suggests that spread complexity should be
    associated with a family of geodesics, echoing the
    ``Complexity=Anything'' proposal.

  \item The $h$-dependence at order $t^6$ is subleading in $1/h$, while the higher harmonics of the first small-$\jmath$ correction rule out a linear combination of Noether charges at leading semiclassical order.

  \item We propose three complementary expansions of the spread
    complexity.  The small-$\jmath$ expansion is asymptotic: at
    each order in $\jmath^2$, the coefficient $C_n(t)$ can be
    computed exactly without truncating the Krylov basis.
    Following the derivation of~\cite{Li:2025fqz}, we find that $C_n(t)$ can be expressed in terms of expectation values of
    polynomials of the Noether-charge observables.  The
    small-$|z|$ and cold-hot expansions, by contrast, appear to converge with a finite radius, according to the numerical evidence.  Together,
    these results show that, although $C_{\rm std}$ is encoded
    in the geodesics, the encoding is nontrivial and nonlinear.
\end{enumerate}

\section{The symmetry-resolved spread complexity}
\label{sec:res}
Since the system possesses an additional conserved charge, we consider the symmetry-resolved spread complexity~\cite{Caputa:2025ozd,Caputa:2025mii}. In Section~\ref{sec:effectAdS2} we also consider the standard spread complexity of a fixed-charge initial state.
Such refinements are particularly welcome here, because it has been observed that the radial momentum may not be directly related to the spread complexity once the probe carries
additional charges~\cite{Nastase:2026lhz,Chatzis:2026ekd,Chatzis:2026oou}.
Indeed, Fig.~\ref{fig:global-equal-complexity-geodesics} already shows
that the radial motion $\rho(t)$ is $\pi$-periodic, whereas $C_{\rm std}$ is $2\pi$-periodic. To disentangle the angular contribution, it is therefore natural to resolve the complexity into charge sectors, or to restrict to a single one. As we show below, this restores the $\pi$-periodicity of the radial motion. The holographic interpretation of these refined quantities is itself an interesting question, to which we return at the end of this section and the next section.

To define symmetry-resolved complexity, we decompose the initial state into fixed-charge sectors of the conserved charge $P$: 
\begin{equation}
  \lvert\psi_0\rangle
  =
  \sum_{p\in\mathbb Z}\sqrt{w_p}\,\lvert\Psi_p\rangle,
  \label{eq:momentum-sector-state-decomposition}
\end{equation}
where $p=n-m$ is the eigenvalue of $P$ and the sector states are
\begin{align}
  \lvert\Psi_p\rangle
  ={}&\frac{1}{\sqrt{w_p}}
  \sum_{\substack{n,m\geq0\\n-m=p}}
  \sqrt{P_+(n)P_-(m)}\,e^{-\ii(n t_++m t_-)}
  \lvert h,n\rangle\otimes\lvert h,m\rangle,
  \nonumber\\
  ={}&\frac{(1-|z_+|^2)^h(1-|z_-|^2)^h}{\sqrt{w_p}}
  \sqrt{\frac{(2h)_{|p|}}{|p|!}}\,
  z_+^{(p+|p|)/2}z_-^{(|p|-p)/2}
  \nonumber\\
  &\times\sum_{n=0}^{\infty}
  \sqrt{\frac{(2h+|p|)_n}{n!}}
  \sqrt{\frac{(2h)_n}{(|p|+1)_n}}\,
  (z_+z_-)^n\lvert n;p\rangle,
  \label{eq:momentum-sector-state}
\end{align}
with
\begin{equation}
  w_p=\sum_{\substack{n,m\geq0\\n-m=p}}P_+(n)P_-(m),
  \qquad
  P_\sigma(n)=\frac{\Gamma(n+2h)}{\Gamma(2h)\Gamma(n+1)}
  \lvert z_\sigma\rvert^{2n}
  \left(1-\lvert z_\sigma\rvert^2\right)^{2h}.
  \label{eq:momentum-sector-weight-definition}
\end{equation}
The second equality in \eqref{eq:momentum-sector-state} follows from
$(2h)_{n+|p|}=(2h)_{|p|}(2h+|p|)_n$ and
$(n+|p|)!=|p|!(|p|+1)_n$, and we have introduced the sector basis
\begin{equation}
  \lvert n;p\rangle
  =\left\lvert h,n+\tfrac{p+|p|}{2}\right\rangle
  \otimes\left\lvert h,n+\tfrac{|p|-p}{2}\right\rangle,
  \qquad n=0,1,2,\ldots .
  \label{eq:fixed-sector-energy-basis}
\end{equation}
The weights $w_p$ can be evaluated in closed form in terms of
$q_1=|z_+|^2$ and $q_2=|z_-|^2$. Their Laurent generating function is
\begin{equation}
  \mathcal W(s)\equiv\sum_{p\in\mathbb Z}w_p s^p
  =\left(\frac{1-q_1}{1-q_1s}\right)^{2h}
  \left(\frac{1-q_2}{1-q_2/s}\right)^{2h},
  \qquad q_2<\lvert s\rvert<q_1^{-1},
  \label{eq:momentum-sector-weight-generating-function}
\end{equation}
and extracting the coefficient of $s^p$ gives
\begin{equation}
  \begin{aligned}
    w_p={}&(1-q_1)^{2h}(1-q_2)^{2h}
    \frac{(2h)_{|p|}}{|p|!}\,
    q_1^{(p+\lvert p\rvert)/2}q_2^{(\lvert p\rvert-p)/2}\\
    &\times{}_2F_1\!\left(2h+\lvert p\rvert,2h;
    \lvert p\rvert+1;q_1q_2\right),
    \qquad p\in\mathbb Z.
  \end{aligned}
  \label{eq:momentum-sector-weight-closed-form}
\end{equation}

Since $[H,P]=0$, the sectors evolve independently. Within each sector, we
construct a Krylov basis and the corresponding fixed-charge spread complexity $C_{\rm std}^{(p)}(t)$. The symmetry-resolved spread complexity is then the weighted average
\begin{equation}
  C_{\rm res}(t)=\sum_{p\in\mathbb Z}w_p\,C_{\rm std}^{(p)}(t).
  \label{eq:resolved-definition}
\end{equation}
Because the Lanczos algorithm does not commute with the sector decomposition, $C_{\rm res}$ does not coincide with $C_{\rm std}$ in
general. It has nevertheless been conjectured~\cite{Caputa:2025ozd,Caputa:2025mii} that
\begin{equation}
  C_{\rm std}(t)\geq C_{\rm res}(t),
  \label{eq:resolved-inequality}
\end{equation}
at least at early times, since $C_{\rm std}$ also accounts for the dispersion between sectors. Indeed, the early-time expansion shows
\begin{equation}
  C_{\rm std}-C_{\rm res}
  =t^2\sum_{p}w_p\left[\langle\Psi_p|H|\Psi_p\rangle
  -\langle\psi_0|H|\psi_0\rangle\right]^2
  +\mathcal O(t^4)\geq 0.
  \label{eq:resolved-early-time-difference}
\end{equation}
Clearly, the difference between $C_{\rm std}$ and $C_{\rm res}$ comes from the fact that the average energy in each sector differs from the average energy in all the sectors.
At late times, by contrast, the two quantities are shown to agree within a time window after saturation in finite-dimensional chaotic systems~\cite{Das:2026gko}.

In general, neither the standard nor the symmetry-resolved spread complexity has a simple analytic closed form. Let us focus first on a single momentum sector, in which the projected state is
\begin{align}
  \lvert\Psi_p\rangle
  &\propto\sum_{n=0}^{\infty}
  \sqrt{\frac{(2h+|p|)_n}{n!}}
  \sqrt{\frac{(2h)_n}{(|p|+1)_n}}\,\xi^n\lvert n;p\rangle,
  \qquad \xi\equiv z_+z_-,
  \label{eq:fixedp}\\
  &\propto\sum_{n=0}^{\infty}
  \sqrt{\frac{(2h)_n}{(|p|+1)_n}}\,\frac{\xi^n}{n!}\,
  \mathcal L_-^n\lvert 0;p\rangle,
\end{align}
where $\mathcal L_n$ generate an effective $\SU(1,1)$ representation of
lowest weight $h_p=(2h+|p|)/2$, normalized so that $\mathcal L_-^n\lvert 0;p\rangle=\sqrt{n!\,(2h_p)_n}\,\lvert n;p\rangle$.
The first square-root factor in \eqref{eq:fixedp} is exactly that of the
coherent-state expansion \eqref{eq:expansion}, while the second, $\sqrt{(2h)_n/(|p|+1)_n}$, is an additional sector-dependent weight. Thus, in the special sector $|p|=2h-1$ (assuming $2h\in\mathbb Z_{>0}$), this factor is unity and the spread complexity is
\begin{equation}
  {\left.C_{\rm std}^{(p)}(t)\right|_{|p|=2h-1}}
  =\frac{4(4h-1)q_1q_2}{(1-q_1q_2)^2}\sin^2t.
  \label{eq:fixed-sector-coherent-spread}
\end{equation}
For generic $p$, the normalized spectral weights are
\begin{equation}
  P_p(E_n)=\frac{(2h+\lvert p\rvert)_n(2h)_n}
  {(\lvert p\rvert+1)_n\,n!}\,
  \frac{(q_1q_2)^n}{\mathcal Z_{|p|}(q_1q_2)},
  \qquad E_n=2h+\lvert p\rvert+2n,
  \label{eq:fixed-sector-spectral-weights}
\end{equation}
with
\begin{equation}
  \mathcal Z_{|p|}(q_1q_2)
  ={}_2F_1(2h+\lvert p\rvert,2h;\lvert p\rvert+1;q_1q_2).
  \label{eq:fixed-sector-normalization}
\end{equation}
The fixed-sector survival amplitude is therefore
\begin{equation}
  \mathcal A_p(t)
  =\langle\Psi_p\rvert e^{-\ii Ht}\lvert\Psi_p\rangle
  =e^{-\ii(2h+\lvert p\rvert)t}
  \frac{\mathcal Z_{|p|}(q_1q_2e^{-2\ii t})}
  {\mathcal Z_{|p|}(q_1q_2)}.
  \label{eq:fixed-sector-survival-amplitude}
\end{equation}
Since $\mathcal A_p$ depends on $p$ only through $|p|$, and on $t$ only through $e^{-2\ii t}$, we immediately obtain
\begin{equation}
  C_{\rm std}^{(p)}(t)=C_{\rm std}^{(-p)}(t),
  \qquad
  C_{\rm std}^{(p)}(t)=C_{\rm std}^{(p)}(t+\pi).
  \label{eq:fixed-sector-periodicity}
\end{equation}
Therefore, $C_{\rm res}$ inherits this
$\pi$-periodicity, in contrast to the $2\pi$-periodic $C_{\rm std}$. This is precisely the period of the radial motion $\rho(t)$ noted at the beginning of this section.

\begin{figure}[t]
  \centering
  \includegraphics[width=\linewidth]{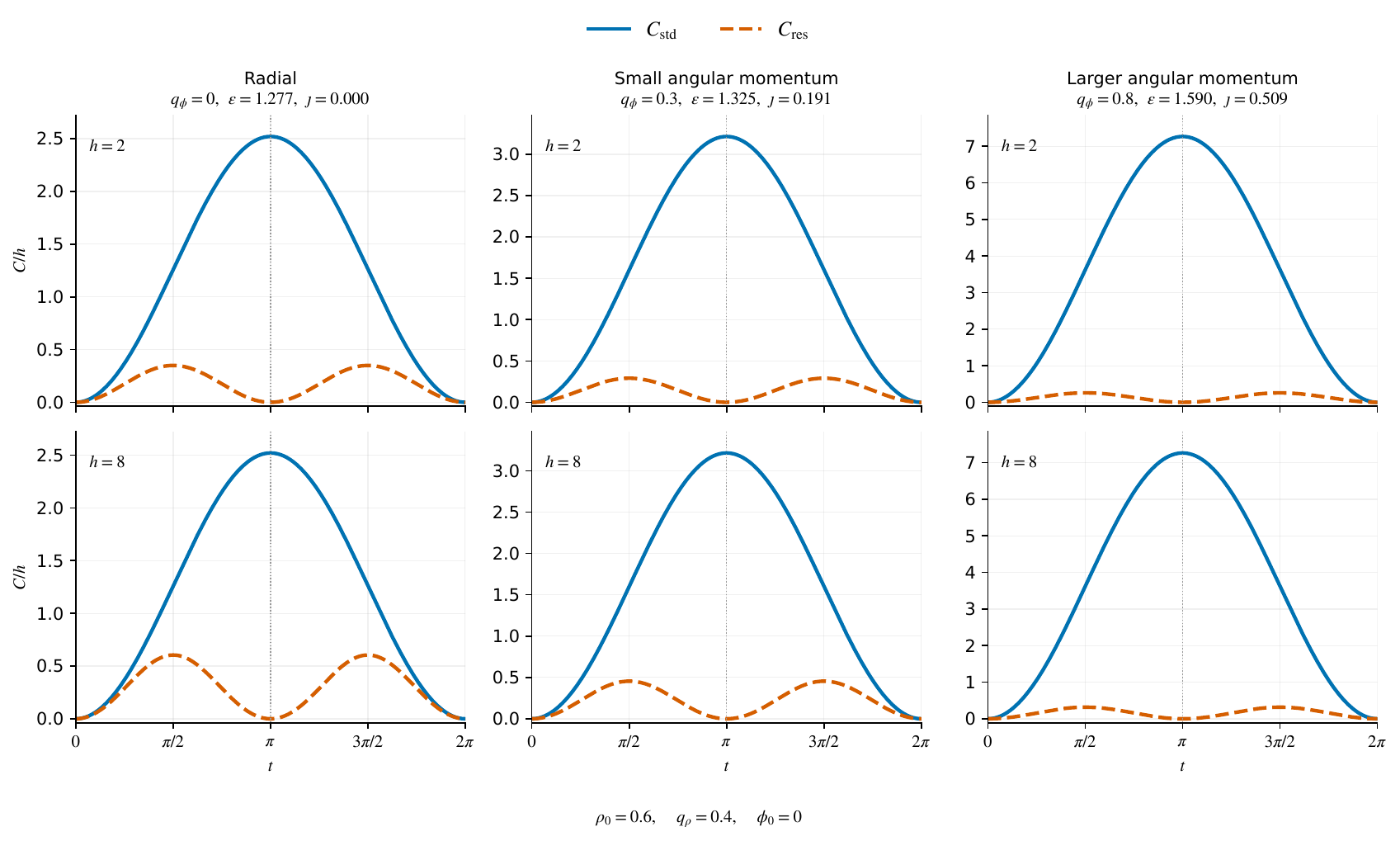}
  \caption{Comparison of $C_{\rm std}/h$ (solid blue) and $C_{\rm res}/h$ (dashed orange) for $h=2$ (top row) and $h=8$ (bottom row). The initial data are $\rho_0=0.6$, $q_\rho=0.4$, and $\phi_0=0$, with $q_\phi=0$, $0.3$, and $0.8$ from left to right.}
  \label{fig:global-standard-resolved-comparison}
\end{figure}

Fig.~\ref{fig:global-standard-resolved-comparison} compares the normalized complexities $C_{\rm std}/h$ and $C_{\rm res}/h$. Among the cases shown, their difference is smallest for purely radial motion ($q_\phi=0$). Increasing $h$ brings the two normalized complexities closer at early times, so that they remain indistinguishable for longer. The large-$h$ behavior of $C_{\rm res}$, including a sharp time scale at which the two first separate, will be discussed elsewhere.

Since neither $C_{\rm res}$ nor $C_{\rm std}^{(p)}$ admits a simple closed form, their holographic interpretations are difficult to pin down. Their $\pi$-periodicity nonetheless suggests that they capture the radial motion rather than the full geodesic. In the next two sections we introduce further variants of spread complexity whose holographic interpretations are transparent.

\section{The chiral spread complexity}
\label{sec:chiral}
Since the Hilbert space admits a tensor product structure, in analogy with the symmetry-resolved complexity, we define the chiral spread complexity as 
\be 
C_{\rm ch}=C_++C_-=2h(\sinh^2\rho_++\sinh^2\rho_-)\sin^2\frac{t}{2},
\ee 
where $C_\sigma$ denotes the spread complexity of a single chiral sector, computed from the pair
$(L_0^\sigma,|z_\sigma\rangle)$ or $(G_\sigma,|h,0\rangle_\sigma)$.
Each chiral contribution can be expressed directly in terms of the particle's initial data:
\bea 
C_\sigma(t)
&=&2h\sinh^2\rho_\sigma\sin^2\frac{t}{2}
\nonumber\\
&=&2h\left[
q_\rho^2+
\left(s_{\rho_0}\gamma_0-\sigma c_{\rho_0}q_\phi\right)^2
\right]\sin^2\frac{t}{2},
\qquad \sigma=\pm1.
\label{eq:chiral-complexity-particle-initial-data}
\eea 
To identify the corresponding Noether observables, we work in the original frame. In each chiral sector, the Krylov level operator is
\begin{equation}
 \begin{aligned}
 \widehat N_\sigma
 &=g_\sigma(L_0^\sigma-h)g_\sigma^{-1}\\
 &=\cosh\rho_\sigma L_0^\sigma
 -\frac{\sinh\rho_\sigma}{2}
 \left(e^{\ii t_\sigma}L_{+1}^\sigma
       +e^{-\ii t_\sigma}L_{-1}^\sigma\right)-h,
 \qquad
 C_{\rm ch}(t)=\sum_{\sigma=\pm1}
 \langle\psi(t)|\widehat N_\sigma|\psi(t)\rangle .
 \end{aligned}
 \label{eq:chiral-complexity-physical-level-operator}
\end{equation}
Applying the Heisenberg equation of motion directly yields the growth rate of the chiral complexity:
\begin{equation}
 \dot C_{\rm ch}(t)
 =\sum_{\sigma=\pm1}
 \left\langle\ii[H,\widehat N_\sigma]\right\rangle_t
 =\frac{\ii}{2}\sum_{\sigma=\pm1}\sinh\rho_\sigma
 \left\langle e^{\ii t_\sigma}L_{+1}^\sigma
             -e^{-\ii t_\sigma}L_{-1}^\sigma\right\rangle_t,
 \label{eq:chiral-complexity-physical-rate-operator}
\end{equation}
where $\langle\cdot\rangle_t \equiv \langle\psi(t)|\cdot|\psi(t)\rangle$ denotes the expectation value in the time-evolved physical state.  Using the dictionary in Appendix~\ref{app:embedding-charge-dictionary}, we get
\begin{equation}
 {\begin{aligned}
 C_{\rm ch}(t)
 &=h\sum_{\sigma=\pm1}\Bigg[
 \cosh\rho_\sigma
 \left(L_{-1,0}^{\mathrm{bulk}}(0)-\sigma L_{12}^{\mathrm{bulk}}(0)\right)
 -\sinh\rho_\sigma
 \bigg[
 \left(L_{01}^{\mathrm{bulk}}(t)
       +\sigma L_{-1,2}^{\mathrm{bulk}}(t)\right)\cos t_\sigma
 \\
 &\hspace{42mm}
 -\left(\sigma L_{02}^{\mathrm{bulk}}(t)
       -L_{-1,1}^{\mathrm{bulk}}(t)\right)\sin t_\sigma
 \bigg]-1\Bigg],\\
 \dot C_{\rm ch}(t)
 &=-h\sum_{\sigma=\pm1}\sinh\rho_\sigma
 \bigg[
 \left(L_{01}^{\mathrm{bulk}}(t)
       +\sigma L_{-1,2}^{\mathrm{bulk}}(t)\right)\sin t_\sigma
 \\
 &\hspace{42mm}
 +\left(\sigma L_{02}^{\mathrm{bulk}}(t)
       -L_{-1,1}^{\mathrm{bulk}}(t)\right)\cos t_\sigma
 \bigg].
 \end{aligned}}
 \label{eq:chiral-complexity-noether-observables}
\end{equation}

\begin{figure}[tbp]
  \centering
  \includegraphics[width=0.98\linewidth]{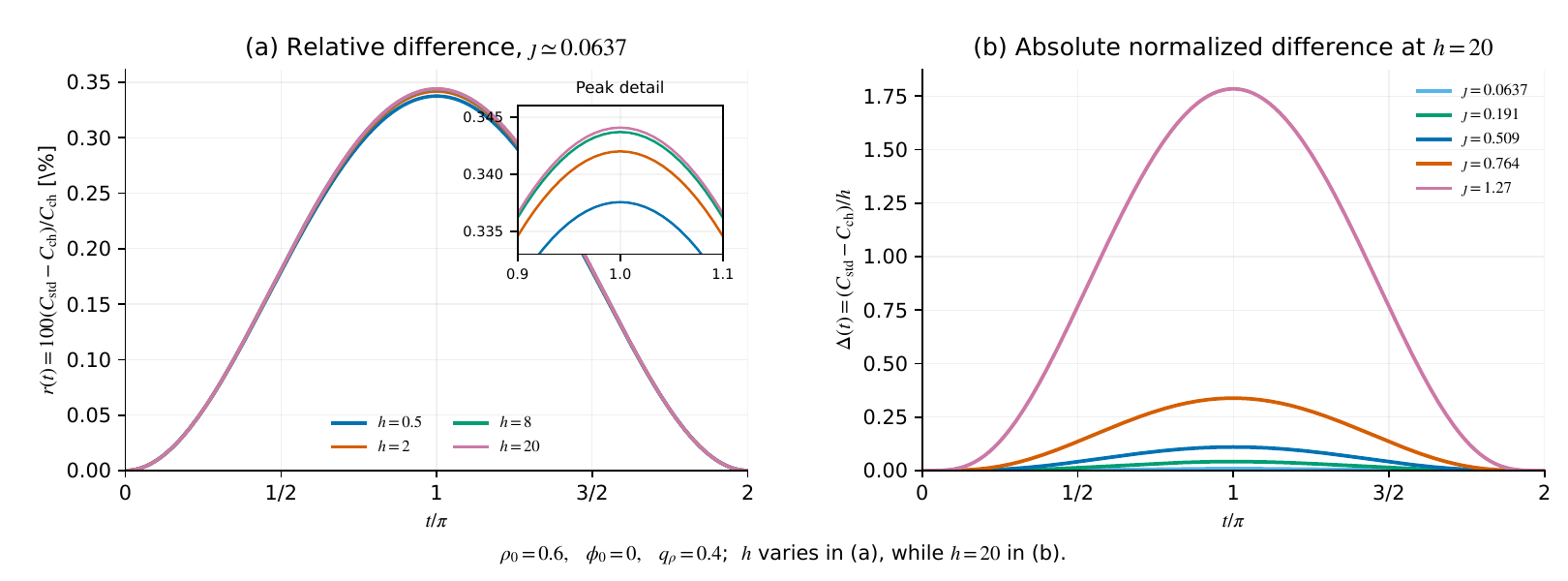}
  \caption{Comparison between the standard and chiral spread complexities for $\rho_0=0.6$, $\phi_0=0$, and $q_\rho=0.4$. 
  (a) Relative difference $r(t) = 100\,[C_{\mathrm{std}}(t)-C_{\mathrm{ch}}(t)]/C_{\mathrm{ch}}(t)$ (in percent) at small angular momentum $\jmath \simeq 0.0637$ for $h = 0.5, 2, 8, 20$; the inset resolves the peak region. 
  (b) Normalized absolute difference $\Delta(t) = [C_{\mathrm{std}}(t)-C_{\mathrm{ch}}(t)]/h$ at fixed $h=20$ for increasing values of $\jmath$.}
  \label{fig:standard-chiral-closeness-j-scan}
\end{figure}

Fig.~\ref{fig:standard-chiral-closeness-j-scan} compares the numerically evaluated standard spread complexity $C_{\mathrm{std}}(t)$ with its chiral counterpart $C_{\mathrm{ch}}(t)$. 
At small angular momentum $\jmath$, the two quantities agree closely across different weights $h$. As $\jmath$ increases, the normalized absolute difference grows with $|\jmath|$. This behavior can be understood analytically in the large-$|\jmath|$ regime. With $\mathcal{J} = |\jmath|$, $\epsilon = \mathcal{J} + \mathcal{D}$, and $u = \sin^2(t/2)$, the chiral complexity takes the exact closed form
\begin{equation}
  \frac{C_{\mathrm{ch}}(t)}{h}
  = \left[8\mathcal{J}^2 + 8\mathcal{D}\mathcal{J} + 4(\mathcal{D}^2 - 1)\right]u.
  \label{eq:chiral-large-j-fixed-gap-exact}
\end{equation}
For fixed $h$ and gap parameter $\mathcal{D} > 1$, expanding $C_{\mathrm{std}}(t)$ at large $\mathcal{J}$ yields the normalized absolute difference
\begin{equation}
  \frac{C_{\mathrm{std}}(t) - C_{\mathrm{ch}}(t)}{h}
  = 4(\mathcal{D} - 1)\mathcal{J}u {-(\mathcal D-1)+\mathcal{O}(\mathcal{J}^{-1})},
  \label{eq:standard-chiral-large-j-absolute-difference}
\end{equation}
which explains the growth observed in Fig.~\ref{fig:standard-chiral-closeness-j-scan}(b). The relative difference scales as
\begin{equation}
  \frac{C_{\mathrm{std}}(t) - C_{\mathrm{ch}}(t)}{C_{\mathrm{ch}}(t)}
  = \frac{\mathcal{D} - 1}{2\mathcal{J}}
  {-\frac{\mathcal D-1}{2\mathcal J^2}\left(\mathcal D+\frac{1}{4u}\right)}
  + \mathcal{O}(\mathcal{J}^{-3}),
  \label{eq:standard-chiral-large-j-relative-difference}
\end{equation}
vanishing asymptotically as $\mathcal{O}(\mathcal{J}^{-1})$ in the large-angular-momentum limit. These expansions hold at fixed $u\gg 1/\mathcal J$ and are non-uniform near $t=0$, where $C_{\rm std}-C_{\rm ch}$ starts at order $t^4$ with a nonnegative coefficient.

\section{Effective AdS$_2$ dynamics: fixed-charge spread complexity}
\label{sec:effectAdS2}
The preceding sections used the CFT Krylov construction to relate spread complexity to bulk Noether charges; in particular, our perturbative analysis revealed a nonlinear dependence in the standard construction. Here we adopt a complementary, bulk-oriented approach~\cite{Li:2026pdh}. Specifically, we quantize the radial theory within a sector of fixed angular momentum, construct coherent states associated with the resulting effective AdS$_2$ dynamics, and evaluate their spread complexity. This construction directly connects spread complexity to the classical radial trajectory and the lapse function of the background geometry.

Rather than quantizing a worldline particle directly, we start from a massive scalar field obeying the Klein--Gordon equation,
\begin{equation}
  (\square_3-m^2)\Psi=0,
  \qquad \Psi=e^{-\ii E t+\ii\ell\phi}\psi_{E\ell}(\rho).
  \label{eq:KG-ansatz}
\end{equation}
Within a fixed-angular-momentum sector, the radial wave equation reduces to
\begin{equation}
  \left[
  \frac{1}{\cosh\rho\,\sinh\rho}
  \partial_{\rho}\!\left(\cosh\rho\,\sinh\rho\,\partial_{\rho}\right)
  +\frac{E^{2}}{\cosh^{2}\rho}
  -m^{2}
  -\frac{\ell^{2}}{\sinh^{2}\rho}
  \right]\psi_{E\ell}=0,
  \label{eq:radial-KG}
\end{equation}
which governs the reduced effective AdS$_2$ dynamics. Normalizability selects the discrete energy spectrum
\begin{equation}
  E_{n,\ell}=2h+|\ell|+2n,\qquad n=0,1,2,\dots,
  \qquad 2h=1+\sqrt{1+m^2}.
  \label{eq:AdS3-spectrum}
\end{equation}
The spectrum in \eqref{eq:AdS3-spectrum} is exact. The heavy-field approximation $2h\simeq m$ enters the classical matching below. With the orientation of $P$ adopted above, a bulk mode $e^{\ii\ell\phi}$ belongs to the CFT sector $p=-\ell$. Each fixed-$\ell$ sector thus forms an equally spaced energy tower starting from $\bar E_\ell\equiv 2h+|\ell|$. We introduce ladder operators $\hat A_\pm$ connecting adjacent levels: 
\begin{equation}
  \hat A_\pm\,\lvert\bar E_\ell,n\rangle
  \propto\lvert\bar E_\ell,n\pm1\rangle,
  \qquad
  [H,\hat A_\pm]=\pm2\hat A_\pm,
  \qquad
  H=L_0^++L_0^-.
  \label{eq:ladder-action}
\end{equation}
Up to a normalization, these operators can be realized by the paired chiral generators $\hat A_\pm\propto L_\mp^+L_\mp^-$, which preserve the angular momentum $\ell$. Choosing the normalization appropriately yields a closed $\SU(1,1)$ algebra:
\begin{equation}
  \begin{gathered}
    A_0=\frac{H}{2},
    \qquad
    A_+=L_-^+L_-^-\,f_\ell(\hat N),
    \qquad
    A_-=f_\ell(\hat N)\,L_+^+L_+^-,\\[2pt]
    \hat N=\frac{H-2h-|\ell|}{2},
    \qquad
    [A_0,A_\pm]=\pm A_\pm,
    \qquad
    [A_+,A_-]=-2A_0,
  \end{gathered}
  \label{eq:su11-algebra}
\end{equation}
where $\hat N\,\lvert\bar E_\ell,n\rangle = n\,\lvert\bar E_\ell,n\rangle$ and
\begin{equation}
  f_\ell(n)=\frac{1}{\sqrt{(n+2h)(n+|\ell|+1)}}.
  \label{eq:fl-def}
\end{equation}
The corresponding normalized Perelomov coherent state is given by
\begin{equation}
  \lvert\xi\rangle
  =(1-|\xi|^2)^{\bar E_\ell/2}\sum_{n=0}^\infty\frac{\xi^n(A_+)^n}{n!}
  \lvert\bar E_\ell,0\rangle
  =(1-|\xi|^2)^{\bar E_\ell/2}\sum_{n=0}^\infty
  \sqrt{\frac{(2h+|\ell|)_n}{n!}}\,\xi^n\lvert\bar E_\ell,n\rangle .
  \label{eq:AdS2-coherent}
\end{equation}
The coherent state $|\xi\rangle$ generally differs from the projection of the AdS$_3$ coherent state $|\psi_0\rangle$ onto the CFT sector $p=-\ell$. By \eqref{eq:fixedp}, the projected state has the parameter $\xi_{\rm proj}=z_+z_-$ and carries the additional factor $\sqrt{(2h)_n/(|\ell|+1)_n}$. For nonzero coherent parameters, the two states coincide up to an overall phase only if $|\ell|=2h-1$ and $\xi=\xi_{\rm proj}$ in the same phase convention. Accordingly, the corresponding spread complexity generally differs from $C^{(p=-\ell)}_{\rm std}$ of Section~\ref{sec:res}.
In analogy with the free AdS$_2$ particle studied in~\cite{Li:2026pdh}, the coherent state $\lvert\xi\rangle$  parametrizes a point in the effective two-dimensional phase space labeled by the initial data $(\rho_0,t_0)$. Although our construction begins in the bulk, the AdS/CFT dictionary directly maps these states and ladder operators to the dual CFT~\cite{Balasubramanian:1998sn,Balasubramanian:1998de}, allowing the resulting spread complexity to be equivalently formulated as a boundary Krylov problem.

We now match the coherent-state parameter $\xi$ to a classical radial trajectory. Using time-translation invariance, we set $t_0=0$ at the outer radial turning point $\rho_0$ (satisfying $s_{\rho_0}^2\ge|\jmath|$), where $p_\rho(0)=m q_\rho(0)=0$. This choice renders $\xi\ge 0$ real and uniquely determined by $\rho_0$. Because angular motion in AdS$_3$ generates a centrifugal barrier, the projected radial motion is not a free AdS$_2$ geodesic; instead, it follows
\begin{equation}
  \tanh^2\rho(t)=\frac{\epsilon^2-1+\jmath^2
  +\sqrt{\left[\epsilon^2-(1+|\jmath|)^2\right]\left[\epsilon^2-(1-|\jmath|)^2\right]}\cos(2t)}
  {2\epsilon^2},
  \label{eq:radial-trajectory}
\end{equation}
where $\epsilon\equiv E_{\rm cl}/m$ and $\jmath\equiv\ell/m$ denote the classical specific energy and angular momentum, respectively. Matching the expectation value of $H=2A_0$ in the coherent state to the classical energy,
\begin{equation}
  \langle\xi|2A_0|\xi\rangle=E_{\rm cl}=m\epsilon,
  \label{eq:energy-matching}
\end{equation}
gives $\bar E_\ell(1+\xi^2)/(1-\xi^2)=m\epsilon$, or $\xi^2=(m\epsilon-\bar E_\ell)/(m\epsilon+\bar E_\ell)$ when $m\epsilon\geq\bar E_\ell$. At leading order in the heavy-field limit $2h\simeq m$, this becomes
\begin{equation}
  \xi^2=\frac{\epsilon-|\jmath|-1}{\epsilon+|\jmath|+1}.
  \label{eq:xi-from-energy}
\end{equation}
Substituting the relations $\epsilon=c_{\rho_0}\gamma_0$ and $\jmath=s_{\rho_0}q_\phi$, we obtain
\begin{equation}
  \xi^2=
  \frac{c_{\rho_0}\sqrt{1+\jmath^2/s_{\rho_0}^2}-1-|\jmath|}
  {c_{\rho_0}\sqrt{1+\jmath^2/s_{\rho_0}^2}+1+|\jmath|}.
  \label{eq:xi-rho0}
\end{equation}
For $\jmath=0$, this expression reduces to the free-particle relation
\begin{equation}
  \xi=\tanh\frac{\rho_0}{2}.
  \label{eq:xi-free}
\end{equation}

Let $C_{\rm eff}(t)$ denote the spread complexity of the time-evolved state $e^{-\ii Ht}\lvert\xi\rangle$. The effective $\SU(1,1)$ symmetry gives $C_{\rm eff}(t)=4\bar E_\ell\xi^2\sin^2t/(1-\xi^2)^2$. The following global AdS$_3$ complexity and lapse formulas use the leading heavy-field approximation. In this limit, \eqref{eq:xi-rho0} gives
\begin{equation}
  C_{\rm eff}(t)
  =m\,\frac{(s_{\rho_0}^2-|\jmath|)^2}{(1+|\jmath|)s_{\rho_0}^2}\sin^2t.
  \label{eq:Ceff}
\end{equation}
Because the initial states and energy spacing are different, $C_{\rm eff}$ at $\jmath=0$ does not reduce to $C_{\rm std}=4h s_{\rho_0}^2\sin^2\frac{t}{2}$ in \eqref{eq:cr}, the result for a purely radial geodesic. In particular, the period is different. Comparing its time derivative with the canonical radial momentum,
\begin{equation}
  p_\rho(t)=-m\,\frac{\sqrt{\left[\epsilon^2-(1+|\jmath|)^2\right]
  \left[\epsilon^2-(1-|\jmath|)^2\right]}\sin(2t)}{2\epsilon\tanh\rho(t)},
  \label{eq:prho}
\end{equation}
we find
\begin{equation}
  \frac{\dd C_{\rm eff}}{\dd t}
  =-\frac{2\epsilon}{1+|\jmath|}
  \frac{s_{\rho_0}^2-|\jmath|}{s_{\rho_0}^2+|\jmath|}
  \,\tanh\rho(t)\,p_\rho(t).
  \label{eq:dCeff-dt}
\end{equation}
Unlike the original conjecture, which assumes purely free motion, the complexity growth rate for the effective AdS$_2$ dynamics is no longer proportional to $p_\rho$ alone.

The geometric origin of the extra $\tanh\rho(t)$ factor becomes transparent once we introduce the lapse function $N(\rho)$. Expressing the global AdS$_3$ metric~\eqref{eq:global-metric} as
\begin{equation}
  \dd s^2=-N(\rho)^2\dd t^2+\dd\rho^2+\cdots,
  \qquad
  N(\rho)=\cosh\rho,
  \label{eq:lapse-def}
\end{equation}
we can recast the spread complexity~\eqref{eq:Ceff} directly in terms of the lapse function along the trajectory:
\begin{equation}
  C_{\rm eff}(t)
  =\frac{m\epsilon^2}{1+|\jmath|}
  \frac{s_{\rho_0}^2-|\jmath|}{s_{\rho_0}^2+|\jmath|}
  \left(\sech^2\rho(t)-\sech^2\rho_0\right)
  \propto N^{-2}(t)-N^{-2}(0).
  \label{eq:Ceff-lapse}
\end{equation}
Along any radial trajectory in a static metric of the form~\eqref{eq:lapse-def}, energy conservation implies $\dot\rho = N^2 p_\rho/(m\epsilon)$, so that
\begin{equation}
  \frac{\dd C_{\rm eff}(t)}{\dd t}
   \propto-(\partial_\rho\log N)\,p_\rho .\label{eq:dCeff}
\end{equation}
The same relations \eqref{eq:Ceff-lapse} and \eqref{eq:dCeff} also hold  in the Poincar\'e and two-sided Rindler patches\footnote{
For Poincar\'e ($\epsilon_P>|j|$),
\[
 C_P(t_P)\simeq m(\epsilon_P-|j|)^2\,t_P^2,\qquad
 N_P=\frac1z,\qquad
 z^2(t_P)=\frac1{\epsilon_P^2-j^2}
             +\left(1-\frac{j^2}{\epsilon_P^2}\right)t_P^2 .
\]
For two-sided Rindler, let $b_R=1-(1+j^2)/\epsilon_R^2$ and $D_R=\sqrt{b_R^2+4/\epsilon_R^2}$, 
\[
 C_R(t_R)\simeq m(\epsilon_R^2+1)\,\sinh^2t_R,\qquad
 N_R=\sinh\rho_R,\qquad
 \operatorname{csch}^2\rho_R(t_R)=\frac{D_R-b_R}{2}+D_R\sinh^2t_R .
\]}.

We note why $C_{\rm eff}(t)$ differs from the full AdS$_3$ spread complexity analyzed earlier. From the three-dimensional perspective, time evolution in both setups is generated by the same Hamiltonian $H=L_0^++L_0^-$. However, the coherent state used here is restricted to a fixed-angular-momentum sector, rather than superposing different $\ell$ modes, reflecting the well-known sensitivity of Krylov spread complexity to the choice of initial state. Therefore, $C_{\rm eff}$ is not a newly defined complexity measure, but rather the standard spread complexity evaluated for a different initial state.

\section{Summary and Outlook}
\label{sec:summary}
The primary objective of this work has been to investigate the holographic dual of spread complexity, serving as a tractable prototype of quantum complexity, within the framework of probe dynamics. Central to our inquiry is identifying the precise nature of bulk dynamics that spread complexity encodes. Investigating probe dynamics offers a distinct advantage: upon quantization, the resulting Hilbert space relates directly to that of the boundary CFT, and the semiclassical limit provides a transparent bridge between quantum observables and classical bulk trajectories. Nevertheless, this approach has an inherent limitation, as probe dynamics is largely insensitive to the backreaction or the intrinsic evolution of the spacetime geometry itself, such as the growth of the wormhole volume. Building upon our earlier studies~\mbox{\cite{Li:2025fqz,Li:2026pdh}}, we conclude that the conventional identification \mbox{$\dot{C}\sim p_\rho$} should not be generalized uncritically. Different initial states can yield distinct holographic interpretations; for instance, we have identified an alternative relation, \mbox{$C\sim N^{-2}(t)-N^{-2}(0)$}. Furthermore, to gain sharper analytic control, we developed several perturbative schemes leveraging the theory of orthogonal polynomials. These expansions reveal rich analytic structures, most notably that the small-\mbox{$\jmath$} regime is governed by an asymptotic double expansion. Understanding the bulk mechanism dual to these analytic properties presents a compelling direction for future work, and the perturbative toolkit developed here should prove valuable for analyzing Krylov and spread complexities across broader setups. Finally, we initiated the study of holographic interpretations for refined variants of spread complexity, including symmetry-resolved, fixed-charge, and chiral sectors. Their distinct dynamical features highlight a rich structure that warrants further exploration.
\section*{Acknowledgments}
This work was supported by a talent program of Shanxi Province and by Shanxi University.

\noindent\textbf{Statement of AI use.} We acknowledge the use of GPT Codex 5.5/6.1/6.2 (OpenAI), Claude Opus 5.5 (Anthropic), and K3.1 Pro (Kimi) to help develop some of the ideas, carry out computations, edit the manuscript, and correct typographical errors. The authors conceived the study and take sole responsibility for the accuracy of this work.

\appendix
\section{Bulk Noether charges and chiral generator expectation values}
\label{app:embedding-charge-dictionary}
In this appendix, we detail the exact dictionary relating the mass-normalized bulk Noether charges to the normalized expectation values of the dual CFT symmetry generators.

The mass-normalized bulk $\SU(1,1)_+\times \SU(1,1)_-$ Noether charges are defined in embedding coordinates as
\begin{equation}
  L_{AB}^{\mathrm{bulk}}
  \equiv
  X_A\frac{\dd X_B}{\dd\tau}
  -X_B\frac{\dd X_A}{\dd\tau},
  \qquad
  X_A=\eta_{AB}X^B,
  \label{eq:appendix-bulk-embedding-charges}
\end{equation}
which are conserved along the geodesic. More generally, they can be viewed as observables on the bulk phase space (or moment maps). Evaluating them at
\(\tau=0\) in terms of  the initial data \eqref{eq:bulk-initial-data} yields
\begin{align}
  L_{-1,0}^{\mathrm{bulk}}
  &={}
  c_{\rho_0}\gamma_0,
  &
  L_{12}^{\mathrm{bulk}}
  &={}
  s_{\rho_0}q_\phi,
  \label{eq:appendix-bulk-charge-longitudinal}
  \\
  L_{-1,1}^{\mathrm{bulk}}
  &={}
  -q_\rho\cos\phi_0+c_{\rho_0}q_\phi\sin\phi_0,
  &
  L_{-1,2}^{\mathrm{bulk}}
  &={}
  -q_\rho\sin\phi_0-c_{\rho_0}q_\phi\cos\phi_0,
  \label{eq:appendix-bulk-charge-minus-one}
  \\
  L_{01}^{\mathrm{bulk}}
  &={}
  s_{\rho_0}\gamma_0\cos\phi_0,
  &
  L_{02}^{\mathrm{bulk}}
  &={}
  s_{\rho_0}\gamma_0\sin\phi_0.
  \label{eq:appendix-bulk-charge-zero}
\end{align}
In particular, \(L_{-1,0}^{\mathrm{bulk}}=E/m\) and
\(L_{12}^{\mathrm{bulk}}=p_\phi/m\).

On the CFT side, we normalize the generator expectation values in the normalized chiral coherent state $|z_\sigma\rangle$ by the conformal weight $h_\sigma=h=\bar h$ for the scalar primary:
\begin{equation}
  \ell_n^\sigma
  \equiv
  \frac{\langle z_\sigma|L_n^\sigma|z_\sigma\rangle}{h_\sigma},
  \qquad \sigma=\pm1,
  \label{eq:appendix-normalized-chiral-expectation-values}
\end{equation}
which evaluate explicitly to
\be 
  \ell_0^\sigma
  =
  \frac{1+|z_\sigma|^2}{1-|z_\sigma|^2},\qquad 
  \ell_{+1}^\sigma
  =
  \frac{2z_\sigma}{1-|z_\sigma|^2},\qquad 
  \ell_{-1}^\sigma
  =
  \frac{2\bar z_\sigma}{1-|z_\sigma|^2}.
  \label{eq:appendix-chiral-coherent-state-expectations}
\ee
Using the matching condition \eqref{eq:identification}, we can express them in terms
of the bulk initial data:
\begin{align}
  \ell_0^\sigma
  &={}
  c_{\rho_0}\gamma_0-\sigma s_{\rho_0}q_\phi,
  \nonumber\\
  \ell_{+1}^\sigma
  &={}
  e^{\ii\sigma\phi_0}
  \left(
    s_{\rho_0}\gamma_0
    -\sigma c_{\rho_0}q_\phi
    +\ii q_\rho
  \right),
  \nonumber\\
  \ell_{-1}^\sigma
  &={}
  e^{-\ii\sigma\phi_0}
  \left(
    s_{\rho_0}\gamma_0
    -\sigma c_{\rho_0}q_\phi
    -\ii q_\rho
  \right).
  \label{eq:appendix-chiral-expectations-initial-data}
\end{align}
Comparing the bulk charges \eqref{eq:appendix-bulk-charge-longitudinal}--\eqref{eq:appendix-bulk-charge-zero} with \eqref{eq:appendix-chiral-expectations-initial-data}
establishes the dictionary:
\begin{equation}
  \begin{aligned}
    \ell_0^\sigma
    &={}
    L_{-1,0}^{\mathrm{bulk}}-\sigma L_{12}^{\mathrm{bulk}},
    \\
    \ell_{+1}^\sigma
    &={}
    L_{01}^{\mathrm{bulk}}+\ii\sigma L_{02}^{\mathrm{bulk}}
    -\ii\left(
      L_{-1,1}^{\mathrm{bulk}}
      +\ii\sigma L_{-1,2}^{\mathrm{bulk}}
    \right),
    \\
    \ell_{-1}^\sigma
    &={}
    L_{01}^{\mathrm{bulk}}-\ii\sigma L_{02}^{\mathrm{bulk}}
    +\ii\left(
      L_{-1,1}^{\mathrm{bulk}}
      -\ii\sigma L_{-1,2}^{\mathrm{bulk}}
    \right).
  \end{aligned}
  \label{eq:appendix-chiral-bulk-charge-dictionary}
\end{equation}
Since $\ell_{-1}^\sigma=(\ell_{+1}^\sigma)^*$, the six real quantities $\ell_0^\pm$, $\operatorname{Re}\ell_{+1}^\pm$, and $\operatorname{Im}\ell_{+1}^\pm$ uniquely determine all six bulk charges via the inverse relations:
\begin{equation}
  \begin{aligned}
    L_{-1,0}^{\mathrm{bulk}}
    &={}
    \frac{\ell_0^++\ell_0^-}{2},
    &
    L_{12}^{\mathrm{bulk}}
    &={}
    \frac{\ell_0^--\ell_0^+}{2},
    \\
    L_{01}^{\mathrm{bulk}}
    &={}
    \frac{\operatorname{Re}\ell_{+1}^+
      +\operatorname{Re}\ell_{+1}^-}{2},
    &
    L_{-1,2}^{\mathrm{bulk}}
    &={}
    \frac{\operatorname{Re}\ell_{+1}^+
      -\operatorname{Re}\ell_{+1}^-}{2},
    \\
    L_{02}^{\mathrm{bulk}}
    &={}
    \frac{\operatorname{Im}\ell_{+1}^+
      -\operatorname{Im}\ell_{+1}^-}{2},
    &
    L_{-1,1}^{\mathrm{bulk}}
    &={}
    -\frac{\operatorname{Im}\ell_{+1}^+
      +\operatorname{Im}\ell_{+1}^-}{2}.
  \end{aligned}
  \label{eq:appendix-inverse-chiral-bulk-charge-dictionary}
\end{equation}
As a consistency check, the classical chiral Casimir functions of these normalized expectation values satisfy
\begin{equation}
  \left(\ell_0^\sigma\right)^2
  -\ell_{+1}^\sigma\ell_{-1}^\sigma
  =1,
  \qquad \sigma=\pm1.
  \label{eq:appendix-normalized-chiral-casimir}
\end{equation}
In terms of the coherent-state parameters, this identity simply reflects $\cosh^2\rho_\sigma-\sinh^2\rho_\sigma=1$; in terms of the bulk initial data~\eqref{eq:appendix-chiral-expectations-initial-data}, it reduces to $\gamma_0^2-q_\rho^2-q_\phi^2=1$, which is precisely the proper-time normalization of the geodesic.

The relations derived above hold at the initial CFT time $t=0$, corresponding to bulk proper time $\tau=0$. Under the time evolution
$|\psi(t)\rangle=e^{-\ii tH}|\psi_0\rangle$, the chiral expectation values evolve as
\begin{equation}
  \ell_0^\sigma(t)=\ell_0^\sigma(0),
  \qquad
  \ell_{+1}^\sigma(t)=e^{-\ii t}\ell_{+1}^\sigma(0),
  \qquad
  \ell_{-1}^\sigma(t)=e^{\ii t}\ell_{-1}^\sigma(0).
  \label{eq:appendix-chiral-expectation-time-evolution}
\end{equation}
Accordingly, we can define time-dependent bulk charge observables $L_{AB}^{\mathrm{bulk}}(t)$ at global time $t$ via the right-hand sides of Eq.~\eqref{eq:appendix-inverse-chiral-bulk-charge-dictionary}, with $\ell_n^\sigma$ promoted to $\ell_n^\sigma(t)$. 
With the initial charges denoted by $L_{AB}\equiv L_{AB}^{\mathrm{bulk}}(0)$, their time evolution is 
\begin{equation}
  \begin{aligned}
    L_{0a}^{\mathrm{bulk}}(t) &= L_{0a}\cos t - L_{-1,a}\sin t, \\
    L_{-1,a}^{\mathrm{bulk}}(t) &= L_{-1,a}\cos t + L_{0a}\sin t,
  \end{aligned}
  \qquad (a=1,2), \label{eq:timebulk}
\end{equation}
while $L_{-1,0}^{\mathrm{bulk}}=E/m$ and $L_{12}^{\mathrm{bulk}}=p_\phi/m$ remain time-independent. Geometrically, $L_{AB}^{\mathrm{bulk}}(t)$ corresponds to the charges evaluated on a geodesic shifted backward in global time by $t$, and thus varies across this family of geodesics; the charges remain conserved along each individual geodesic. On the initial-data phase space at fixed reference time, this active isometry flow is governed by Hamilton's equation $\dot{L}_{AB}^{\mathrm{bulk}}=\{L_{AB}^{\mathrm{bulk}},E\}_{\rm can}$, where $E=mL_{-1,0}^{\mathrm{bulk}}$ and $\{\cdot,\cdot\}_{\rm can}$ is the canonical Poisson bracket associated with the particle action.

\section{Negative-binomial measure and Meixner polynomials}
\label{app:nb-meixner}
In this appendix, we collect several useful properties of the negative binomial distribution and the associated Meixner polynomials.

The spectral measure of a single chiral sector, as well as that of every reference state
used in the perturbative expansions, follows a negative binomial distribution:
\begin{equation}
 \mathrm{NB}(\kappa,q):\qquad
 \nu(m)=(1-q)^\kappa\,c_m\,q^m,\qquad
 c_m=\frac{(\kappa)_m}{m!},\qquad m=0,1,2,\ldots,
 \label{eq:app-nb-weight}
\end{equation}
with $\kappa>0$ and $0<q<1$. Its moment generating function, mean, and variance are given by
\begin{equation}
 \sum_{m\geq0}\nu(m)e^{pm}=\left(\frac{1-q}{1-qe^p}\right)^\kappa,\qquad
 \langle m\rangle=\frac{\kappa q}{1-q},\qquad
 \langle m^2\rangle-\langle m\rangle^2=\frac{\kappa q}{(1-q)^2}.
 \label{eq:app-nb-moments}
\end{equation}
Furthermore, the weight $\nu(m)$ obeys the simple recurrence relation
\begin{equation}
 \frac{\nu(m-1)}{\nu(m)}=\frac{m}{q\,(m+\kappa-1)},\qquad m\geq1 .
\end{equation}
The simplest way to see that the spectral measure of the pair $(H=L_0,|z\rangle)$ is governed by this distribution is to evaluate the moment generating function $\langle z|e^{p(L_0-h)}|z\rangle$, which reproduces that of $\mathrm{NB}(2h,|z|^2)$. Similarly, two aligned chiral sectors sharing the same modulus $|z|^2$ correspond to the convolution $\mathrm{NB}(4h,|z|^2)$.
With respect to the measure $\mathrm{NB}(\kappa,q)$, the orthonormal polynomials take the form 
\begin{equation}
 p_n(m)=(-1)^n\sqrt{c_n\,q^n}\;M_n(m),
 \label{eq:app-meixner-orthonormal}
\end{equation}
where $M_n(m)$ denotes the Meixner polynomial of degree $n$:
\begin{equation}
 M_n(m)\equiv M_n(m;\kappa,q)={}_2F_1\!\left(-n,-m;\kappa;1-\frac1q\right),
 \label{eq:app-meixner-def}
\end{equation}
with generating function
\be 
\sum_{n\geq0}c_nM_n(m)w^n=\left(1-\frac wq\right)^m(1-w)^{-m-\kappa}.
\ee 
Shifting $m\to m+1$ in the generating function multiplies it by
$(1-w/q)/(1-w)=1-\frac{1-q}{q}\sum_{k\geq1}w^k$. Matching powers of $w$ yields the shift identity:
\begin{equation}
 c_nM_n(m+1)=c_nM_n(m)-\frac{1-q}{q}\sum_{j=0}^{n-1}c_jM_j(m),
 \label{eq:app-meixner-shift}
\end{equation}
while the forward difference lowers the degree and raises $\kappa$,
\begin{equation}
 M_n(m+1;\kappa,q)-M_n(m;\kappa,q)=-\frac{1-q}{q}\,\frac{n}{\kappa}\,
 M_{n-1}(m;\kappa+1,q).
 \label{eq:app-meixner-difference}
\end{equation}
From the standard three-term recurrence relation of the Meixner polynomials, one obtains 
\be 
m\,p_n=b_{n+1}p_{n+1}+a_np_n+b_np_{n-1},
\ee
from which we read off the Lanczos coefficients:
\begin{equation}
 a_n=\frac{(1+q)\,n+\kappa q}{1-q},\qquad
 b_n=\frac{\sqrt{q\,n(n+\kappa-1)}}{1-q} .
 \label{eq:app-nb-lanczos}
\end{equation}
If the energy is $E_0+m$, the $a_n$ are shifted by $E_0$.

Under the time evolution $e^{-\ii tm}$ (the shift $E_0$ only contributes a phase),
the generating function gives the Krylov transition amplitudes in closed form:
\begin{equation}
 \Phi_n(t)=\sum_{m\geq0}\nu(m)\,p_n(m)\,e^{-\ii tm}
 =(-1)^n\sqrt{c_n\,q^n}\left(\frac{1-q}{1-qe^{-\ii t}}\right)^{\kappa}
 \left(\frac{1-e^{-\ii t}}{1-qe^{-\ii t}}\right)^{n}.
 \label{eq:app-nb-amplitudes}
\end{equation}
The resulting Krylov probability distribution $|\Phi_n(t)|^2$ remains negative binomial
$\mathrm{NB}(\kappa,x_t)$ with the time-dependent parameter
\begin{equation}
 x_t=\frac{4qu}{(1-q)^2+4qu},\qquad u=\sin^2\frac t2.
\end{equation}
The spread complexity is given directly by its mean:
\begin{equation}
 C(t)=\sum_{n\geq0}n\,|\Phi_n(t)|^2=\frac{\kappa x_t}{1-x_t}
 =\frac{4\kappa q}{(1-q)^2}\,u .
 \label{eq:app-nb-complexity}
\end{equation}

\section{Perturbation theory for spread complexity}
\label{app:per}
This appendix sets up perturbation theory for the spread complexity.
\subsection*{Quantum perturbation theory}
\label{app:qpt}
Let a real coupling \(g\) deform both the Hamiltonian and the normalized
initial state,
\begin{align}
 H(g)&=H_0+gH_1+g^2H_2+O(g^3),\nonumber\\
 |\psi(g)\rangle&=|\psi_0\rangle+g|\psi_1\rangle
                  +g^2|\psi_2\rangle+O(g^3).
 \label{eq:pert-input}
\end{align}
The Lanczos recursion starts from \(|K_0(g)\rangle=|\psi(g)\rangle\)
and \(b_0(g)=0\):
\begin{align}
 a_n(g)&=\langle K_n(g)|H(g)|K_n(g)\rangle,\nonumber\\
 |R_{n+1}(g)\rangle
   &=[H(g)-a_n(g)]|K_n(g)\rangle-b_n(g)|K_{n-1}(g)\rangle,\nonumber\\
 b_{n+1}(g)&=\big\||R_{n+1}(g)\rangle\big\|,\qquad
 |K_{n+1}(g)\rangle=\frac{|R_{n+1}(g)\rangle}{b_{n+1}(g)}.
 \label{eq:pert-lanczos-exact}
\end{align}
We expand \(|K_n(g)\rangle=\sum_{k\geq0}g^k|K_n^{(k)}\rangle\), and
likewise \(|R_n(g)\rangle\), \(a_n(g)\) and \(b_n(g)\). Collecting powers
of \(g\), with \(H_s=0\) for \(s>2\), gives
\begin{align}
 a_n^{(k)}&=\sum_{r+s+v=k}
    \langle K_n^{(r)}|H_s|K_n^{(v)}\rangle,\nonumber\\
 |R_{n+1}^{(k)}\rangle
   &=\sum_{s=0}^k(H_s-a_n^{(s)})|K_n^{(k-s)}\rangle
     -\sum_{s=0}^k b_n^{(s)}|K_{n-1}^{(k-s)}\rangle.
 \label{eq:pert-lanczos-convolution}
\end{align}
Through second order, the Lanczos coefficients are
\begin{align}
 a_n^{(1)}&={}\langle K_n^{(0)}|H_1|K_n^{(0)}\rangle
 +2\operatorname{Re}\langle K_n^{(1)}|H_0|K_n^{(0)}\rangle,\nonumber\\
 a_n^{(2)}&={}\langle K_n^{(0)}|H_2|K_n^{(0)}\rangle
 +2\operatorname{Re}\langle K_n^{(1)}|H_1|K_n^{(0)}\rangle\nonumber\\
 &\quad+2\operatorname{Re}\langle K_n^{(2)}|H_0|K_n^{(0)}\rangle
 +\langle K_n^{(1)}|H_0|K_n^{(1)}\rangle,\nonumber\\
 b_{n+1}^{(0)}&={}\big\||R_{n+1}^{(0)}\rangle\big\|,\qquad
 b_{n+1}^{(1)}=\frac{\operatorname{Re}\langle R_{n+1}^{(0)}|R_{n+1}^{(1)}\rangle}
                       {b_{n+1}^{(0)}},\nonumber\\
 b_{n+1}^{(2)}&={}\frac{\big\||R_{n+1}^{(1)}\rangle\big\|^2
    +2\operatorname{Re}\langle R_{n+1}^{(0)}|R_{n+1}^{(2)}\rangle
    -(b_{n+1}^{(1)})^2}{2b_{n+1}^{(0)}},
 \label{eq:pert-offdiagonal-orders}
\end{align}
and the normalization \(|R_{n+1}\rangle=b_{n+1}|K_{n+1}\rangle\) gives
the Krylov vectors
\begin{align}
 |K_{n+1}^{(0)}\rangle&=\frac{|R_{n+1}^{(0)}\rangle}{b_{n+1}^{(0)}},\nonumber\\
 |K_{n+1}^{(1)}\rangle&=\frac{|R_{n+1}^{(1)}\rangle
                  -b_{n+1}^{(1)}|K_{n+1}^{(0)}\rangle}{b_{n+1}^{(0)}},\nonumber\\
 |K_{n+1}^{(2)}\rangle&=\frac{|R_{n+1}^{(2)}\rangle
                  -b_{n+1}^{(1)}|K_{n+1}^{(1)}\rangle
                  -b_{n+1}^{(2)}|K_{n+1}^{(0)}\rangle}{b_{n+1}^{(0)}}.
 \label{eq:pert-basis-orders}
\end{align}
The expansion requires \(b_{n+1}^{(0)}\neq0\), that is, the unperturbed
Krylov chain must not terminate before level \(n+1\).

For the time evolution, write
\(e^{-\ii tH(g)}=\sum_k g^k\,\mathcal U_k[H](t)\). The Dyson series gives
\begin{align}
 \mathcal U_0[H](t)&=e^{-\ii tH_0},\nonumber\\
 \mathcal U_1[H](t)&=-\ii\int_0^t\dd s\,
      e^{-\ii(t-s)H_0}H_1e^{-\ii sH_0},\nonumber\\
 \mathcal U_2[H](t)&=-\ii\int_0^t\dd s\,
      e^{-\ii(t-s)H_0}H_2e^{-\ii sH_0}\nonumber\\
 &\quad-\int_0^t\dd s\int_0^s\dd s'\,
      e^{-\ii(t-s)H_0}H_1e^{-\ii(s-s')H_0}H_1e^{-\ii s'H_0},
 \label{eq:pert-dyson-template}
\end{align}
so that the evolved state
\(|\Psi(g,t)\rangle=e^{-\ii tH(g)}|\psi(g)\rangle\) has the coefficients
\begin{align}
 |\Psi_0\rangle&=\mathcal U_0[H]|\psi_0\rangle,\nonumber\\
 |\Psi_1\rangle&=\mathcal U_0[H]|\psi_1\rangle
                    +\mathcal U_1[H]|\psi_0\rangle,\nonumber\\
 |\Psi_2\rangle&=\mathcal U_0[H]|\psi_2\rangle
                    +\mathcal U_1[H]|\psi_1\rangle
                    +\mathcal U_2[H]|\psi_0\rangle.
 \label{eq:pert-physical-state}
\end{align}
Together with \eqref{eq:pert-basis-orders}, the Krylov amplitudes
\(\Phi_n(g,t)=\langle K_n(g)|\Psi(g,t)\rangle\) expand as
\begin{align}
 \Phi_n^{(0)}&=\langle K_n^{(0)}|\Psi_0\rangle,\nonumber\\
 \Phi_n^{(1)}&=\langle K_n^{(1)}|\Psi_0\rangle
                   +\langle K_n^{(0)}|\Psi_1\rangle,\nonumber\\
 \Phi_n^{(2)}&=\langle K_n^{(2)}|\Psi_0\rangle
                   +\langle K_n^{(1)}|\Psi_1\rangle
                   +\langle K_n^{(0)}|\Psi_2\rangle,
 \label{eq:pert-amplitudes}
\end{align}
and the spread complexity
\(C(g,t)=\sum_n n|\Phi_n(g,t)|^2
=C^{(0)}(t)+gC^{(1)}(t)+g^2C^{(2)}(t)+O(g^3)\) has
\begin{align}
 C^{(0)}&=\sum_n n|\Phi_n^{(0)}|^2,\nonumber\\
 C^{(1)}&=2\operatorname{Re}\sum_n n(\Phi_n^{(0)})^*\Phi_n^{(1)},\nonumber\\
 C^{(2)}&=\sum_n n\left[|\Phi_n^{(1)}|^2
        +2\operatorname{Re}(\Phi_n^{(0)})^*\Phi_n^{(2)}\right].
 \label{eq:pert-spread-coefficients}
\end{align}
Equivalently, one can expand the complexity operator
\(\mathcal N(g)=\sum_n n|K_n(g)\rangle\langle K_n(g)|\),
\begin{align}
 \mathcal N_0&=\sum_n n|K_n^{(0)}\rangle\langle K_n^{(0)}|,\nonumber\\
 \mathcal N_1&=\sum_n n\left(|K_n^{(1)}\rangle\langle K_n^{(0)}|
                          +|K_n^{(0)}\rangle\langle K_n^{(1)}|\right),\nonumber\\
 \mathcal N_2&=\sum_n n\left(|K_n^{(2)}\rangle\langle K_n^{(0)}|
                          +|K_n^{(0)}\rangle\langle K_n^{(2)}|
                          +|K_n^{(1)}\rangle\langle K_n^{(1)}|\right),
 \label{eq:pert-level-operator}
\end{align}
which gives
\begin{align}
 C^{(1)}={}&\langle\Psi_0|\mathcal N_1|\Psi_0\rangle
       +2\operatorname{Re}\langle\Psi_0|\mathcal N_0|\Psi_1\rangle,\nonumber\\
 C^{(2)}={}&\langle\Psi_1|\mathcal N_0|\Psi_1\rangle
       +2\operatorname{Re}\langle\Psi_0|\mathcal N_0|\Psi_2\rangle\nonumber\\
       &+2\operatorname{Re}\langle\Psi_0|\mathcal N_1|\Psi_1\rangle
       +\langle\Psi_0|\mathcal N_2|\Psi_0\rangle.
 \label{eq:pert-spread-operator-form}
\end{align}

The expansion is simpler in the fixed Krylov frame, where
\(|K_n(g)\rangle\) is identified with \(|n\rangle\) for every \(g\). The
initial state is then \(|0\rangle\), the coupling enters only through
the Lanczos coefficients, and the evolution is generated by the Jacobi
operator, while the level operator does not depend on \(g\):
\begin{align}
 J(g)&=\sum_n a_n(g)|n\rangle\langle n|
  +\sum_n b_{n+1}(g)\left(|n+1\rangle\langle n|+|n\rangle\langle n+1|\right),
  \nonumber\\
 \mathcal{N}&=\sum_n n|n\rangle\langle n|.
 \label{eq:app-pert-jacobi-definition}
\end{align}
Write \(J(g)=J_0+gJ_1+g^2J_2+O(g^3)\), where \(J_r\) is built from the
coefficients \(a_n^{(r)}\) and \(b_{n+1}^{(r)}\) of
\eqref{eq:pert-offdiagonal-orders}, and define
\(J_{r,I}(s)=e^{\ii sJ_0}J_re^{-\ii sJ_0}\),
\(\mathcal{N}_I(t)=e^{\ii tJ_0}\mathcal{N}e^{-\ii tJ_0}\) and
\(\langle A\rangle_0=\langle0|A|0\rangle\). Expanding
\(C(g,t)=\langle0|e^{\ii tJ(g)}\mathcal N e^{-\ii tJ(g)}|0\rangle\) then gives
\begin{align}
 C^{(1)}(t)&=\ii\int_0^t\dd s\,
       \langle[J_{1,I}(s),\mathcal{N}_I(t)]\rangle_0,\nonumber\\
 C^{(2)}(t)&=\ii\int_0^t\dd s\,
       \langle[J_{2,I}(s),\mathcal{N}_I(t)]\rangle_0\nonumber\\
 &\quad-\int_0^t\dd s\int_0^s\dd s'\,
       \langle[J_{1,I}(s'),[J_{1,I}(s),\mathcal{N}_I(t)]]\rangle_0.
 \label{eq:pert-kubo-response}
\end{align}
In this form, each correction to the spread complexity is an expectation
value of nested commutators in the unperturbed state. Such expectation
values can often be reduced to generator expectation values, which
Appendix~\ref{app:embedding-charge-dictionary} converts into bulk
charges; this is what gives the corrections a bulk interpretation.

\subsection*{Jacobi flow}
\label{app:spectral-flow}
Following the standard treatment for measure deformations (see, e.g.,~\cite{VanAssche:2022toda}), we derive here the flow equation for the Jacobi matrix $J(g)$, from which the expansion $J(g)=\sum_{r\ge0} g^r J_r$ can be constructed recursively.

We consider the energy $E$ representation. The real orthonormal polynomials $p_n^{(g)}(E)$ associated with the measure $\mu_g$  satisfy
\bea 
 &&\sum_m\mu_g(m)\,p_k^{(g)}(E_m)\,p_n^{(g)}(E_m)=\delta_{kn},\nn
 &&E\,p_n^{(g)}(E)=b_{n+1}(g)p_{n+1}^{(g)}(E)+a_n(g)p_n^{(g)}(E)+b_n(g)p_{n-1}^{(g)}(E),
 \label{eq:spec-op}
\eea 
so that the Jacobi matrix $J(g)$ represents multiplication by $E$ in the basis $\{p_n^{(g)}\}$. To track the $g$-dependence within the fixed reference Hilbert space $L^2(\mu_0)$, we absorb the measure deformation into the basis functions by defining
\begin{equation}
 \varphi_n(g;m)=\sqrt{\frac{\mu_g(m)}{\mu_0(m)}}\;p_n^{(g)}(E_m),\label{eq:varphin}
\end{equation}
which remain orthonormal with respect to $\mu_0$ for all $g$, i.e., $\langle\varphi_k,\varphi_n\rangle_0=\langle p_k^{(g)},p_n^{(g)}\rangle_g=\delta_{kn}$. Assuming completeness, the $g$-evolution of $\varphi_n$ is generated by an antisymmetric matrix $\mathsf B(g)$:
\be 
\partial_g\varphi_n(g;m)=-\sum_k\mathsf B_{kn}(g)\,\varphi_k(g;m),\qquad \varphi_n(0;m)=p_n^{(0)}(E_m),
\ee 
with matrix elements
\bea 
 \mathsf B_{kn}(g)&=&-\langle\varphi_k,\partial_g\varphi_n\rangle_0
 =-\frac12\,\mathsf L_{kn}(g)-\langle p^{(g)}_k,\dot p^{(g)}_n\rangle_g,\\
 \mathsf L_{kn}(g)&\equiv&\sum_m\mu_g(m)\,p_k^{(g)}(E_m)p_n^{(g)}(E_m)\,
 \partial_g\log\mu_g(m),
 \label{eq:spec-B-two-pieces}
\eea 
where $\dot p_n^{(g)}\equiv\partial_gp_n^{(g)}$. Indeed, differentiating $\langle\varphi_k,\varphi_n\rangle_0=\delta_{kn}$ with respect to $g$ gives
\begin{equation}
\langle\dot p^{(g)}_k,p^{(g)}_n\rangle_g+\langle p^{(g)}_k,\dot p^{(g)}_n\rangle_g+\mathsf L_{kn}
=\langle\dot{\varphi}_k,\varphi_n\rangle_0+\langle\varphi_k,\dot{\varphi}_n\rangle_0=0,
\label{eq:spec-orth-derivative}
\end{equation}
confirming that $\mathsf B_{kn}(g)=-\mathsf B_{nk}(g)$ and in particular $\mathsf B_{nn}(g)=0$. Moreover, because $\dot p_n^{(g)}$ is a polynomial of degree at most $n$, it is orthogonal to $p_k^{(g)}$ whenever $k>n$. Thus $\langle p_k^{(g)},\dot p_n^{(g)}\rangle_g=0$ for $k>n$, which fixes the strictly lower triangular entries to be $\mathsf B_{kn}=-\frac12\mathsf L_{kn}$ ($k>n$). By antisymmetry, the strictly upper triangular entries ($k<n$) are $\mathsf B_{kn}=+\frac12\mathsf L_{kn}$. In matrix notation,
\begin{equation}
 \mathsf B(g)=\frac12\,\mathcal S\big(\mathsf L(g)\big)\equiv \frac{1}{2}\big(\mathsf L_+-\mathsf L_-\big),
 \label{eq:spec-flow-generator}
\end{equation}
where $\mathcal S$ denotes the antisymmetrization projector.

Since multiplication by $E$ acts on $\varphi_n$ as $E_m\varphi_n(g;m)=\sum_l J_{ln}(g)\varphi_l(g;m)$, differentiating the matrix element $J_{kn}(g)=\langle\varphi_k, E\varphi_n\rangle_0$ with respect to $g$ yields
\begin{equation}
 \frac{\dd J_{kn}}{\dd g}
 =\langle\partial_g\varphi_k,E\varphi_n\rangle_0+\langle\varphi_k,E\partial_g\varphi_n\rangle_0
 =\sum_l\big(-\mathsf B_{lk}J_{ln}-J_{kl}\mathsf B_{ln}\big)
 =\big[\mathsf B(g),J(g)\big]_{kn},
\end{equation}
where we used $\mathsf B_{lk}=-\mathsf B_{kl}$ and the symmetry $J_{lk}=J_{kl}$. We thus arrive at the Lax equation
\begin{equation}
 \frac{\dd J(g)}{\dd g}=\big[\mathsf B(g),J(g)\big].
 \label{eq:spec-flow}
\end{equation}

For exponential deformations $\mu_g=e^{gV(E)}\mu_0$, \eqref{eq:spec-flow} reduces to the standard Toda lattice when $V(E)$ is linear in $E$, while higher-degree polynomials $V(E)$ generate the higher flows of the Toda hierarchy~\cite{Aptekarev:1997toda,VanAssche:2022toda}. In the context of Krylov complexity, the linear case governs the Euclidean-time evolution of Lanczos coefficients~\cite{Dymarsky:2019toda}, whereas reweighting the initial state by $e^{-f(H)/2}$ with a quadratic $f(H)$ triggers the first two Toda flows~\cite{Takahashi:2025toda}. To leading order in $g=\jmath^2$, our deformation falls into this quadratic class because $\mathcal V_1$ in \eqref{eq:spec-small-j-V1} is quadratic in $E$; however, this correspondence holds only infinitesimally along the tangent direction at $g=0$, as the exponentiated weight $\mu_0 e^{g\mathcal V_1}$ itself ceases to be normalizable for $g>0$. Beyond leading order, $\log(\mu_g/\mu_0)$ is neither polynomial in $E$ nor linear in $g$, meaning that \eqref{eq:spec-flow} traces a non-autonomous trajectory in the space of Toda times.

To solve \eqref{eq:spec-flow} perturbatively, we expand the log-likelihood ratio in powers of $g$:
\begin{equation}
 \log\frac{\mu_g(m)}{\mu_0(m)}=\sum_{q\ge1}g^q\,\mathcal V_q(E_m)
 \quad\Longrightarrow\quad
 \partial_g\log\mu_g(m)=\sum_{q\ge1}q\,g^{q-1}\mathcal V_q(E_m).
 \label{eq:spec-log-weight}
\end{equation}
Because $E_m^j\varphi_n(g;m)=\sum_l [J(g)^j]_{ln}\varphi_l(g;m)$, each polynomial (or power series) $\mathcal V_q(E_m)$ translates in the $\varphi_n$ basis into the matrix function $\mathcal V_q(J(g))$, giving
\begin{equation}
 \mathsf L(g)=\sum_{q\ge1}q\,g^{q-1}\,\mathcal V_q\big(J(g)\big).
 \label{eq:spec-L-of-J}
\end{equation}
Note that any $E$-independent constant in $\mathcal V_q$ (arising, for instance, from the normalization of $\mu_g$) produces a multiple of the identity matrix, which drops out under the projection $\mathcal S$. Substituting the power series $J(g)=\sum_{r\ge0}g^rJ_r$ and $\mathsf B(g)=\sum_{s\ge0}g^s\mathsf B_s$ into \eqref{eq:spec-flow} and matching powers of $g^r$, we obtain the recursion relations
\begin{equation}
 J_{r+1}=\frac1{r+1}\sum_{s=0}^r\big[\mathsf B_s,J_{r-s}\big],\qquad
 \mathsf B_s=\frac12\,\mathcal S\Big([g^s]\sum_{q\ge1}q\,g^{q-1}
 \mathcal V_q\big(J(g)\big)\Big),
 \label{eq:spec-recursion}
\end{equation}
where $[g^s](\cdots)$ extracts the coefficient of $g^s$. In $[g^s]\mathsf L(g)$, a given term $q\,g^{q-1}\mathcal V_q(J(g))$ contributes only for $q\le s+1$ via the $\mathcal O(g^{s-q+1})$ piece of $\mathcal V_q(J(g))$, which depends solely on $J_0,\dots,J_{s-q+1}$. Consequently, $\mathsf B_s$ is completely determined by $J_0,\dots,J_s$, allowing \eqref{eq:spec-recursion} to be solved order by order. Explicitly, for the first two orders we have $[g^0]\mathsf L=\mathcal V_1(J_0)$ and $[g^1]\mathsf L=\mathcal V_1^{(1)}+2\mathcal V_2(J_0)$, where $\mathcal V_1^{(1)}\equiv [g^1]\mathcal V_1(J(g))$, yielding
\begin{align}
 \mathsf B_0&=\frac12\,\mathcal S\big(\mathcal V_1(J_0)\big),&
 J_1&=[\mathsf B_0,J_0],\nonumber\\
 \mathsf B_1&=\frac12\,\mathcal S\big(\mathcal V_1^{(1)}+2\mathcal V_2(J_0)\big),&
 J_2&=\frac12\big([\mathsf B_0,J_1]+[\mathsf B_1,J_0]\big).
 \label{eq:spec-first-orders}
\end{align}
\section{Rational spectral transformations}
\label{app:christoffel}
In this appendix, we briefly review the rational spectral transformations of orthogonal polynomials. We focus on polynomial multipliers, known as Christoffel transformations; general rational multipliers can be treated by combining Christoffel transformations with their inverses. For a comprehensive treatment of these constructions and other measure deformations, we refer the reader to~\cite{Ismail_2005}.

Let $\mu$ be a positive probability measure with finite moments, and let $\Pi_n(m)=m^n+\cdots$ be its monic orthogonal polynomials. Their squared norms and the corresponding orthonormal polynomials are given by
\begin{equation}
 \mathcal N_n=\int \Pi_n(m)^2\,\dd\mu(m),\qquad
 p_n(m)=\frac{\Pi_n(m)}{\sqrt{\mathcal N_n}},\qquad \mathcal N_0=1.
 \label{eq:christoffel-reference}
\end{equation}
Given a real polynomial ${\mathcal M}(m)$ that is nonnegative on the support of $\mu$, we define the deformed measure
\begin{equation}
 \dd\widehat\mu(m)={\mathcal M}(m)\,\dd\mu(m) .
 \label{eq:christoffel-measures}
\end{equation}
Assuming the roots in ${\mathcal M}(m)=\ell\prod_{i=1}^{M}(m-m_i)$ are distinct, we define the $M\times M$ determinant
\begin{equation}
 D_n=\det\big[\Pi_{n+j-1}(m_i)\big]_{i,j=1}^{M} .
 \label{eq:christoffel-root-determinant}
\end{equation}
Whenever $D_n\neq0$, the transformed monic orthogonal polynomial is given by the multi-point Christoffel formula:
\begin{equation}
 \widehat\Pi_n(m)=\frac{1}{D_n\prod_{i=1}^{M}(m-m_i)}
 \begin{vmatrix}
 \Pi_n(m_1)&\Pi_{n+1}(m_1)&\cdots&\Pi_{n+M}(m_1)\\
 \vdots&\vdots&&\vdots\\
 \Pi_n(m_M)&\Pi_{n+1}(m_M)&\cdots&\Pi_{n+M}(m_M)\\
 \Pi_n(m)&\Pi_{n+1}(m)&\cdots&\Pi_{n+M}(m)
 \end{vmatrix} .
 \label{eq:christoffel-multiple}
\end{equation}
The determinant in the numerator vanishes at each root $m=m_i$, rendering the ratio a polynomial of degree $n$. 
Expanding the last row shows that dividing the numerator by $D_n$ yields the linear combination
\begin{equation}
 \frac{1}{\ell}{\mathcal M}(m)\widehat\Pi_n(m) = \Pi_{n+M}(m)+\sum_{j=0}^{M-1}c_{n,j}\Pi_{n+j}(m).
 \label{eq:christoffel-numerator}
\end{equation}
For any polynomial $s(m)$ of degree less than $n$, multiplying $\widehat\Pi_n(m)$ by ${\mathcal M}(m)$ thus produces a linear combination of $\Pi_{n},\ldots,\Pi_{n+M}$, which is automatically orthogonal to $s(m)$ with respect to $\dd\mu$. Furthermore, since $\widehat\Pi_n-\Pi_n$ has degree less than $n$, orthogonality with respect to $\dd\widehat\mu$ allows us to replace one factor of $\widehat\Pi_n$ by $\Pi_n$ in the transformed norm $\widehat{\mathcal N}_n=\int \widehat\Pi_n(m)^2\,\dd\widehat\mu(m)=\int \widehat\Pi_n(m)\Pi_n(m){\mathcal M}(m)\,\dd\mu(m)$. Substituting \eqref{eq:christoffel-numerator} and cofactor-expanding the determinant along the last row gives
\begin{equation}
 {\widehat{\mathcal N}_n=\ell c_{n,0}\mathcal N_n
 =\ell(-1)^M\frac{D_{n+1}}{D_n}\mathcal N_n ,}
 \label{eq:christoffel-multiple-norm}
\end{equation}
where $(-1)^M$ accounts for the cofactor sign of the $\Pi_n(m)$ entry in the bottom-left corner of the determinant. 

Finally, if a root $m_i$ has multiplicity $k>1$, the corresponding evaluation rows in both determinants are replaced by derivatives of orders $0,1,\ldots,k-1$ at $m_i$. When ${\mathcal M}(m)$ has complex roots, they occur in conjugate pairs; interchanging a conjugate pair flips the signs of both determinants, leaving $\widehat\Pi_n(m)$ real and invariant. Degenerate cases where $D_n$ or $\widehat{\mathcal N}_n$ vanishes require separate treatment, which we omit here.

\bibliographystyle{unsrturl}
\bibliography{freeref}

\end{document}